\documentclass[submission, PhysLectNotes]{SciPost}

\hypersetup{
    colorlinks,
    linkcolor={red!50!black},
    citecolor={blue!50!black},
    urlcolor={blue!80!black}
}

\usepackage{caption}
\usepackage{subcaption}

\usepackage{amssymb}
\usepackage{graphicx}
\usepackage{xcolor}
\usepackage{physics}
\usepackage{enumitem}
\usepackage{hyperref}
\usepackage{dsfont}
\usepackage[most]{tcolorbox}
\tcbuselibrary{breakable,skins}
\usepackage{tabularx,booktabs}
\usepackage{siunitx}
\usepackage{tikz}

\newcommand{\br}{\mathbf{r}}
\newcommand{\bk}{\mathbf{k}}
\newcommand{\tp}{\mathrm{p}}
\newcommand{\tb}{\mathrm{B}}
\newcommand{\tH}{\mathrm{H}}

\newcommand{\trr}{\mathrm{res}}

\DeclareSymbolFont{usualmathcal}{OMS}{cmsy}{m}{n}
\DeclareSymbolFontAlphabet{\mathcal}{usualmathcal}

\fancypagestyle{SPstyle}{
\fancyhf{}
\lhead{\colorbox{scipostblue}{\bf \color{white} ~SciPost Physics Lecture Notes }}
\rhead{{\bf \color{scipostdeepblue} ~Submission }}

\fancyfoot[C]{\textbf{\thepage}}
}

\begin{document}
\pagestyle{SPstyle}
\begin{center}{\Large \textbf{Acoustic horizons and the Hawking effect in polariton fluids of light}}\end{center}

\begin{center}
Maxime J. Jacquet\textsuperscript{1$\star$},
Elisabeth Giacobino\textsuperscript{1}
\end{center}

\begin{center}
{\bf 1} Laboratoire Kastler Brossel, Sorbonne Universit\'{e}, CNRS, ENS-Universit\'{e} PSL, Coll\`{e}ge de France, Paris 75005, France
\\

${}^\star$ {\small \sf maximjacquet@gmail.com}
\end{center}

\begin{center}
\today
\end{center}

\section*{Abstract}
These lecture notes develop polariton fluids of light as programmable simulators of quantum fields on tailored curved spacetimes, with emphasis on acoustic horizons and the Hawking effect.
After introducing exciton-polariton physics in semiconductor microcavities, we detail the theoretical tools to study the mean field and the quantum hydrodynamics of this driven-dissipative quantum system.
We derive the mapping to relativistic field theories and cast horizon physics as a pseudounitary stationary scattering problem.
We present the Gaussian optics circuit that describes observables and fixes detection weights for the horizon modes in near‑ and far‑field measurements.
We provide a practical experimental toolkit (phase‑imprinted flows, coherent pump–probe spectroscopy, balanced and homodyne detection) and a step‑by‑step workflow to extract amplification, quadrature squeezing, and entanglement among correlations.
Finally, we discuss the potential of this platform to investigate open questions in quantum field theory in curved spacetime, such as near horizon effects and quasinormal modes, as well as other phenomena universal to rotating geometries, from rotational superradiance to dynamical instabilities.
We further outline the interplay between rotational superradiance and the Hawking effect, proposing to spatially resolve measurements as a roadmap for `dumb hole spectroscopy' and the study of entanglement dynamics in curved spacetimes.

\vspace{10pt}
\emph{Based on the lectures given by Elisabeth Giacobino at the Summer School "Analogue Gravity" in Benasque, Spain, in May 2023. MJJ expanded, reworked, and compiled the content based on original derivations and calculations. All figures are directly adapted from MJJ's HDR manuscript~\cite{jacquet_superradiant_2025}, except where explicitly stated otherwise.}
\newpage

\tableofcontents
\newpage
\section{Introduction}

Quantum fields evolving on curved spacetimes occupy a special place in modern physics. They furnish the language of black hole thermodynamics and early–universe cosmology, yet most of their predictions are experimentally remote. Analogue gravity proposes a complementary route: emulate the kinematics of fields on curved geometries within a controllable medium, and then interrogate those fields with the full arsenal of the laboratory. In these notes, we develop and use exciton–polariton fluids of light as a platform for this.

Planar semiconductor microcavities confine photons between distributed Bragg reflectors. When quantum wells are embedded at the antinodes, the cavity photon strongly couples to the bright exciton resonance. Normal modes are lower and upper exciton-polaritons, mixed light-matter quasiparticles with ultralight mass in the plane (set by the photonic component), and short-range interactions (set by the excitonic component). Their dispersion is approximately parabolic at small in–plane wave vectors, and, near-resonance, their driven–dissipative dynamics inherit the coherence of the pump laser while remaining sensitive to interactions and loss. As a result, polaritons offer a rare combination in condensed matter physics: a quantum fluid of light whose macroscopic phase and flow are directly set by the amplitude and phase of the driving field, yet which displays quantum hydrodynamic phenomena -- superfluidity, sound, vortices -- traditional to interacting bosons~\cite{carusotto_quantum_2013}.

The lower polariton field obeys a driven–dissipative Gross–Pitaevskii equation. 
When coherently pumped near resonance in a locally uniform configuration, the steady state is phase locked to the pump, while small fluctuations comprise the familiar Bogoliubov doublet of particle and hole-like modes.
At low wave vectors and close to the resonant working point, these excitations behave as phonons with a linear dispersion; away from resonance, a finite low–frequency gap opens and the spectrum becomes ``massive''. Pump detuning therefore plays the same role here that compressibility plays in conservative quantum fluids: it fixes the low–energy content of fluctuations and, with it, the relevant light–cone speed for information propagation~\cite{barcelo_causal_2004}.

Spatially varying the pump phase imprints a tailored flow on the polariton fluid.
Linearising around a stationary background yields a scalar phase fluctuation that obeys a Klein–Gordon equation on an effective spacetime whose metric is determined by the local flow velocity and the sound speed of the fluid~\cite{Unruh,visser_acoustic_1998}, and its null cones define the causal structure for long–wavelength excitations.
Where the component of the flow normal to a surface equals the local sound speed, an acoustic (Killing) horizon forms.
This kinematic mapping is the heart of the analogy: it guarantees Doppler shifts, horizons, and the mixing between positive and negative norm modes that underlies the Hawking effect, while leaving intact the intrinsically non-equilibrium character of the polariton medium~\cite{falque_polariton_2025}.

Polaritons provide three strategic assets. First, \emph{direct optical access}: the in–plane momentum maps to the far–field emission angle, enabling \emph{in situ} measurement of both near–field density correlations and momentum–space covariance. Second, \emph{programmable flows}: spatial light modulators and simple relay optics let one sculpt pump amplitude and phase, and hence set the background density and velocity profile that define the metric. Third, \emph{high frequency resolution}: By adjusting detuning and power, one can move between gapless and massive regimes, controlling the Hawking frequency window and the group velocities of the participating modes. Weak probe beams with controllable noise statistics can be used to monochromatically study scattering processes. These advantages complement platforms based on atomic Bose–Einstein condensates or superfluid helium, and they make microcavity polaritons a natural setting for quantum–optical probes of curved–spacetime kinematics.

Historically, the field was enabled by the observation of strong light–matter coupling in semiconductor microcavities~\cite{weisbuch_observation_1992} and the subsequent demonstrations of condensation~\cite{kasprzak_boseeinstein_2006}, superfluid flow~\cite{amo_superfluidity_2009}, and quantised vortices~\cite{lagoudakis_quantized_2008} and dark solitons~\cite{amo_polariton_2011} in exciton-polariton systems.
In parallel, the programme of analogue gravity established that small fluctuations in a moving, compressible medium can be described by a relativistic scalar field on an effective space-time, with horizons where the flow becomes transcritical~\cite{Unruh,visser_acoustic_1998}.
See~\cite{almeida_analogue_2023} for a recent historical review of the field, and~\cite{barcelo_analogue_2011} for an exhaustive review of publications.
Fluids of light then emerged as an optical realisation of these ideas in the 2000s~\cite{schutzhold_hawking_2005,Marino_proposal_2008}.
Early optical demonstrations included fibre–based horizons~\cite{philbin_fiber-optical_2008} and wave–optical Laval nozzles~\cite{elazar_all-optical_2012}; in microcavities, proposals showed how a driven polariton flow would create an acoustic horizon and yield Hawking–type mode conversion detectable in spatial correlations~\cite{Solnyshkov,Gerace,Grissins,jacquet_analogue_2022}. Subsequent experiments created stationary hydrodynamic horizons for polaritons~\cite{Nguyen,jacquet_polariton_2020,falque_polariton_2025} and developed the measurement toolkit required to access the relevant correlation functions~\cite{jacquet_superradiant_2025}.

Section~\ref{sec:polaritons-planar} assembles the lower polariton fluid from its single–particle and many–body ingredients, establishes notation, and derives the hydrodynamic and Bogoliubov descriptions. We then show how, in the long–wavelength regime, phase fluctuations realise a Klein–Gordon field on an acoustic spacetime. Section~\ref{sec:horizons-hawking} introduces horizons in tailored flows, constructs the stationary scattering problem, and explains the Hawking effect both in field–theoretic terms and in the language of Gaussian quantum optics. Section~\ref{sec:experimental-methods} details the experimental methods -- phase imprinting, pump–probe spectroscopy, balanced and homodyne detection, and camera–based covariance metrology -- linking each observable to the underlying Bogoliubov mode content. Throughout, the emphasis is on a practical bridge between the concepts of quantum field theory in curved spacetimes and the measurable quantum optics of polariton fluids of light.

\newpage
\newtcolorbox{notationbox}[1][]{
  enhanced, breakable,
  colback=white, colframe=black!50,
  boxrule=0.5pt, arc=1pt,
  left=1em, right=1em, top=0.6em, bottom=0.6em,
  fonttitle=\bfseries, title={Notation \& acronyms},
  #1
}

\begin{notationbox}

\textbf{Conventions.} Frequencies are in the pump’s rotating frame; time-harmonic fields use $e^{-i\omega t}$. We work in 2D in-plane coordinates; $\hbar$ kept explicit. Positive/negative “norm” refers to the Klein–Gordon (KG) inner product in the hydrodynamic limit.

\vspace{0.4em}
\textbf{Symbols (single–particle, mean field, hydrodynamics).}

\begin{tabularx}{\linewidth}{@{}>{\raggedright\arraybackslash}p{2.9cm}X@{}}
\toprule
\textbf{Symbol} & \textbf{Meaning} \\
\midrule
$m^\ast$ & Lower–polariton (LP) effective mass. \\
$m_{\rm ph}$ & Cavity–photon in‑plane mass (parabolic at small $k$). \\
$E_0,\ \omega_0$ & LP energy/frequency at $k=0$. \\
$\Omega_\mathrm{R}$ & Vacuum Rabi frequency (photon–exciton coupling). \\
$C_k,\ X_k$ & Hopfield photon/exciton amplitudes, $C_k^2+X_k^2=1$. \\
$\omega_p,\ k_p$ & Pump frequency and in‑plane wave vector. \\
$n_0,\ n_{\rm res}$ & LP density; dark‑exciton reservoir density. \\
$g,\ g_{\rm res}$ & LP–LP and LP–reservoir interaction strengths. \\
$\gamma,\ \gamma_{\rm res},\ \gamma_{\rm in}$ & LP radiative loss; reservoir decay; LP $\to$ reservoir scattering. \\
$\delta(k_p),\ \delta(v)$ & Effective detuning in the plane‑wave / local (LPA) description. \\
$\boldsymbol v_0$ & Background flow, $\boldsymbol v_0=\hbar\nabla\phi_p/m^\ast$. \\
$c_B$ & Acoustic (``light‑cone'') speed; at resonance reduces to $c_s$. \\
$\xi$ & Healing length, $\xi=\sqrt{\hbar/[m^\ast(2gn_0-\delta)]}$. \\
$m_{\rm det}$ & Mass parameter setting the low–$k$ gap; $\Delta\equiv m_{\rm det}c_B^2/\hbar$. \\
$D_t$ & Convective derivative, $D_t\equiv \partial_t+\boldsymbol v_0\!\cdot\nabla$. \\
\bottomrule
\end{tabularx}

\vspace{0.4em}
\textbf{Linear response, BdG and effective geometry.}

\begin{tabularx}{\linewidth}{@{}>{\raggedright\arraybackslash}p{2.9cm}X@{}}
\toprule
$u, v$ & Bogoliubov amplitudes; mode spinor $w=(u,v)^{\mathsf T}$. \\
$L,\ L^\ddagger$ & BdG operator and its $\sigma_3$‑adjoint; biorthogonal modes. \\
$\sigma_3$ & Symplectic metric, $\mathrm{diag}(1,-1)$. \\
$q^{\mu\nu}$ & Inverse acoustic metric density (see Eq.~(36) in the main text). \\
$\kappa$ & Surface gravity at the acoustic horizon (Eq.~(39)). \\
$\omega_{\min},\ \omega_{\max}$ & Hawking window: mass threshold and dispersive cut‑off. \\
\bottomrule
\end{tabularx}

\vspace{0.4em}
\textbf{Scattering channels and detection weights.}

\begin{tabularx}{\linewidth}{@{}>{\raggedright\arraybackslash}p{2.9cm}X@{}}
\toprule
H, P, W & Outgoing channels: \emph{H} (Hawking radiation, upstream, $+$ norm), \emph{P} (partner, downstream, $-$ norm), \emph{W} (witness, downstream, $+$ norm). \\
$S(\omega)$ & Stationary scattering matrix; pseudo‑unitary with respect to $\sigma_3$. \\
$D$ & Density (near‑field) weight $D\equiv u+v$. \\
$N$ & Number (far‑field) weight $N\equiv |u|^2+|v|^2$. \\
$A_{ij}$ & Anomalous pair weight $A_{ij}\equiv u_i v_j+v_i u_j$. \\
$T(\omega)$ & Greybody transmission to the far field. \\
$r(\omega)$ & Two–mode squeezing parameter (Gaussian optics map). \\
\bottomrule
\end{tabularx}

\vspace{0.4em}
\textbf{Acronyms.}

\begin{tabularx}{\linewidth}{@{}>{\raggedright\arraybackslash}p{2.9cm}X@{}}
\toprule
LP/UP & Lower/upper polariton. \\
ddGPE & driven–dissipative Gross–Pitaevskii equation. \\
BdG & Bogoliubov–de Gennes. \\
KG & Klein–Gordon (effective field on the acoustic metric). \\
WKB & Wentzel–Kramers–Brillouin (local analysis). \\
LPA & Local Phase‑Locking Approximation. \\
TE–TM (LT) & Longitudinal–transverse polarisation splitting. \\
SLM & Spatial light modulator. \\
PSD/SNU & Power spectral density / shot‑noise unit. \\
LO/RF (RBW/VBW) & Local oscillator / radio frequency; analyser bandwidths. \\
EFT/QFT & Effective / quantum field theory. \\
\bottomrule
\end{tabularx}

\end{notationbox}
\newpage
\section{The polariton fluid of light}
\label{sec:polaritons-planar}

In this section, we build the notion of a polariton quantum fluid of light in planar semiconductor microcavities, starting from the single–particle optics of confined cavity photons and the 2D phenomenology of quantum-well excitons, then introduce strong coupling and the polariton basis. We set the notation and kinematics, and develop the driven–dissipative polariton dynamics: the equation of state, its hydrodynamic form, and the linear Bogoliubov theory. Finally, we recast collective excitations as a Klein–Gordon field propagating on an acoustic space-time.

A good reference for Sections~\ref{subsec:cavity-photon-mass} and~\ref{subsec:excitons-strong-coupling} is the book~\cite{kavokin_cavity_2003}.
Section~\ref{subsec:ddgpe} is well covered in the review~\cite{carusotto_quantum_2013} while the collective excitations~\ref{subsec:bogoliubov} (Bogoliubov treatment~\ref{subsubsec:collective-excitations} and effective field theory~\ref{subsubsec:KG-analogy}) are treated in the most modern way in~\cite{falque_polariton_2025} and~\cite{guerrero_multiply_2025}.
We will not repeatedly cite these references when results are canonical as in Sections~\ref{subsec:cavity-photon-mass},~\ref{subsec:excitons-strong-coupling} and~\ref{subsec:ddgpe}.

\subsection{Cavity photons}
\label{subsec:cavity-photon-mass}

We consider a single transverse optical mode confined between two planar Bragg mirrors of effective refractive index $n_0$ and separated by a distance $\ell_z$. Quantisation along $z$ fixes $q_z=\pi M/\ell_z$ for integer $M$. For a given longitudinal index $M$, the in–plane photon dispersion reads
\begin{equation}
E_\gamma(k)=\frac{\hbar c}{n_0}\sqrt{q_z^2+k^2}
\simeq E_0 + \frac{\hbar^2 k^2}{2\,m_{\rm ph}},
\qquad
E_0\equiv \frac{\hbar c\,q_z}{n_0},
\label{eq:cav-disp}
\end{equation}
which is parabolic for $k\ll q_z$. Confinement endows the photon with an in–plane \emph{effective mass}
\begin{equation}
m_{\rm ph}=\frac{\hbar q_z}{c/n_0}=\frac{E_0}{(c/n_0)^2},
\label{eq:ph-mass}
\end{equation}
typically a factor $10^{-5}$--$10^{-4}$ of the free-electron mass in GaAs platforms. Neglecting the small TE--TM splitting, the second–quantised Hamiltonian is
\begin{equation}
\hat H_\gamma=\int d^2\mathbf r\,\hat a^\dagger(\mathbf r)
\Big(E_0-\frac{\hbar^2\nabla_{\!\mathbf r}^2}{2m_{\rm ph}}\Big)\hat a(\mathbf r),
\qquad
\hat a(\mathbf r)=\int\!\frac{d^2\mathbf k}{(2\pi)^2}e^{i\mathbf k\cdot\mathbf r}\hat a_{\mathbf k}.
\label{eq:H-cav}
\end{equation}

\subsection{Quantum well excitons and strong coupling: Hopfield diagonalisation}
\label{subsec:excitons-strong-coupling}

\paragraph{2D exciton phenomenology.}
In a single quantum well (QW), the relative electron--hole motion is effectively two–dimensional. Using material parameters $\mu$ (reduced mass) and dielectric constant $\varepsilon$, the effective Bohr radius and Rydberg are
\begin{equation}
a_B^\ast=\frac{4\pi\varepsilon\hbar^2}{\mu e^2},\qquad
{\rm Ry}^\ast=\frac{\mu e^4}{2(4\pi\varepsilon)^2\hbar^2}.
\label{eq:exc-hydrogenic}
\end{equation}
The bound states are hydrogenic with energies
\begin{equation}
E_{n}=-\frac{{\rm Ry}^\ast}{(n+\tfrac{1}{2})^2},\qquad n=0,1,2,\dots,
\label{eq:2D-series}
\end{equation}
so that the $1s$ binding energy is $E_B=4\,{\rm Ry}^\ast$ and the in–plane extent is set by $a_B^\ast$ (up to order–unity QW–confinement factors).
The lowest bright exciton (centre–of–mass mass $m_X$) provides the dominant optical resonance in typical III--V QWs~\cite{ANDREANI1991641,SAVONA1995733}.

\paragraph{Linear light--matter coupling.}
Near resonance, the cavity photon couples to the $1s$ QW exciton with vacuum Rabi frequency $\Omega_\mathrm{R}$. In the weak–excitation limit, the linear Hamiltonian is
\begin{equation}
\hat H_{\rm lin}=\sum_{\mathbf k}\!\left[
E_X\,\hat b^\dagger_{\mathbf k}\hat b_{\mathbf k}
+E_\gamma(k)\,\hat a^\dagger_{\mathbf k}\hat a_{\mathbf k}
+\frac{\hbar\Omega_\mathrm{R}}{2}\big(\hat a^\dagger_{\mathbf k}\hat b_{\mathbf k}+\hat b^\dagger_{\mathbf k}\hat a_{\mathbf k}\big)
\right],
\label{eq:H-lin}
\end{equation}
where $\hat a_{\mathbf k}$ and $\hat b_{\mathbf k}$ destroy a cavity photon and a QW exciton, respectively. Diagonalisation via the Hopfield transformation \cite{Hopfield1958} yields lower/upper polaritons (LP/UP)
\begin{equation}
\begin{pmatrix}\hat p_{\mathbf k}\\ \hat u_{\mathbf k}\end{pmatrix}
=\begin{pmatrix} C_k & X_k\\ X_k & -C_k\end{pmatrix}
\begin{pmatrix}\hat a_{\mathbf k}\\ \hat b_{\mathbf k}\end{pmatrix},
\qquad C_k^2+X_k^2=1,
\label{eq:Hopfield}
\end{equation}
with dispersions~\cite{arakawa_polariton_1992}
\begin{equation}
E_{\rm LP/UP}(k)=\frac{E_X+E_\gamma(k)}{2}\mp
\frac{1}{2}\sqrt{\big[E_\gamma(k)-E_X\big]^2+\hbar^2\Omega_\mathrm{R}^2}.
\label{eq:LPUP}
\end{equation}
The photon (exciton) fractions $C_k$ ($X_k$) are fixed by detuning and $\Omega_\mathrm{R}$. At small $k$ and zero detuning, the LP inherits a light mass $m^\ast\simeq m_{\rm ph}/|C_0|^2$, while its matter fraction controls interactions and loss channels \cite{carusotto_quantum_2013,Ciuti1998,TassoneYamamoto1999}.

In all situations considered here, the pump is tuned quasi‑resonantly to the lower polariton (LP) at small in‑plane wave vectors, and the relevant energy scales kinetic energies \(\hbar^2k^2/2m_{\rm ph}\), mean field shifts \(\hbar g n\), and linewidths remain much smaller than the Rabi splitting \(\hbar\Omega_\mathrm{R}\) and than the instantaneous LP–UP separation \(E_{\mathrm{UP}}(k)-E_{\mathrm{LP}}(k)\).
Inter‑branch mixing is then negligible and the upper branch remains essentially unpopulated, while the LP combines a light effective mass (from its photonic fraction) with appreciable non-linearity (from its excitonic fraction).
Hereafter, we therefore project the dynamics onto the LP manifold, which we refer to as "the polariton field", and drop UP operators.

\begin{figure*}[t]
  \centering
  \begin{subfigure}[c]{0.5\textwidth}
    \centering
    \includegraphics[width=\linewidth]{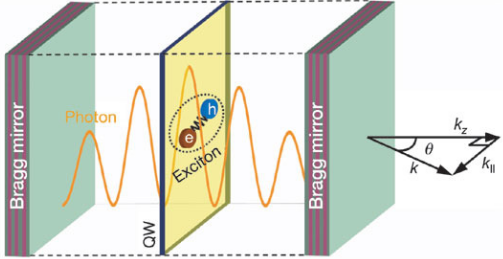}
  \end{subfigure}\hfill
  \begin{subfigure}[c]{0.50\textwidth}
    \centering
    \includegraphics[width=\linewidth]{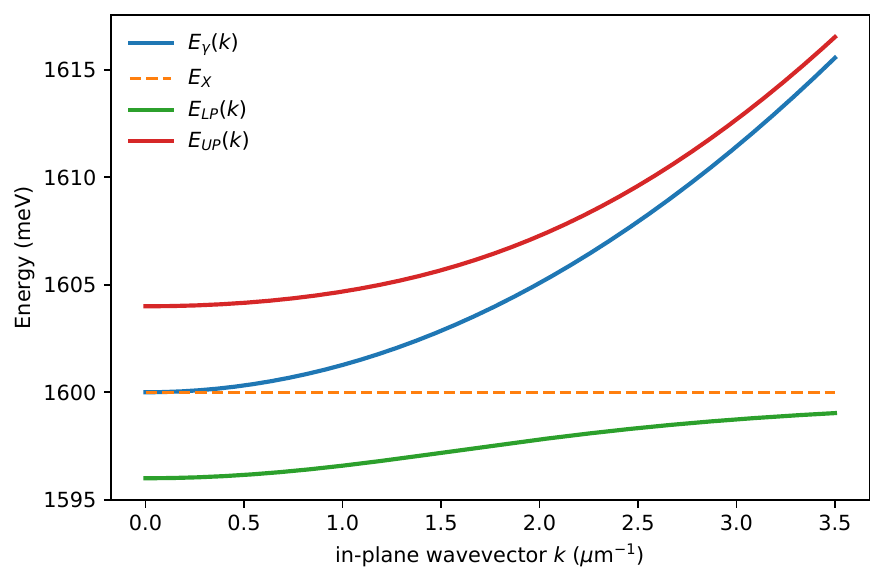}
  \end{subfigure}\\[4pt]
  \begin{subfigure}[c]{0.5\textwidth}
    \centering
    \includegraphics[width=\linewidth]{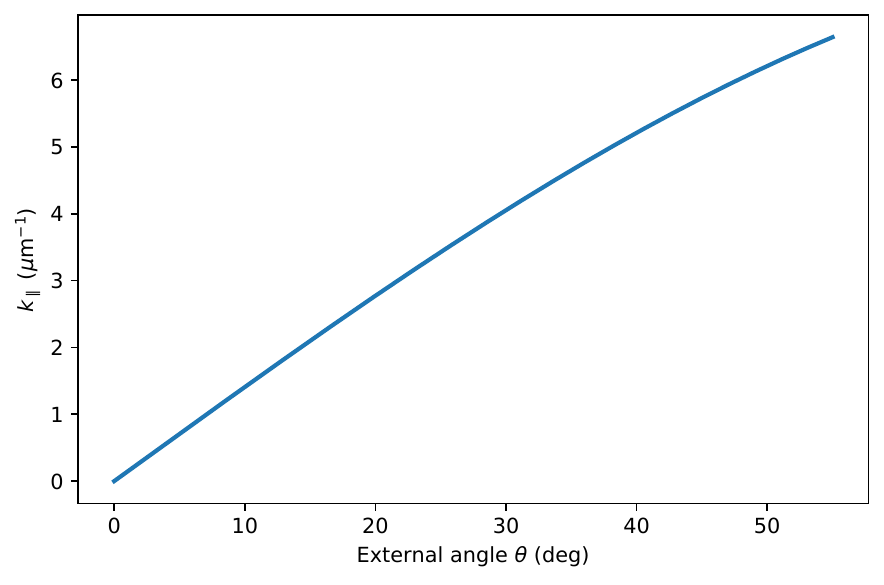}
  \end{subfigure}\hfill
    \begin{subfigure}[c]{0.5\textwidth}
      \centering
      \includegraphics[width=\linewidth]{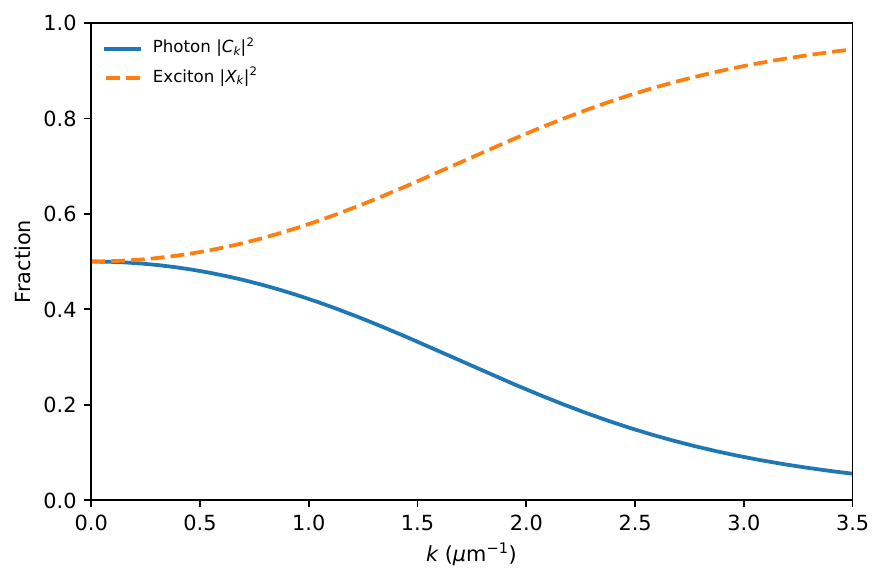}  \end{subfigure}
  \caption{\textbf{Platform and kinematics}.
  (a) Planar microcavity with DBR mirrors, embedded QWs, and a pump incident at angle $\theta$ [from~\cite{kasprzak_boseeinstein_2006}].
  (b) Dispersion overlay showing the cavity photon $E_\gamma(k)$ [Eq.~\eqref{eq:cav-disp}], the QW exciton $E_X$, and the lower/upper polariton branches $E_{LP/UP}(k)$ from Hopfield diagonalisation [Eq.~\eqref{eq:Hopfield}].
  Parameters are representative GaAs near zero detuning (consistent with App.~\ref{app:units}): $E_0\approx E_X=\SI{1.6}{\electronvolt}$, $\hbar\Omega_\mathrm{R}=\SI{8}{\milli\electronvolt}$, $m_\mathrm{ph}\sim 3\times 10^{-5}m_e$.
  (c) Angle–momentum mapping $k_\parallel=(n_{\rm ext}\omega/c)\sin\theta$ used for excitation and detection [Eq.~\eqref{eq:k-from-theta} with $n_\mathrm{ext}=1$ (air) and $\hbar\omega=\SI{1.6}{\electronvolt}$]. Representative GaAs parameters [App.~\ref{app:units}] are used for scale.}
  \label{fig:platform_kinematics}
\end{figure*}

\newtcolorbox{anglebox}[1][]{
  enhanced, breakable,
  colback=white, colframe=black!50,
  boxrule=0.5pt, arc=1pt,
  left=1em, right=1em, top=0.6em, bottom=0.6em,
  fonttitle=\bfseries, title={Incidence angle \(\leftrightarrow\) in‑plane momentum.},
  #1
}

\begin{anglebox}
Because the in‑plane component of the wave vector is conserved at a planar interface, a monochromatic plane‑wave pump at frequency \(\omega_\tp\) and external incidence angle \(\theta_{\mathrm{ext}}\) injects cavity polaritons at an in‑plane wave vector
\begin{equation}
\mathbf{k}_\tp = k_\tp\,\hat{\mathbf{e}}_\parallel,\qquad
k_\tp = \frac{n_{\mathrm{ext}}\omega_\tp}{c}\,\sin\theta_{\mathrm{ext}}
      = \frac{n_{0}\,\omega_\tp}{c}\,\sin\theta_{\mathrm{cav}},
\label{eq:k-from-theta}
\end{equation}
where \(n_{\mathrm{ext}}\) is the refractive index of the incident medium (typically~\(\simeq 1\) in air), \(n_0\) is the effective cavity index, and \(\theta_{\mathrm{cav}}\) is the internal angle, related to \(\theta_{\mathrm{ext}}\) by Snell’s law \(n_{\mathrm{ext}}\sin\theta_{\mathrm{ext}} = n_{0}\sin\theta_{\mathrm{cav}}\).
Normal incidence corresponds to \(k_\tp=0\) and thus a spatially homogeneous drive.
The same kinematic relation is used in detection: the emission at external angle \(\theta\) maps to an in‑plane momentum \(k_\parallel=(n_{\mathrm{ext}}\omega/c)\sin\theta\).
For a compact formulation within the Langevin formalism, one may write the drive as
\begin{equation}\label{eq:pump}
    F_\tp(\mathbf{r},t)=F_0\,e^{-i\omega_\tp t}\,e^{i\mathbf{k}_\tp\cdot\mathbf{r}},
\end{equation}
with \(\mathbf{k}_\tp\) the in‑plane projection of the pump wave vector.
\end{anglebox}

\paragraph{Scaling of LP--LP interactions.}
At low densities, the LP mean–field blueshift is $\Delta E_{\rm LP}\simeq g_{\rm LP}\,n_{\rm LP}$ with an effective 2D contact constant
\begin{equation}
g_{\rm LP}\;\simeq\;|X_{k_\tp}|^4\,g_{XX},
\qquad
g_{XX}\sim \alpha\,E_B\,a_B^{\ast\,2},
\label{eq:g-scaling}
\end{equation}
where $k_\tp$ is the relevant (pumped) in–plane wavevector and $\alpha=\mathcal O(1)$ encapsulates microscopic exchange and phase–space–filling contributions for $1s$ excitons in 2D \cite{Ciuti1998,TassoneYamamoto1999}. Finite–density corrections scale as $n\,a_B^{\ast\,2}$ and produce a weak renormalisation $g_{\rm LP}\!\to\!g_{\rm LP}^{\rm eff}(n)$, while additional blue–shift channels (e.g.\ saturation nonlinearity and off–resonant dark states) can be folded phenomenologically into $g_{\rm LP}^{\rm eff}$ when comparing to experiment \cite{stepanov_dispersion_2019,amelio_galilean_2020,claude2021highresolution}.

\subsection{Driven--dissipative model for the polariton fluid}
\label{subsec:ddgpe}

\subsubsection{Mean-field dynamics}
Throughout these notes we work at the level of a single, scalar lower‑polariton (LP) field \(\psi(\mathbf r,t)\) with effective mass \(m^\ast\) and short‑range repulsion \(g>0\). In the presence of continuous coherent driving and radiative losses, a convenient starting point is the driven–dissipative Gross–Pitaevskii/Langevin equation  (ddGPE) ~\cite{carusotto_parametric_threshold_2005,busch_spectrum_2014}
\begin{equation}
i\hbar\,\partial_t \psi
=
\Big[E_{\rm LP}(-i\nabla)-\hbar\omega_\tp-\tfrac{i\hbar\gamma}{2}\Big]\psi
+\hbar g\,|\psi|^2\psi
+\hbar g_{\rm res}\,n_{\rm res}\,\psi
+i\hbar\,F_\tp(\mathbf r,t)
+\hat{\zeta}(\mathbf r,t),
\label{eq:ddgpe}
\end{equation}
written in a frame rotating at the pump frequency \(\omega_\tp\). Here \(E_{\rm LP}(-i\nabla)\simeq \hbar\omega_0-\hbar^2\nabla^2/(2m^\ast)\) near \(k{=}0\) (correspondingly, $\omega_0$) is the kinetic energy of polaritons, \(\gamma\) is the spatially homogeneous decay rate set by the mirror losses~\footnote{The effective lifetime of polaritons is angle dependent and the excitonic contribution to it can become non-negligible at high angles, but for the sake of these lecture notes, we will always assume that they are dominated by photonic losses and thus spatially homogeneous and angle independent.}, \(F_\tp\) is the complex intracavity pump envelope~\eqref{eq:pump}, and \(\hat{\zeta}\) denotes the quantum noise associated with the coupling to the external electromagnetic bath (we will neglect \(\hat\zeta\) except if specified otherwise).

$g_\trr n_\trr=\beta\times gn$ ($\beta=cst\geq0$) phenomenologically accounts for possible modifications to the interaction energy under the effect of a long-lived exciton reservoir not coupled to the pump field~\cite{amelio_galilean_2020}.
We have $\partial_t n_\trr = -\gamma_\trr n_\trr + \gamma_\mathrm{in} n$. 
where $\gamma_\trr$ is the long-lived exciton reservoir decay rate and $\gamma_\mathrm{in}$ the decay rate of polaritons into the reservoir.
Taking the steady state of the reservoir rate equation shows that $n_\trr$ and $n$ are proportional through $\gamma_\trr n_\trr =  \gamma_\mathrm{in}n$.

Equation~\eqref{eq:ddgpe} is the LP‑only reduction of the standard input–output formulation for planar microcavities and will be our workhorse.

\begin{figure*}[ht]
    \centering
    \includegraphics[width=.75\linewidth]{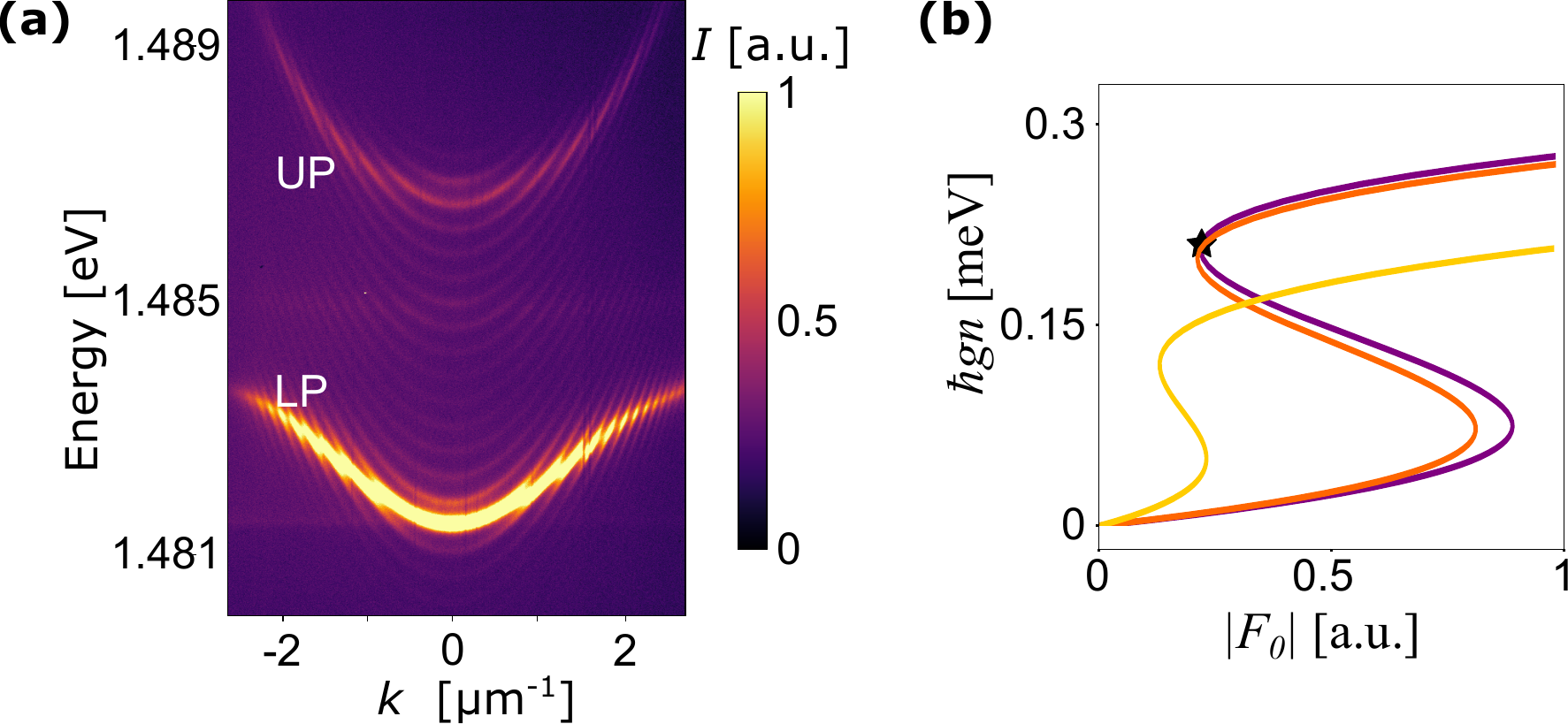}
    \caption{
    \textbf{Eigenstates and optical bistability}. In this figure, $\delta(0)= \SI{56}{\giga\hertz} \rightarrow \hbar\delta(0)=\SI{0.2}{\milli\electronvolt}$.
    \textbf{(a)} \textbf{Polariton eigenstates} in the strong coupling regime. UP, upper polaritons; LP, lower polaritons.
    \textbf{(b)} \textbf{Optical bistability of the polariton density}~\eqref{eq:eos}.
    Pump incident on the cavity at different $k_\mathrm{p}$: purple, $k_\mathrm{p}=\SI{0}{\per\micro\meter}$ (the
    black star, marks the point for which
    $gn=\delta(0)$); orange, $k_\mathrm{p}=\SI{0.11}{\per\micro\meter}$; yellow, $k_\mathrm{p}=\SI{0.55}{\per\micro\meter}$.
    Data and parameters from~\cite{falque_polariton_2025}.
    \label{fig:fig2}}
\end{figure*}

When the cavity is driven by a plane wave of in‑plane wavevector \(\mathbf k_\tp\) and frequency \(\omega_\tp\) of the form~\eqref{eq:pump}, the meanfield can be expressed as [see paragraph~\ref{subsubsec:hydro-ddgpe}]
\begin{equation}\label{eq:madelung}
\psi(\mathbf r,t)=\sqrt{n}\,e^{i(\mathbf k_\tp\!\cdot\!\mathbf r-\omega_\tp t)},
\end{equation}
Eq.~\eqref{eq:ddgpe} gives the algebraic state equation
\begin{equation}
\Big[\delta(\mathbf{k_\tp})-\hbar g\,n-\hbar g_{\rm res}n_{\rm res}+i\tfrac{\hbar\gamma}{2}\Big]\sqrt{n}
= i\hbar F_0,
\label{eq:eos}
\end{equation}
with the effective pump-polariton detuning
\begin{equation}
    \label{eq:detuning}
    \delta(\mathbf{k_\tp})\equiv \hbar\omega_\tp-\hbar\omega_0-\frac{\hbar^2 (\mathbf{k_\tp})^2}{2m^\ast}.
\end{equation}
The nonlinear character of the equation of state~\eqref{eq:eos} underlies a phenomenon known as optical bistability that arises when a coherently driven, lossy and $\chi^{(3)}$-type nonlinear cavity reaches a steady state in which the pump‑induced influx of polaritons balances radiative decay while interactions blueshift the resonance; as a result, Eq.~\eqref{eq:eos} can admit multiple steady amplitudes for the same drive~\cite{baas_bista_2004}.
For sufficiently blue detuning and above a threshold input power, the input–output curve becomes S‑shaped with two stable solutions (a dim and a bright state) separated by an unstable one, so that the system can reside on either branch at fixed pump frequency [see Fig.~\ref{fig:fig2} (b)].
When the pump power is swept, the interaction‑induced blueshift pulls the mode into or out of resonance with the fixed‑frequency drive, producing hysteresis (branch switching at the turning points) in coherently pumped polariton microcavities.

\subsubsection{A quantum fluid of light}
\label{subsubsec:hydro-ddgpe}

We write the lower-polariton field as~\eqref{eq:madelung} and
define the velocity field $\mathbf v=(\hbar/m^\ast)\nabla\theta$.
Neglecting the dark exciton reservoir and going to the frame rotating at $\omega_\tp$, the ddGPE~\eqref{eq:ddgpe} with a general coherent drive~\eqref{eq:pump} reads~\cite{carusotto_quantum_2013}
\begin{equation}
i\hbar\,\partial_t\psi=
\Big[-\frac{\hbar^2\nabla^2}{2m^\ast}-\hbar\delta(\mathbf{k_\tp})+\hbar g\,n
-i\frac{\hbar\gamma}{2}\Big]\psi
+i\hbar\,F_\tp(\mathbf r).
\label{eq:ddgpe-hydro-start}
\end{equation}
Separating real and imaginary parts gives the continuity and Euler-like equations
\begin{align}
\partial_t n+\nabla\!\cdot(n\,\mathbf v)
&=2\,|F_\tp(\mathbf r)|\,\sqrt{n}\,\cos\Delta\theta-\gamma\,n,
\label{eq:continuity}\\[4pt]
\hbar\,\partial_t\theta+\frac{m^\ast}{2}|\mathbf v|^2+\hbar g\,n-\hbar\delta
-\frac{\hbar^2}{2m^\ast}\frac{\nabla^2\!\sqrt{n}}{\sqrt{n}}
&=-\,\hbar\,|F_\tp(\mathbf r)|\,\frac{\sin\Delta\theta}{\sqrt{n}},
\label{eq:Euler}
\end{align}
where the pump--phase mismatch is
\begin{equation}
\Delta\theta(\mathbf r,t)\equiv \theta(\mathbf r,t)-\phi_\tp(\mathbf r)-\mathbf k_\tp\!\cdot\!\mathbf r.
\end{equation}
Equation~\eqref{eq:continuity} shows that the drive injects particles at a rate
$\propto\cos\Delta\theta$, while losses deplete them at rate $\gamma$~\footnote{For a spatially homogeneous pump ($|F_\tp|=\mathrm{const}$ and $\mathbf k_\tp=\mathrm{const}$), a stationary
solution has $n(\mathbf r,t)=n_0$ and $\theta(\mathbf r,t)=\mathbf k_\tp\!\cdot\!\mathbf r-\omega_\tp t+\theta_0$,
with $\Delta\theta=\mathrm{const}$ fixed by the balance of drive and loss.}.
Equation~\eqref{eq:Euler} contains the interaction strength $\hbar g\,n$, the detuning shift $-\hbar\delta$, and the
quantum pressure term $Q[n]\equiv-\frac{\hbar^2}{2m^\ast}\frac{\nabla^2\!\sqrt{n}}{\sqrt{n}}$, via which dispersive corrections enter.
$Q[n]$ adds an energy cost $\propto\abs{\nabla \sqrt{n}}^2$ to sharp density gradients, providing a dispersive regularisation that smooths variations on scales lower or equal to the healing length~\footnote{There are two characteristic lengths in the driven case: (i) the dispersive cutoff $\xi$, which organises the $k\xi\ll1$ vs $k\xi\gg1$ regimes; (ii) the mass length $\ell_m=\hbar/(m_{\rm det}c_B)$ when the spectrum is gapped. When $gn+g_\trr n_\trr=\delta(\mathbf{ k_\tp})$, $m_{\rm det}=0$ and $\ell_m\to\infty$, while $\xi$ remains finite and continues to set the crossover to the single‑particle branch.}
\begin{equation}
    \label{eq:healinglength}
    \xi=\sqrt{\hbar/m^*(2gn_0-\delta(k_\tp))}.
\end{equation}

\subsection{Collective excitations}\label{subsec:bogoliubov}
\subsubsection{The Bogoliubov-de Gennes problem}
\label{subsubsec:collective-excitations}
Given the coherent pump~\eqref{eq:pump}, the polariton meanfield is a stationary solution of Eq.~\eqref{eq:ddgpe-hydro-start} with flow velocity
\begin{equation}
    \label{eq:polaritonv}
    \mathbf v_0(\mathbf r)=\hbar\nabla\phi_0/m^\ast
\end{equation}
and local effective detuning~\eqref{eq:detuning}.
Writing the total field as
\(\psi=e^{i[\phi_0(\mathbf r)-\omega_\tp t]}\!\left(\sqrt{n_0}+e^{-i\gamma/2}\delta\psi\right)\)
and linearising in \(\delta\psi\)~\footnote{The factor $e^{-\gamma t/2}$ ensures the fluctuation decays due to loss, but it plays no essential role in the linearisation.}, one obtains the Bogoliubov–de Gennes (BdG) equations
\begin{equation}
i\hbar\,\partial_t
\begin{pmatrix}
\delta\psi\\[2pt]\delta\psi^\ast
\end{pmatrix}
=
\begin{pmatrix}
\mathcal A & \hbar g\,n_0\\[2pt]
-\hbar g\,n_0 & -\mathcal A^\ast
\end{pmatrix}
\begin{pmatrix}
\delta\psi\\[2pt]\delta\psi^\ast
\end{pmatrix},
\label{eq:BdG-inhom}
\end{equation}
with $\mathcal A \equiv
-\hbar\delta(\mathbf k_\tp)
+2\hbar g\,n_0(\mathbf r)
+\hbar g_{\rm res}\,n_{\rm res}(\mathbf r)
+\mathcal D$ and the differential operator~\cite{falque_polariton_2025}
\begin{equation}
\mathcal D
\equiv
-\frac{\hbar^2}{2m^\ast}\nabla^2
- i\hbar\,\mathbf v_0(\mathbf r)\!\cdot\!\nabla
- \frac{i\hbar}{2}\,\nabla\!\cdot\!\mathbf v_0(\mathbf r).
\label{eq:D-operator}
\end{equation}
where $\frac{\hbar^2}{2m^\ast}\nabla^2$ and $i\hbar\,\mathbf v_0(\mathbf r)\!\cdot\!\nabla$ are the standard kinetic energy and Doppler shift.
The last term, which is proportional to $\nabla\!\cdot\!\mathbf v_0(\mathbf r)$, is characteristic of driven–dissipative systems and does not have a conservative counterpart~\footnote{In conservative fluids the quantum pressure is the only dispersive term at mean‑field level; in driven dissipative polariton fluids the same \(Q[n]\) appears, but pump and loss make the BdG operator non‑Hermitian, adding convective and local divergence terms that also contribute to the dynamics when the meanfield varies significantly~\cite{guerrero_multiply_2025}.}.

Let $\mathcal L$ be the BdG operator in~\eqref{eq:BdG-inhom} and define right/left modes by~\cite{castin_lecture_notes}
\begin{align}
\mathcal L\,w_j=\hbar\omega_j\,w_j,\qquad
&\mathcal L^\ddagger\,\tilde w_j=\hbar\omega_j\,\tilde w_j,\qquad
w_j=\begin{pmatrix}u_j\\ v_j\end{pmatrix},\quad
\tilde w_j=\begin{pmatrix}\tilde u_j\\ \tilde v_j\end{pmatrix},\nonumber\\
&\mathcal L^\ddagger\equiv \sigma_3\,\mathcal L^\dagger\,\sigma_3,\ \ \sigma_3=\mathrm{diag}(1,-1).
\end{align}
Drive and loss make $\mathcal L$ non‑Hermitian, so we use the biorthogonal normalisation and completeness,
\begin{equation}
\int d^2\mathbf r\;\tilde w_i^\dagger(\mathbf r)\,\sigma_3\,w_j(\mathbf r)=\delta_{ij},
\qquad
\sum_j w_j(\mathbf r)\,\tilde w_j^\dagger(\mathbf r')\,\sigma_3=\delta(\mathbf r-\mathbf r')\,\mathds{1},
\label{eq:biorthonorm-KG}
\end{equation}
which reduces to the conservative symplectic norm $\int (|u|^2-|v|^2)=1$ when $\gamma\to0$.

\paragraph{Hydrodynamic limit and Klein-Gordon product.}
In the long‑wavelength limit $k\xi<1$ the coupled density/phase fluctuations obey the continuity and Euler‑like equations (\S\ref{subsubsec:hydro-ddgpe}). Eliminating $\delta n$ with the linearised Euler relation (quantum pressure neglected and adiabatic elimination of the dark exciton reservoir) we obtain the barotropic closure~\cite{falque_polariton_2025}
\begin{equation}
g\,\delta n \simeq -\,\hbar\,(\partial_t+\mathbf v_0\!\cdot\!\nabla)\,\delta\theta.
\label{eq:barotropic}
\end{equation}
One finds that the BdG symplectic form for two (right) solutions $w_{1,2}$ can be written purely in terms of the phase fields:
\begin{equation}
\int d^2\mathbf r\,\Big(u_1^\ast u_2 - v_1^\ast v_2\Big)
\ \propto\
\frac{i}{g}\int d^2\mathbf r\,
\delta\theta_1^\ast\,\overleftrightarrow{D_t}\,\delta\theta_2,
\qquad
D_t\equiv \partial_t+\mathbf v_0\!\cdot\!\nabla,
\label{eq:symp-to-KG}
\end{equation}
where $f\,\overleftrightarrow{D_t}g\equiv f(D_t g)-(D_t f)\,g$.
Upon rescaling $\Phi\equiv(\hbar/\sqrt{g})\,\delta\theta$, Eq.~\eqref{eq:symp-to-KG} becomes (up to a constant factor) the Klein–Gordon inner product of relativistic bosonic field theories~\cite{WaldBook}~\footnote{Conceptually, the BdG linearisation is a Hamiltonian (canonical) system for the conjugate pair ($\delta n,\,\delta\theta$); the resulting indefinite inner product (the “symplectic norm”) is the discrete version of the KG inner product of a relativistic scalar field. In our driven-dissipative case, the BdG operator is non‑Hermitian, so we use left/right (biorthogonal) modes; but in the weak loss, long-wavelength limit — where the fluctuations obey a (massive) KG equation in the acoustic metric — the biorthogonal product reduces to the KG product (up to a constant rescaling).}
\begin{equation}
(\Phi_1,\Phi_2)_{\mathrm{KG}}
=\ i\!\int d^2\mathbf r\;\Phi_1^\ast\,\overleftrightarrow{D_t}\,\Phi_2,
\label{eq:KG-inner}
\end{equation}
which is conserved in the lossless limit and whose indefinite sign distinguishes positive from negative norm modes. 
Note that, strictly speaking, in the driven-dissipative case, \eqref{eq:biorthonorm-KG} is the appropriate extension; locally and for $\gamma\ll\Re\,\omega$, the KG product \eqref{eq:KG-inner} remains an accurate organisation principle.

\subsubsection{Spectrum of collective excitations}\label{subsubsec:bogospectrum}
For slowly varying backgrounds, a local (Wentzel-Kramers-Brillouin, WKB) analysis gives the laboratory-frame dispersion of elementary excitations at position \(\mathbf r\) and wavevector \(\mathbf k\)~\cite{falque_polariton_2025}:
\begin{equation}
\omega_\pm(\mathbf r,\mathbf k)
=
\mathbf v_0(\mathbf r)\!\cdot\!\mathbf k
\;\pm\;
\sqrt{\Big[\tfrac{\hbar^2 k^2}{2m^\ast}-\delta(\mathbf k_\tp)+2g\,n_0(\mathbf r)+g_{\rm res}\,n_{\rm res}(\mathbf r)\Big]^2
-\big[g\,n_0(\mathbf r)\big]^2}.
\label{eq:WKB-local}
\end{equation}
Equation~\eqref{eq:WKB-local} gives two branches $\omega_+(\mathbf r,\mathbf k)$ and $\omega_-(\mathbf r,\mathbf k)$.
In the weak‑loss hydrodynamic limit, these have positive and negative norm, respectively.
Note that all frequencies here are measured relative to the pump~\footnote{Remark that, because the coherent drive explicitly breaks the global U(1) symmetry, there is no Nambu-Goldstone zero mode: at generic detuning the spectrum is gapped at $k{=}0$ (finite $\Re\,\Omega_\pm$), and even at the resonant, gapless working point $g n_0+g_{\rm res}n_{\rm res}=\delta(\mathbf k_\tp)$ the $k\to0$ limit has $\Re\,\omega_\pm\!\to\!0$ \emph{but} a finite damping $\Im\,\omega_\pm=-\gamma/2$~\cite{wouters_goldstone_2007,claude_observation_2025}.
Thus, ``$\omega{=}0$'' denotes a static pattern in the pump frame, not a freely evolving undamped excitation~\cite{claude_spectrum_2023}.}.

When $g n_0+g_{\rm res}n_{\rm res}=\delta(\mathbf k_\tp)$, the dispersion~\eqref{eq:WKB-local} reduces to the familiar linear sound cone at low $k$~\cite{barcelo_causal_2004,claude_spectrum_2023}:~\footnote{Here $c_\mathrm{s}$ denotes the hydrodynamic sound speed at the resonant working point; in general we write $c_\tb$ as in Eq.~\eqref{eq:cB}.}
\begin{equation}
\omega_B^\pm \simeq \mathbf v\!\cdot\!\mathbf k \pm c_\mathrm{s} \abs{\mathbf k}\,,\qquad c_\mathrm{s}=\sqrt{\frac{\hbar g n}{m^\star}}.
\label{eq:sound}
\end{equation}

Outside the hydrodynamic regime, that is, when $g n + g_\mathrm{res}n_\mathrm{res}>\delta(\mathbf k_\tp)$, the spectrum is massive~\cite{falque_polariton_2025}.
It is then useful to rewrite~\eqref{eq:WKB-local} in a ``relativistic’’ form
\begin{equation}
\omega_B^\pm= \mathbf v\!\cdot\!\mathbf k \pm \sqrt{\left(\frac{\hbar k^2}{2m^\star}\right)^{\!2} + c_\tb^2\left(k^2+\frac{m_{\rm det}^2}{\hbar^2}\right)},
\label{eq:Bogo-rel}
\end{equation}
with
\begin{equation}
c_\tb=\sqrt{\frac{\hbar\,(2g n-\delta(k_\tp)+g_\mathrm{res}n_\mathrm{res})}{m^\star}}
\label{eq:cB}
\end{equation}
and
\begin{equation}
    \label{eq:mdet}
    m_{\rm det}= m^\star \sqrt{\frac{\big(g n-\delta(k_\tp)+g_\mathrm{res}n_\mathrm{res}\big)\,\big(3g n-\delta(k_\tp)+g_\mathrm{res}n_\mathrm{res}\big)}{\,2g n-\delta(k_\tp)+g_\mathrm{res}n_\mathrm{res}}}
\end{equation}
Eq.~\eqref{eq:Bogo-rel} fixes a maximum group velocity $|\partial\omega/\partial k|\le c_\tb$ for excitations.  

The content of $u,v$ changes with $k\xi$.
For $k\xi<1$, $\omega_\pm\simeq \mathbf v\!\cdot\!\mathbf k \pm\sqrt{(c_\tb k)^2+\Delta^2}$ with $\Delta=0$ in the hydrodynamic regime; the content of $u,v$ is phononic (phase only fluctuations), which means that $|u|\simeq|v|$.
For $k\xi>1$, $\omega_\pm\simeq \mathbf v\!\cdot\!\mathbf k \pm \hbar^2 k^2/2m^\ast$ irrespective of detuning; the content of $u,v$ is particle-like ($|u|\to1,\,|v|\to0$), and the dispersion approaches the free parabola shifted by the mean field.

Finally, note that in flowing backgrounds, the Doppler tilt can bring parts of the $\omega_+$ ($\omega_-$) sheet below (above) $\omega{=}0$, creating negative energy modes.

\subsubsection{Effective description}
\label{subsubsec:KG-analogy}
In the long-wavelength regime ($k\xi\ll1$) we neglect quantum pressure and use the linearised hydrodynamics.
Combining Eqs.~\eqref{eq:continuity} and~\eqref{eq:Euler} to eliminate $\delta n$ yields a wave equation for the phase fluctuation $\delta\theta$ (or any rescaled field proportional to it).
Defining the canonical scalar field
\begin{equation}
\Phi \equiv \frac{\hbar}{\sqrt{g}}\,\delta\theta,
\end{equation}
one obtains the covariant Klein--Gordon (KG) equation with a position-dependent mass term [see Appendix~\ref{app:metric}]:
\begin{equation}
\frac{1}{\sqrt{\abs{q}}}\partial_\mu\!\Big(\sqrt{\abs{q}}\,q^{\mu\nu}\partial_\nu \Phi\Big)
\;-\;\frac{m_{\rm det}^2(\mathbf r)}{\hbar^2}\,\Phi \;=\; 0.
\label{eq:KG-covariant}
\end{equation}
Here $q^{\mu\nu}$ is the inverse acoustic metric density (defined up to an overall conformal factor),
\begin{equation}
q^{\mu\nu}(\mathbf r)=
\begin{pmatrix}
-1 & -\mathbf v_0^{\!T}(\mathbf r)\\[2pt]
-\mathbf v_0(\mathbf r) & c_\tb^2(\mathbf r)\,\mathds{1} - \mathbf v_0(\mathbf r)\,\mathbf v_0^{\!T}(\mathbf r)
\end{pmatrix}.
\label{eq:qmunu}
\end{equation}

Up to a conformal factor (irrelevant for null cones and horizons), the corresponding line element can be written as
\begin{equation}
ds^2 \propto c_\tb^2(\mathbf r)\Big[(\mathbf v_0^2-c_\tb^2)\,dt^2 - 2\,\mathbf v_0\!\cdot\!d\mathbf x\,dt + d\mathbf x\!\cdot\!d\mathbf x\Big],
\label{eq:acoustic-metric}
\end{equation}
so that sound cones satisfy $ds^2=0$ and acoustic (Killing) horizons occur where the normal component of the flow equals the local $c_\tb$:
\begin{equation}
|\mathbf v_{0,\perp}(\mathbf r_\tH)|=c_\tb(\mathbf r_\tH)\qquad\Rightarrow\qquad \text{horizon at }\mathbf r_\tH.
\label{eq:horizon}
\end{equation}
The surface gravity (for a stationary, 1D or locally normal flow) reads
\begin{equation}
\kappa \;=\; \frac{1}{2c_\tb}\,\partial_n\!\big(c_\tb^2 - v_{0,\perp}^2\big)\Big|_{\mathbf r=\mathbf r_\tH},
\label{eq:surface-gravity}
\end{equation}
with $\partial_n$ the derivative along the horizon normal.

\paragraph{About the Klein--Gordon mapping}
The fluctuation field that we linearise (phase/density perturbations of the polariton fluid) is not a fundamental relativistic field.
In the hydrodynamic regime $k\xi\ll1$, with barotropic closure and weak losses, its equation of motion reduces to the Klein-Gordon form in the acoustic metric~\eqref{eq:acoustic-metric} so the \emph{kinematics} of long‑wavelength excitations is that of a KG scalar with light‑cone speed $c_\tb$ and drift $\mathbf v_0$.
This is an \emph{analogue} or \emph{effective} description: the underlying polariton medium is Galilean and driven--dissipative.

Null cones ($|\mathbf v_0|=c_\tb$ horizons), Doppler shifts, positive/negative KG norm, mode mixing, and pair creation (the Hawking effect) are all encoded by the KG structure and the biorthogonal product~\eqref{eq:biorthonorm-KG}.
This effective description cuts off at $k\xi\simeq 1$, after which the quantum‑pressure ($k^4$ term) cannot be neglected~\footnote{Additionially, note that far from barotropy the ``pressure'' is retarded ($g_{\rm eff}(\omega)$) and the mass/light‑cone parameters become frequency‑dependent;
dissipation breaks time‑reversal and spoils exact KG conservation (the inner product becomes biorthogonal). Also see Appendix~\ref{app:openEFT}.}.
At higher wavevectors the dispersion becomes superluminal.
The analogy is therefore \emph{kinematic}~\cite{Unruh,visser_acoustic_1998}, not gravitational: There is no Einstein dynamics and no universal coupling to all energy.

\begin{figure*}[ht]
  \centering
  \begin{subfigure}[c]{0.4\textwidth}
    \centering
    \includegraphics[width=\linewidth]{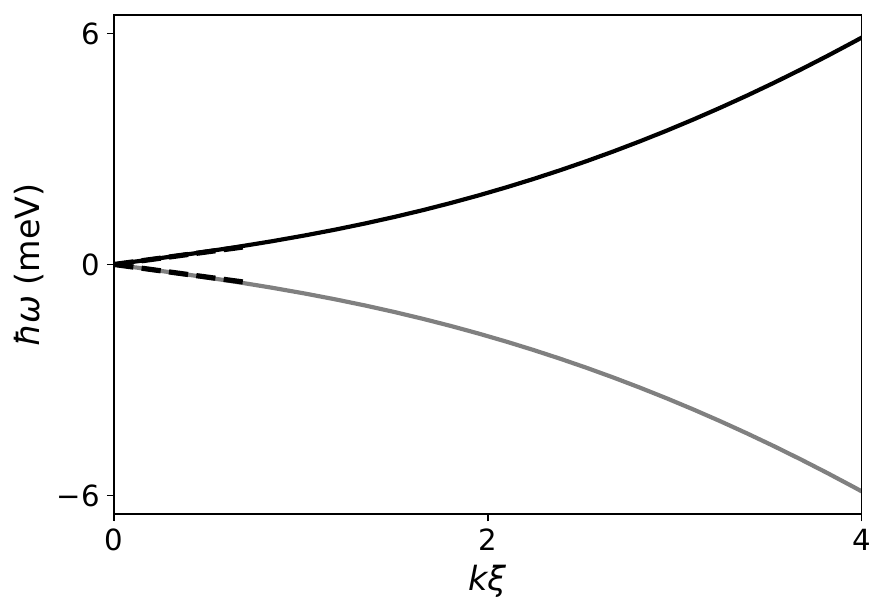}
  \end{subfigure}\hfill
  \begin{subfigure}[c]{0.4\textwidth}
    \centering
    \includegraphics[width=\linewidth]{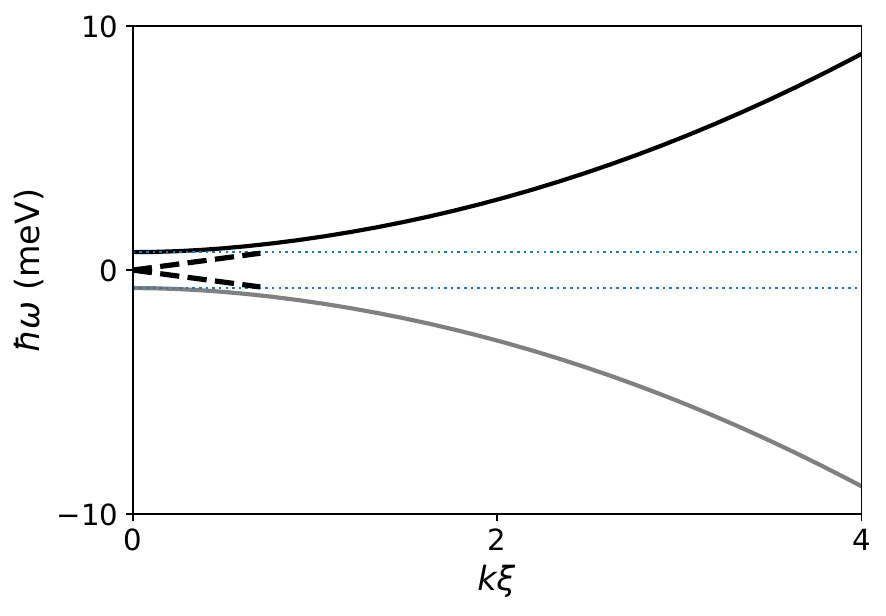}
  \end{subfigure}
  \begin{subfigure}[c]{0.4\textwidth}
    \centering
    \includegraphics[width=\linewidth]{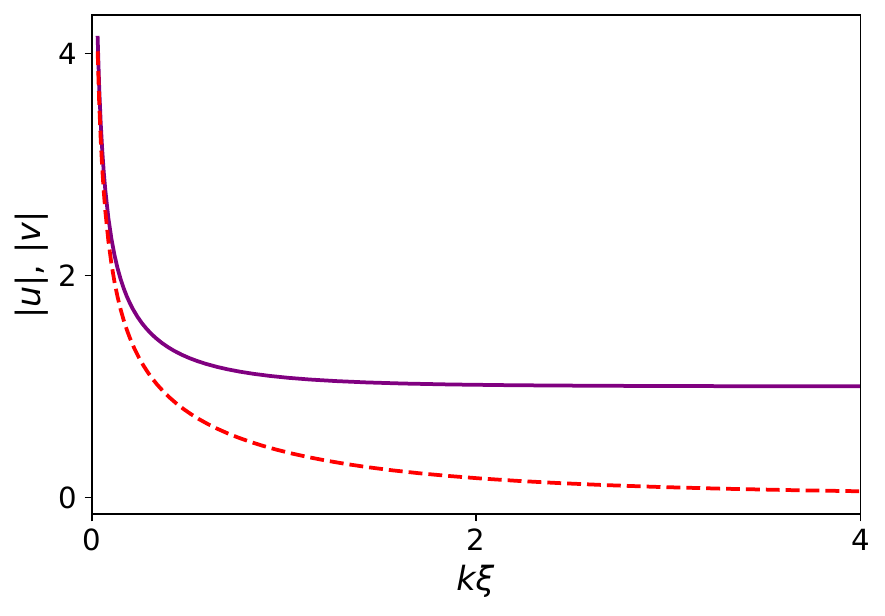}
  \end{subfigure}\hfill
  \begin{subfigure}[c]{0.4\textwidth}
    \centering
    \includegraphics[width=\linewidth]{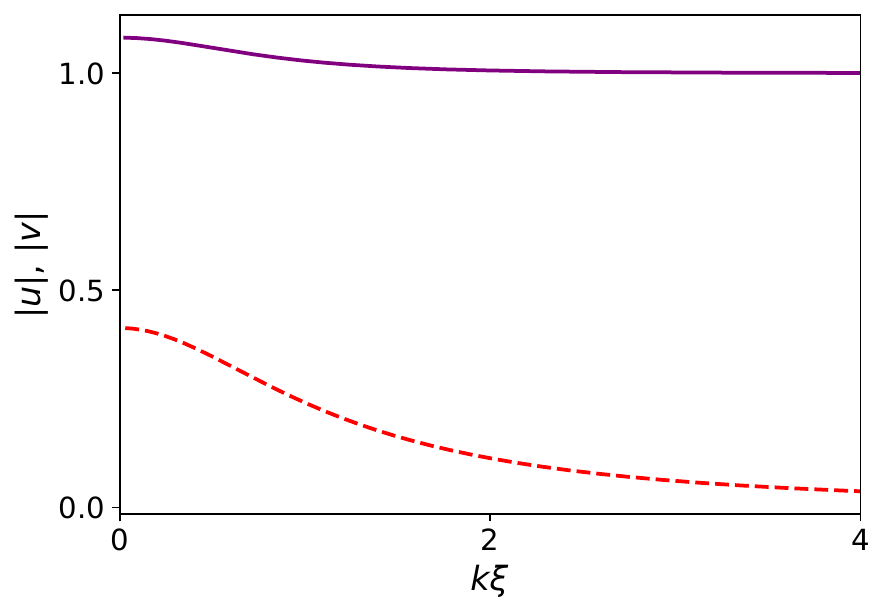}
  \end{subfigure}
  \begin{subfigure}[c]{0.4\textwidth}
    \centering
    \includegraphics[width=\linewidth]{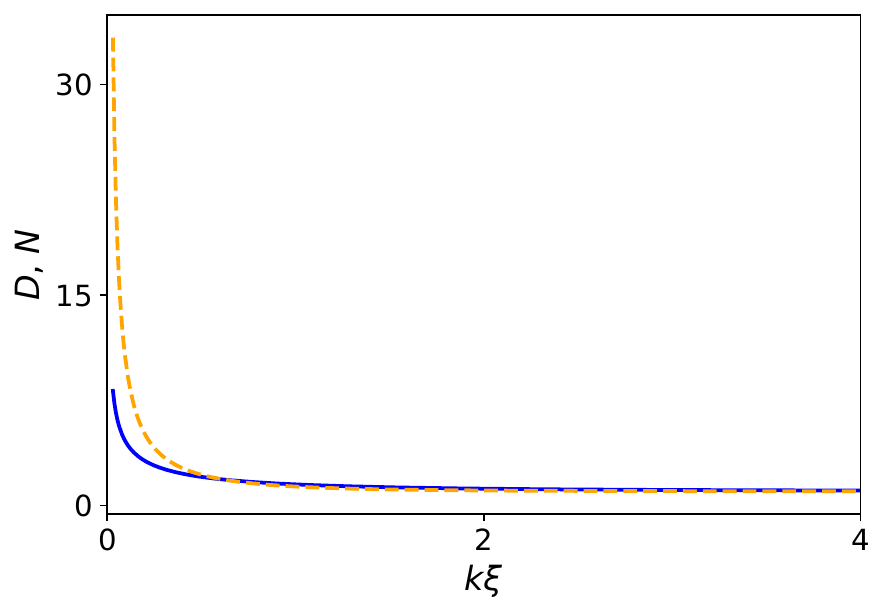}
  \end{subfigure}\hfill
  \begin{subfigure}[c]{0.4\textwidth}
    \centering
    \includegraphics[width=\linewidth]{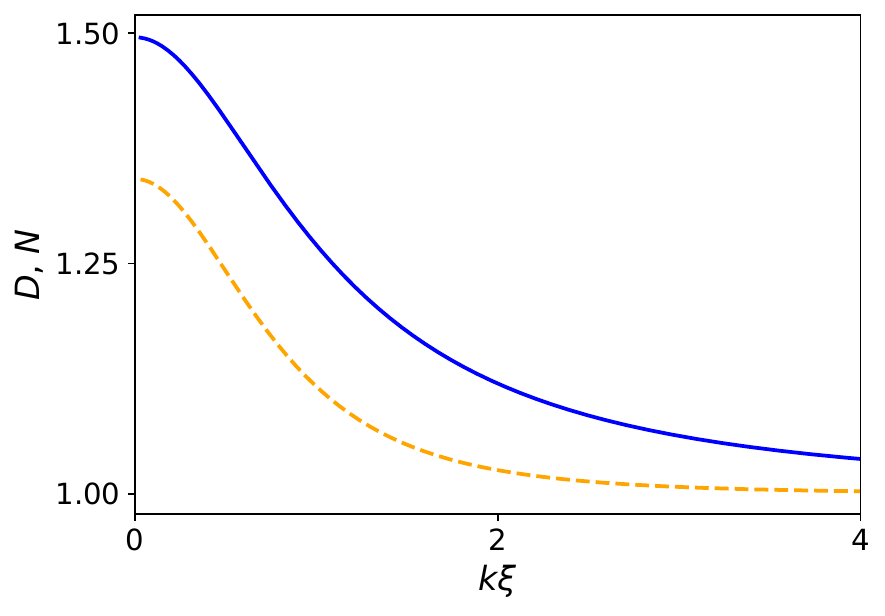}
  \end{subfigure}
  \caption{\textbf{Bogoliubov spectrum and mode content}.
  Black, positive norm; grey, negative norm.
  Dashed lines, sound cones.
  \textbf{(a)} \textbf{Resonant spectrum at $\mathbf{k_\tp}=0$}. $gn_0=\delta(0)$.
  Solid curves: $\omega_\pm(k)$~\eqref{eq:WKB-local} ($\mathbf{v}=0$).
  \textbf{(b)} \textbf{Massive spectrum} for $gn_0>\delta$.
  Solid curves: $\omega_\pm(k)$~\eqref{eq:WKB-local} ($\mathbf{v}=0$) with low-$k$ threshold $\pm\hbar\Delta_0$ (dotted lines).
  \textbf{(c,d)} \textbf{Bogoliubov mode amplitudes} $\abs{u}$ (purple) and $\abs{v}$ (red dashed) in the resonant (c) and massive (d) cases.
  \textbf{(e,f)} \textbf{Detection weights}. Near-field density $D$ (blue) and far-field number $N$ (orange dashed) in the resonant (e) and massive (f) cases.
  \label{fig:fig3}}
\end{figure*}

\subsubsection{Bogoliubov mode structure and observables}
We now characterise the Bogoliubov excitations by their $u$–$v$ (particle–hole) decomposition.
In Fig.~\ref{fig:fig3} we plot the dispersion relation~\eqref{eq:WKB-local} for a fluid at rest in the laboratory frame ($\mathbf{k_\tp}=0\rightarrow v=0$) in the resonant ($gn_0=\delta(0)$) and massive ($gn_0>\delta(0)$) cases.
As discussed in Sec.~\ref{subsubsec:bogospectrum}, the spectrum is then gapless or gapped at low $k$.
The dashed lines in (a) and (b) are the hydrodynamic cones $\hbar\omega=\pm\,\hbar c_{\mathrm{s}/\tb}k$ whose slopes are $\hbar g n_0$ (gapless) and $\hbar(2g n_0-\delta)$ (massive), respectively [Eqs.~\eqref{eq:sound} and \eqref{eq:cB}].
Note that panel~(b) exhibits a finite low‑$k$ gap $\hbar\Delta_0$ (dotted lines) with $\Delta_0=\sqrt{(g n_0-\delta)(3g n_0-\delta)}$, so the group velocity vanishes at $k\to0$ even though $c_\tb$ still fixes the long‑wavelength cones.

The BdG amplitudes~\cite{castin_lecture_notes,busch_spectrum_2014}
\begin{equation}\label{eq:uv}
    \abs{u}=\sqrt{\tfrac{1}{2}\!\left(\frac{\Delta_k}{\Omega}+1\right)},\qquad
\abs{v}=\sqrt{\tfrac{1}{2}\!\left(\frac{\Delta_k}{\Omega}-1\right)},\quad
\Delta_k=\frac{E_k}{\hbar}-\delta+2g n_0,\ \ \Omega=\sqrt{\Delta_k^2-(g n_0)^2}
\end{equation}
are shown in panels~(c) and (d).
In the gapless case, as $k\xi\to0$, $|u|\simeq|v|\propto (k\xi)^{-1/2}$, hence the near and far field detection weights $D=u+v\propto (k\xi)^{-1/2}$ and $N=\abs{u}^2+\abs{v}^2\propto (k\xi)^{-1}$ [panel (e)].
This is the familiar phononic infrared enhancement.
In the massive case, $\Omega\to\Delta_0>0$ at small $k$, so $|u|,|v|,D,N$ remain finite.
There is no $k\to 0$ divergence of the weights; the upstream mode only exists for $\omega>\Delta_0$.
This alone suppresses low‑frequency contrast relative to the gapless case (note that as $v_G\to 0$, the density of state $\propto 1/v_g$ grows).
For $k\xi\gg1$ in either case, $|u|\to1$, $|v|\to0$ and the dispersion tends to the single‑particle parabola shifted by the mean field.

\newtcolorbox{polaritonbox}[1][]{
  enhanced, breakable,
  colback=white, colframe=black!50,
  boxrule=0.5pt, arc=1pt,
  left=1em, right=1em, top=0.6em, bottom=0.6em,
  fonttitle=\bfseries, title={Polaritons at a glance},
  #1
}

\begin{polaritonbox}
\begin{itemize}[leftmargin=1.2em]
\item \textbf{Driven–dissipative dynamics.}
Coherent excitation: the driving phase sets the in–plane momentum in high intensity regions.
We work with a scalar polariton field of mass $m^\ast$ and contact interaction $g>0$.
The polariton wavefunction is the driven-dissipative Gross-Pitaevskii equation
\[i\hbar\,\partial_t\psi=
\Big[-\frac{\hbar^2\nabla^2}{2m^\ast}-\hbar\delta(\mathbf{k_\tp})+\hbar g\,n
-i\frac{\hbar\gamma}{2}\Big]\psi
+i\hbar\,F_\tp(\mathbf r).\]
With the electromagnetic drive $F_\tp(\mathbf{r},t)=F_0\,e^{-i\omega_\tp t}\,e^{i\mathbf{k}_\tp\cdot\mathbf{r}}$.
Polaritons decay into photons exiting the sample at rate $\gamma$.

\item \textbf{Linear response.}
The spectrum in the lab frame is given by
\[
\omega_\pm(\mathbf r,\mathbf k)
=
\mathbf v_0(\mathbf r)\!\cdot\!\mathbf k
\;\pm\;
\sqrt{\Big[\tfrac{\hbar^2 k^2}{2m^\ast}-\delta(\mathbf k_\tp)+2g\,n_0(\mathbf r)+g_{\rm res}\,n_{\rm res}(\mathbf r)\Big]^2
-\big[g\,n_0(\mathbf r)\big]^2},
\]
with the effective detuning $ \delta(\mathbf{k_\tp})= \hbar\omega_\tp-\hbar\omega_0-\frac{\hbar^2 (\mathbf{k_\tp})^2}{2m^\ast}$.

\noindent$\omega_+$ ($\omega_-$) solutions have positive (negative) Klein-Gordon norm $\propto\int (|u|^2-|v|^2)$.
All frequencies are measured in the pump’s rotating frame.

\item \textbf{Degrees of freedom and scales.}
The plane–wave pump fixes the background flow 
\[\mathbf v_0=\frac{\hbar}{m^\ast}\mathbf k_\tp.\]
The sound cones and healing length are
\[
c_\tb=\sqrt{\frac{\hbar\,(2g n-\delta(k_\tp)+g_\mathrm{res}n_\mathrm{res})}{m^\star}},\qquad
\xi=\sqrt{\frac{\hbar}{m^*(2gn_0-\delta(k_\tp))}}\;.
\]
\end{itemize}

\end{polaritonbox}

\section{Horizons and the Hawking effect}
\label{sec:horizons-hawking}
We consider a stationary, inhomogeneous polariton flow that is \emph{transcritical} (crosses the acoustic cone), use it as a case study to introduce the local dispersion and mode content, and then construct the global scattering problem and its quantum description. We work in 1D along the flow, keep the weak‑loss limit for mode classification, and use the long‑wavelength (hydrodynamic) mapping from Sec.~\ref{subsubsec:KG-analogy} to organise the discussion; for quantitative predictions we always refer back to the full Bogoliubov spectrum \eqref{eq:WKB-local} (locally) or \eqref{eq:BdG-inhom} (homogeneous asymptotics). For the sake of simplicity, we drop the reservoir contributions in this section.

\begin{figure*}[h]
  \centering
  \begin{subfigure}[b]{0.32\textwidth}
    \centering
    \includegraphics[width=\linewidth]{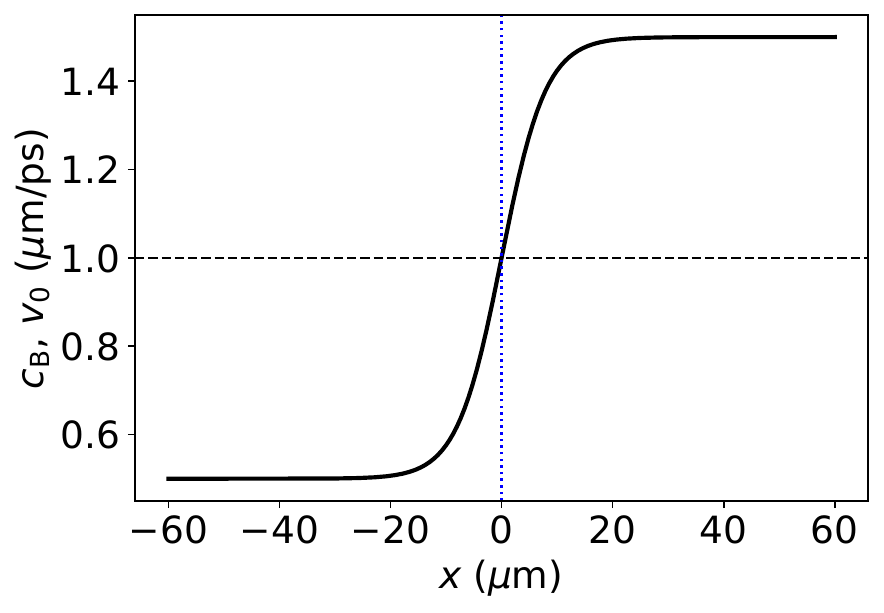}
  \end{subfigure}\hfill
  \begin{subfigure}[b]{0.32\textwidth}
    \centering
    \includegraphics[width=\linewidth]{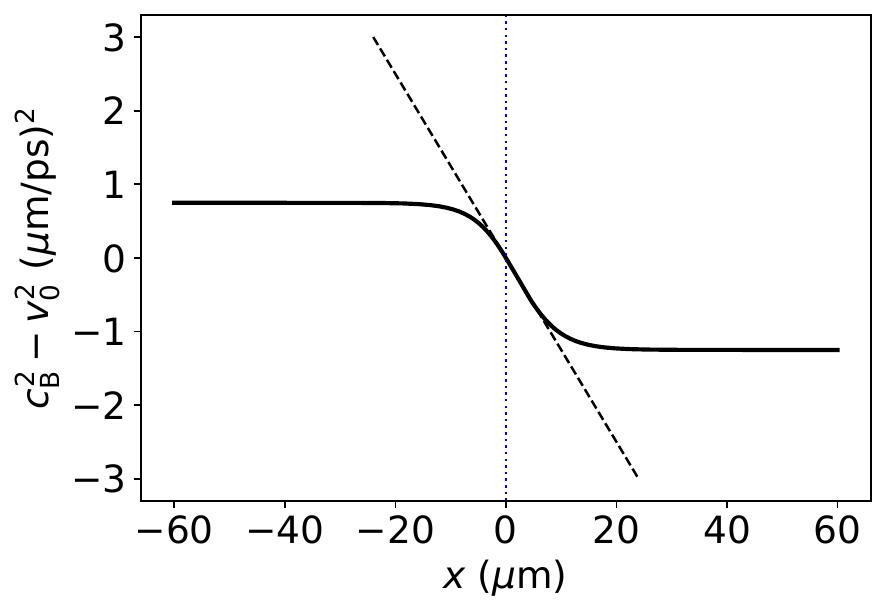}
  \end{subfigure}\hfill
  \begin{subfigure}[b]{0.32\textwidth}
    \centering
    \includegraphics[width=\linewidth]{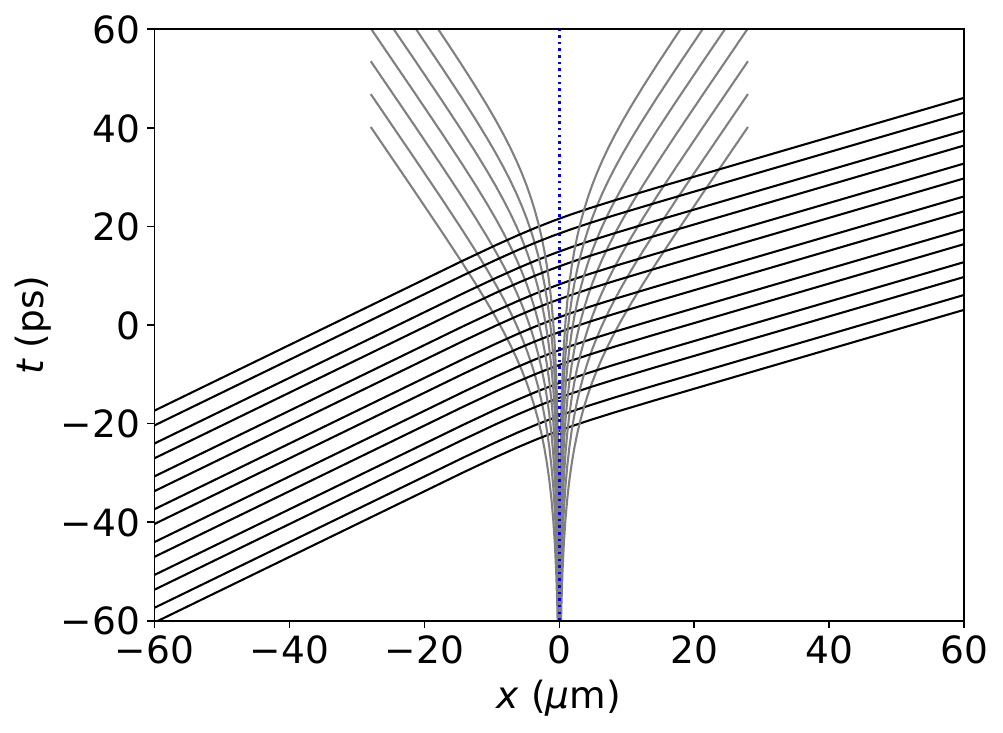}
  \end{subfigure}

  \caption{\textbf{Acoustic metric \& waterfall horizon geometry.}
  The blue dotted line marks the horizon position $x_\tH$.
  \textbf{(a)} Waterfall velocity profile~\eqref{eq:waterfall}.
  Black, \(v_0(x)\); grey dashed \(c_\tb\).
  \textbf{(b)} Acoustic‑metric indicator.
  Black, \(c_\tb^{\,2}-v_0^{\,2}(x)\); grey dashed, local tangent at \(x_\tH\) with slope \(2c_\tb\kappa\), with $\kappa$ defined in Eq.~\eqref{eq:kappa-waterfall}.
  \textbf{(c)} Null characteristics (``light‑cones'') obtained from \(\mathrm{d}x/\mathrm{d}t=v_0(x)\pm c_\tb\): families of right/left‑moving rays across the horizon.
  }
  \label{fig:fig4}
\end{figure*}

\subsection{Case study: the waterfall horizon geometry}
\label{subsec:waterfall-setup}
We take a quasi‑1D stationary background with \emph{constant} acoustic speed \(c_\tb(x)\equiv c\) and a monotonic flow profile (the canonical ``water fall'' geometry~\cite{Macher-HRBE-2009})
\begin{equation}
v_0(x)=v_{\rm d}+\frac{v_{\rm u}-v_{\rm d}}{2}\Big[1-\tanh\!\Big(\frac{x-x_\tH}{\ell}\Big)\Big],
\qquad
\abs{v_{\rm u}}<c<\abs{v_{\rm d}},
\label{eq:waterfall}
\end{equation}
so the upstream (\(x\!\to\!-\infty\)) is \emph{subsonic} with velocity \(v_{\rm u}\), the downstream (\(x\!\to\!+\infty\)) is \emph{supersonic} with velocity \(v_{\rm d}\), and the (hydrodynamic) horizon is at \(x=x_\tH\) where \(|v_0|=c\) [Fig.~\ref{fig:fig4}~(a)]. For constant \(c\) the surface gravity is simply the local shear~\cite{visser_acoustic_1998}
\begin{equation}
\kappa=\frac{1}{2c}\,\partial_x\!\big(c^2-v_0^2\big)\Big|_{x_\tH}= \abs{\partial_x v_0}_{x_\tH}
=\frac{\abs{v_{\rm d}-v_{\rm u}}}{2\ell}\,\mathrm{sech}^2(0)=\frac{\abs{v_{\rm d}-v_{\rm u}}}{2\ell}.
\label{eq:kappa-waterfall}
\end{equation}

\subsection{Local dispersion, branches, and frequency windows}
\label{subsec:local-dispersion}

At fixed \(x\) the local (WKB) lab‑frame dispersion is given by \eqref{eq:WKB-local}; in the present case this simplifies to the standard superluminal Bogoliubov form,
\begin{equation}
\omega_\pm(x,k)= v_0(x)\,k \ \pm\ \Omega(k),\qquad
\Omega(k)=\sqrt{ \big[c^2 k^2 + \Delta^2\big] + \frac{\hbar^2 k^4}{4m^{\ast2}}}\,,
\label{eq:local-disp-waterfall}
\end{equation}
with \(\Delta\equiv m_{\rm det} c^2/\hbar\) the low‑\(k\) gap. We label \(\pm\) by the \emph{comoving} frequency sign, \(\Omega=\pm|\Omega|\); the \(+\) sheet has positive KG norm, the \(-\) sheet negative norm.

\paragraph{Asymptotic branches and channels.}
Let \(x\to-\infty\) (upstream, subsonic \(v_{\rm u}\)) and \(x\to+\infty\) (downstream, supersonic \(v_{\rm d}\)). At a given \(\omega>0\), the real‑\(k\) solutions of \(\omega=v\,k\pm\Omega(k)\) in each asymptotic region define the \emph{propagating channels}. For sufficiently low \(\omega\), one finds:
\begin{itemize}[leftmargin=1.15em]
\item \textbf{Upstream (subsonic):} two real roots on the \(+\) sheet (both positive norm), one outgoing (group velocity $\partial\omega/\partial k$ away from the horizon) and one incoming.
\item \textbf{Downstream (supersonic):} up to three real roots: one on the \(+\) sheet (positive norm) and two on the \(-\) sheet (negative norm), with one of each incoming/outgoing relative to the horizon. In practice a \emph{single} negative‑norm propagating root exists below a maximum frequency \(\omega_{\max}\) (next paragraph).
\end{itemize}
Evanescent roots (complex \(k\)) control near‑horizon matching but do not carry flux asymptotically.

\paragraph{The horizon frequency interval}
In our massive field theory, two kinematic scales organise the spectrum~\cite{jacquet_analogue_2022}:
\begin{enumerate}[leftmargin=1.15em]
\item \textbf{Low‑frequency threshold \(\omega_{\min}\) (mass).}  
When \(\Delta>0\), the comoving energy is bounded below by \(\Delta\). Minimising \(\omega=v_{\rm u}k+\sqrt{c^2k^2+\Delta^2}\) over \(k\) gives the upstream lab‑frame threshold
\begin{equation}
\omega_{\min}^{\rm (u)}=\Delta\,\sqrt{1-\frac{v_{\rm u}^2}{c^2}}\qquad (\Delta>0).
\label{eq:omega-min}
\end{equation}
For \(\omega<\omega_{\min}^{\rm(u)}\) no upstream propagating solution exists (the \emph{mass} blocks long‑wavelength propagation). When \(\Delta=0\), \(\omega_{\min}=0\).
\item \textbf{Upper cutoff \(\omega_{\max}\) (dispersion).}  
Superluminal curvature limits the range where there is a root of negative norms in the downstream. The cutoff is obtained by the tangency conditions (double root)
\begin{equation}
\omega_{\max}= v_{\rm d}\,k_\ast - \Omega(k_\ast),\qquad
v_{\rm d}=\frac{d\Omega}{dk}\Big|_{k_\ast}
=\frac{c^2 k_\ast + \frac{\hbar^2}{2m^{\ast2}}k_\ast^3}{\sqrt{c^2 k_\ast^2+\Delta^2+\frac{\hbar^2}{4m^{\ast2}}k_\ast^4}}.
\label{eq:omega-max}
\end{equation}
Equivalently, in the dimensionless variable \(q\equiv k\xi\), one solves the second equation for \(q_\ast\) and inserts into the first. In the massless limit \(\Delta=0\) this reproduces the standard \(\omega_{\max}\propto (c/\xi)\,(|v_{\rm d}|^2-c^2)^{3/2}/|v_{\rm d}|^2\)~\cite{isoard_departing_2020}.
\end{enumerate}

\newtcolorbox{Horizonfreqbox}[1][]{
  enhanced, breakable,
  colback=white, colframe=black!50,
  boxrule=0.5pt, arc=1pt,
  left=1em, right=1em, top=0.6em, bottom=0.6em,
  fonttitle=\bfseries, title={Horizon frequency interval},
  #1
}
\begin{Horizonfreqbox}
$\rightarrow$ where both an upstream propagating mode and a downstream negative norm mode exist:
\begin{equation}
\omega\in\big(\omega_{\min},\ \omega_{\max}\big).
\label{eq:window}
\end{equation}
\end{Horizonfreqbox}

\begin{figure*}[h]
  \centering
  \includegraphics[width=\textwidth]{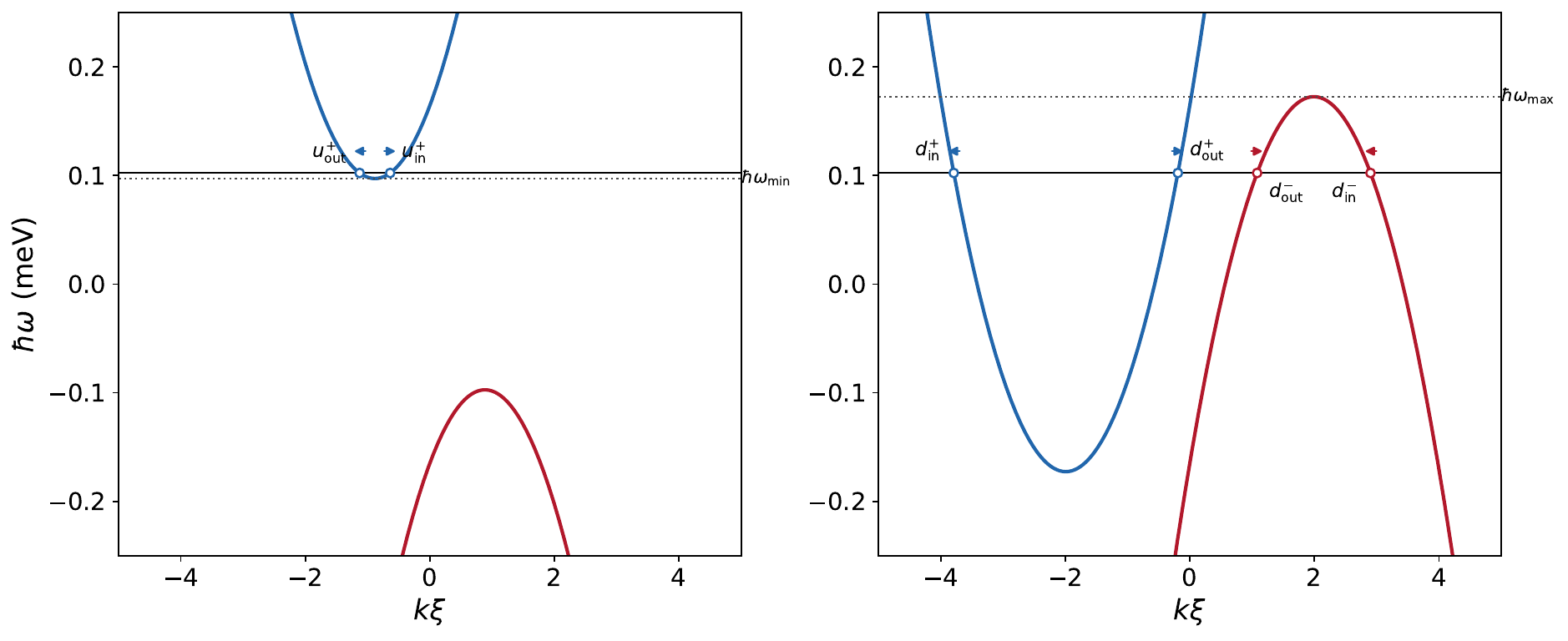}
  \caption{\textbf{Horizon modes}
  Dispersion relation of a transsonic fluid in the laboratory frame~\eqref{eq:WKB-local}.
  Blue, $\omega^+$; red, $\omega^-$.
  \textbf{(a)} subsonic flow (upstream); \textbf{(b)} supersonic flow (downstream).
  Open circles, propagating roots (real $k$) at $\omega\in\{\omega_\mathrm{min},\omega_\mathrm{max}\}$, i.e., local asymptotic modes: upstream $u^{\pm}_{\rm in/out}$ and downstream $d^{\pm}_{\rm in/out}$, where ``in/out'' is defined relative to the horizon (upstream outgoing $\Leftrightarrow v_g<0$; downstream outgoing $\Leftrightarrow v_g>0$).
  The outgoing triplet is mapped to the channel names $H\equiv u^{+}_{\rm out}$, $P\equiv d^{-}_{\rm out}$, and $W\equiv d^{+}_{\rm out}$ for scattering and observables.
  \label{fig:fig5}}
\end{figure*}

\subsection{Global modes and stationary scattering}
\label{subsec:global-scattering}

We now construct the stationary scattering problem at fixed \(\omega>0\). Let \(\{\varphi_j^{\rm in}(\omega;x)\}\) and \(\{\varphi_j^{\rm out}(\omega;x)\}\) denote complete bases of global solutions of the BdG/KG equation [Sec.~\ref{sec:horizons-hawking}], asymptoting to single‑channel plane waves in each far region, normalised to unit KG flux~\cite{macher_black/white_2009}. The channels are labelled by their asymptotic side (upstream / downstream), direction (in / out), and norm sign (\(+\)/\(-\)).

\paragraph{Scattering matrix and pseudo‑unitarity.}
Write the mode expansion (right-moving time dependence \(e^{-i\omega t}\) understood)
\begin{equation}
\varphi_i^{\rm out}(\omega;x)=\sum_j S_{ij}(\omega)\,\varphi_j^{\rm in}(\omega;x),
\label{eq:S-matrix-def}
\end{equation}
with the channel metric \(\eta=\mathrm{diag}(\mathds{1}_{\mathcal P},-\mathds{1}_{\mathcal N})\) distinguishing positive / negative norm subspaces.
Flux conservation implies the Bogoliubov (pseudo‑unitary) constraint
\begin{equation}
S^\dagger(\omega)\,\eta\,S(\omega)=\eta.
\label{eq:pseudounitary}
\end{equation}
In the waterfall case and for \(\omega\in(\omega_{\min},\omega_{\max})\), a minimal description involves \emph{three} asymptotic channels~\cite{recati2009bogoliubov}, see Fig.~\ref{fig:fig5}:
\begin{itemize}[leftmargin=1.2em]
    \item Upstream (subsonic): $u_{\rm in}^{+}$ (pos.\ norm, incoming), $u_{\rm out}^{+}$ (pos.\ norm, outgoing),
    \item Downstream (supersonic): input channels $d_{\rm in}^{+}$ (pos.\ norm) and $d_{\rm in}^{-}$ (neg.\ norm); output channels $d_{\rm out}^{+}$ (pos.\ norm) and $d_{\rm out}^{-}$ (neg.\ norm).
\end{itemize}
Depending on \(\omega\), either \(d_{\rm in}^{+}\) or \(d_{\rm in}^{-}\) is present; the common case at low frequency has the \(\{u_{\rm in}^{+},\,d_{\rm in}^{+},\,d_{\rm in}^{-}\}\to\{u_{\rm out}^{+},\,d_{\rm out}^{+},\,d_{\rm out}^{-}\}\) structure. The matrix \(S(\omega)\) is \(3\times 3\) in that interval and obeys \eqref{eq:pseudounitary}. When \(\omega>\omega_{\max}\) the negative‑norm channel closes and \(S\) reduces to a usual \(2\times 2\) unitary matrix between positive‑norm channels.

\paragraph{Near‑horizon limit and thermal slope.}
Linearising \(c^2-v_0^2\) near \(x_\tH\) gives the Rindler form of the KG equation~\cite{balbinot_nonlocal_2008}; matching to WKB solutions yields (i) the \emph{low‑\(\omega\)} behaviour of the mode‑conversion coefficients,
\begin{equation}
|S_{u^{\rm out}_{+}\leftarrow d^{\rm in}_{-}}(\omega)|^2\ \xrightarrow[\omega\to0]{}\ \frac{1}{e^{\hbar\omega/k_\mathrm{B} T_\tH}-1}\quad
\text{with}\quad k_\mathrm{B} T_\tH=\frac{\hbar\kappa}{2\pi},
\label{eq:thermal-slope}
\end{equation}
modulo grey‑body factors due to smooth variations away from the horizon; and (ii) the frequency cutoff \(\omega_{\max}\) governed by the quartic curvature (Sec.~\ref{subsec:local-dispersion}). A finite mass \(m_{\rm det}\) introduces a soft threshold at \(\omega_{\min}\) [Eq.~\eqref{eq:omega-min}] and reduces the low‑\(\omega\) slope when \(\omega_{\min}\) is not \(\ll \kappa\).

\subsection{Quantum field theory of scattering}
\label{subsec:qft-scattering}

Let \(\hat b_i^{\rm out}(\omega)\) and \(\hat a_j^{\rm in}(\omega)\) be the annihilation operators associated with \(\varphi_i^{\rm out}\) and \(\varphi_j^{\rm in}\) (normalised with the bi‑orthogonal KG product; Sec.~\ref{subsubsec:hydro-ddgpe} and \eqref{eq:biorthonorm-KG}). The mode expansion of the fluctuation field (phase proxy \(\Phi\) or directly \(\delta\psi\)) reads
\begin{equation}
\hat\Phi(x,t)=\sum_i\int_0^\infty\!\frac{d\omega}{2\pi}\,\big[\,
\hat b_i^{\rm out}(\omega)\,\varphi_i^{\rm out}(\omega;x)\,e^{-i\omega t}
+\hat b_i^{\rm out\,\dagger}(\omega)\,\varphi_i^{\rm out\,\ddagger}(\omega;x)\,e^{+i\omega t}\big],
\end{equation}
where \(\ddagger\) denotes the left‑mode associated with the KG product. The in/out ladder operators obey
\begin{equation}
\hat b_i^{\rm out}(\omega)=\sum_{j\in\mathcal P} S_{ij}(\omega)\,\hat a_j^{\rm in}(\omega)\ +\ \sum_{j\in\mathcal N} S_{ij}(\omega)\,\hat a_j^{\rm in\,\dagger}(\omega),
\label{eq:Bogoliubov-transform}
\end{equation}
and the canonical commutators \([\hat a_i^{\rm in}(\omega),\hat a_{i'}^{\rm in\,\dagger}(\omega')]=2\pi\,\delta_{ii'}\delta(\omega-\omega')\) (with the KG norm) imply the pseudo‑unitarity \eqref{eq:pseudounitary} of \(S\)~\cite{WaldBook}. For the vacuum of incoming channels one obtains the spontaneous Hawking flux and pair correlators:
\begin{align}
\langle \hat b_i^{\rm out\,\dagger}(\omega)\hat b_i^{\rm out}(\omega)\rangle
&=\sum_{j\in\mathcal N} \abs{S_{ij}(\omega)}^2,
\label{eq:out-occupancy}\\
\langle \hat b_i^{\rm out}(\omega)\hat b_{i'}^{\rm out}(\omega)\rangle
&=\sum_{j\in\mathcal N} S_{ij}(\omega)\,S_{i'j}(\omega),
\label{eq:out-anomalous}
\end{align}
revealing the two‑mode squeezed nature of the outgoing state.

\subsection{Observables}
\label{subsec:observables}
\subsubsection{Emission channels and pair taxonomy.}
For a transcritical waterfall flow and $\omega\in(\omega_{\min},\omega_{\max})$ there are three asymptotic \emph{outgoing} channels [Fig.~\ref{fig:fig5}]:
\begin{itemize}[leftmargin=1.2em]
\item \textbf{HR (Hawking radiation), $H\equiv u_{\rm out}^{+}$}: upstream, positive norm, subsonic side.
\item \textbf{Partner, $P\equiv d_{\rm out}^{-}$}: downstream, negative norm, supersonic side.
\item \textbf{Witness, $W\equiv d_{\rm out}^{+}$}: downstream, positive norm, the additional propagating branch opened by superluminal dispersion.
\end{itemize}
Vacuum inputs produce a two‑mode squeezed \emph{H–P} pair by mixing positive and negative norm channels at the horizon; pseudo‑unitarity forces additional (weaker) correlations with $W$ because $H$ and $W$ both couple to the same negative‑norm input through the $S$ matrix [Fig.~\ref{fig:fig7}~(a)].

Let $w_i=(u_i,v_i)^{\mathsf T}$ be the mode spinor of outgoing channel $i\in\{H,P,W\}$ at the analysis frequency, normalised (bi‑orthogonally) as in \eqref{eq:biorthonorm-KG}. The detection weights are:
\begin{itemize}[leftmargin=1.2em]
    \item \textbf{density (near field):} $D_i\ \equiv\ u_i+v_i$,
    \item \textbf{number (far field):} $N_i\ \equiv\ \abs{u_i}^2+\abs{v_i}^2$
    \item \textbf{anomalous pair weight:} $A_{ij}\ \equiv\ u_i v_j + v_i u_j$.
\end{itemize}

Panels (e) and (f) of Fig.~\ref{fig:fig3} show the detection weights.
In the gapless case ($gn_0=\delta(\mathbf{k_\tp})$, the low-frequency content is large, revealing the thermal behavioiur of the scattering matrix components~\eqref{eq:thermal-slope} for $\omega\ll\kappa$.
In the massive case, a finite $\omega_\mathrm{\min}$ trims the low‑$\omega$ window and removes the infrared enhancement of $D,N$.
near $\omega_\mathrm{\min}$ the upstream group velocity is small and we see that the per-mode weights remain finite.

\begin{figure*}[h]
  \centering

  \begin{subfigure}[c]{0.49\textwidth}
    \centering
    \includegraphics[width=\linewidth]{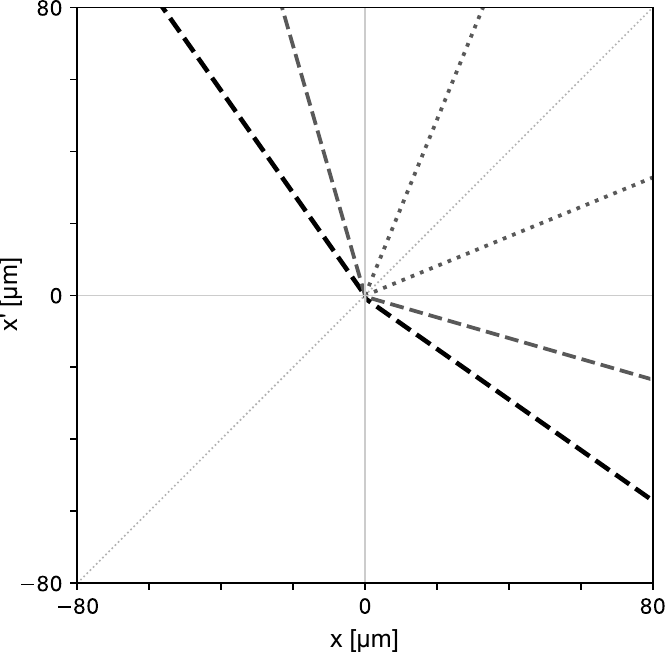}
  \end{subfigure}\hfill
  \begin{subfigure}[c]{0.49\textwidth}
    \centering
    \includegraphics[width=\linewidth]{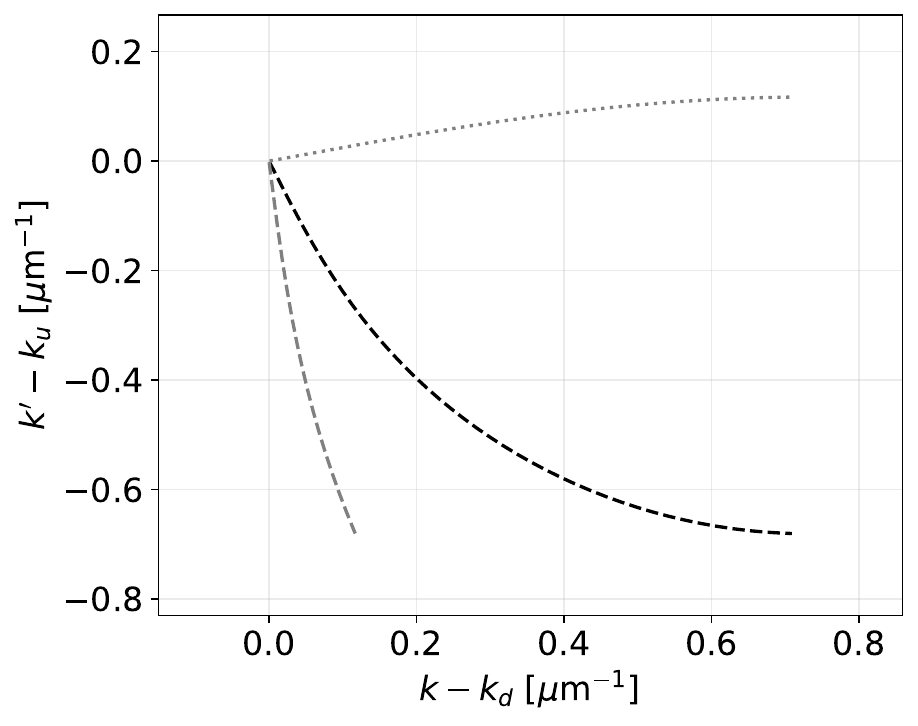}
  \end{subfigure}

  \vspace{0.8em}

  \begin{subfigure}[c]{0.32\textwidth}
    \centering
    \includegraphics[width=\linewidth]{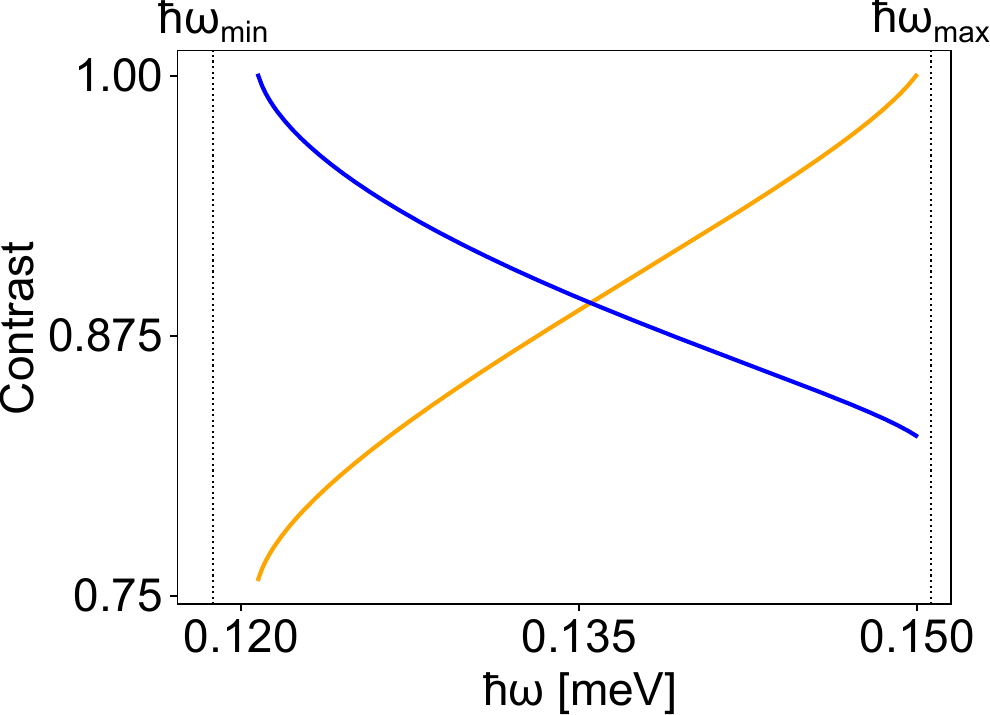}
  \end{subfigure}\hfill
  \begin{subfigure}[c]{0.32\textwidth}
    \centering
    \includegraphics[width=\linewidth]{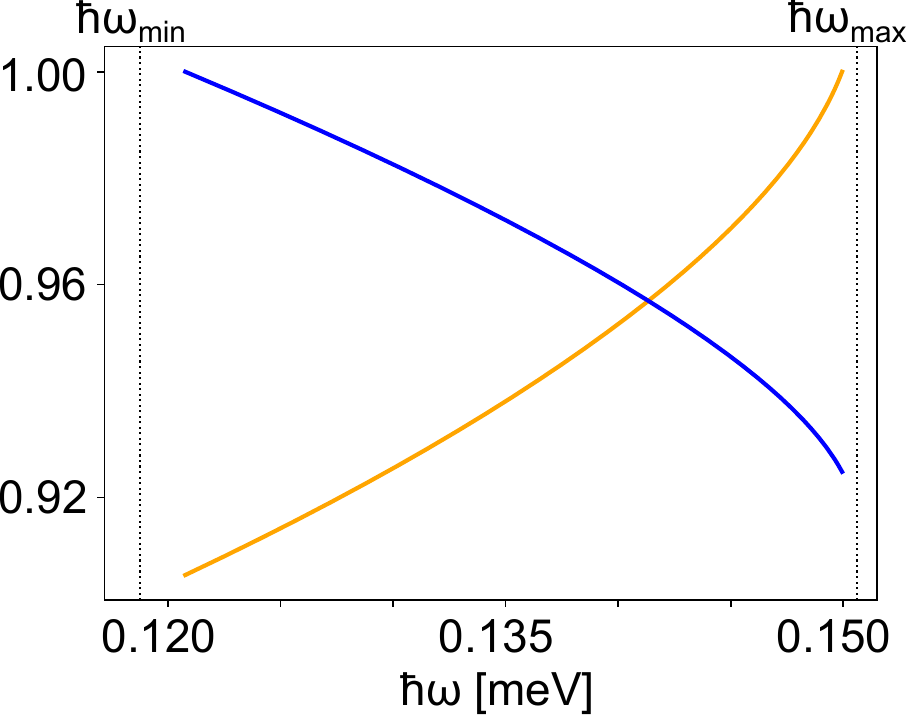}
  \end{subfigure}\hfill
  \begin{subfigure}[c]{0.32\textwidth}
    \centering
    \includegraphics[width=\linewidth]{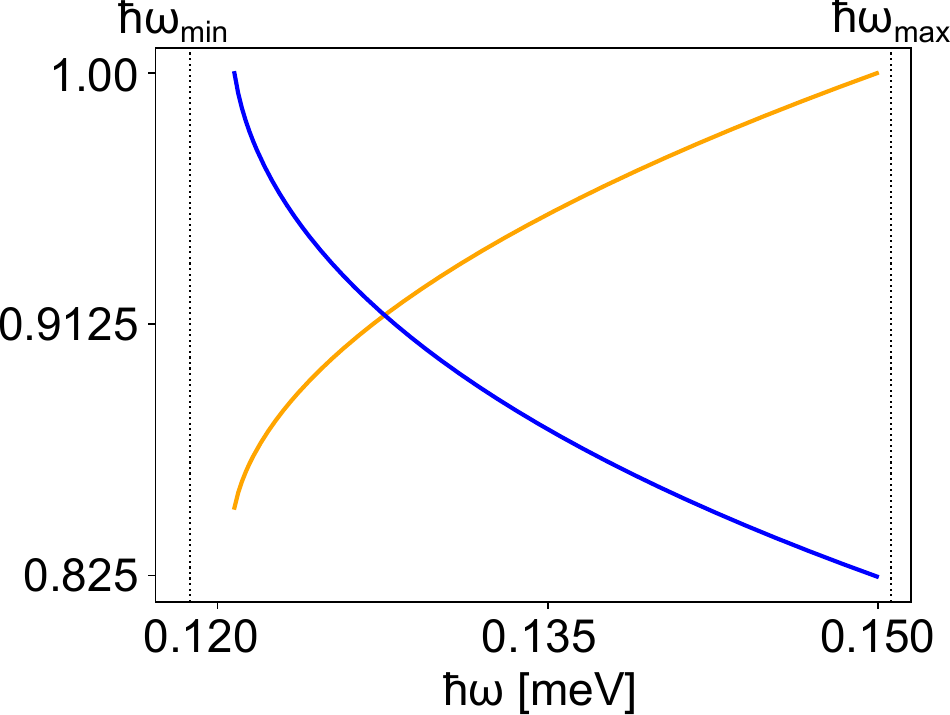}
  \end{subfigure}

  \caption{\textbf{Correlation traces}.
  \textbf{a)} Near-field density-density correlations~\eqref{eq:xcorr}.
  \textbf{b)} Far field number correlations~\eqref{eq:kcorr}.
  Correlation traces: black dashed, $H$–$P^{\ast}$; grey dashed, $W$–$P^{\ast}$; dotted, $W-P$.
  Second row: normalised contrasts in the near‑field, $\abs{D_iD_j}$ (orange); and far‑field,  $\abs{A_{ij}}$ (blue).
  \textbf{c)} $H$–$P^{\ast}$; \textbf{d)} $W$–$P^{\ast}$; \textbf{e)} $H$–$W$.}
  \label{fig:fig6}
\end{figure*}

\subsubsection{Correlation traces}\label{subsubsec:corrtraces}

We can resolve equal‑time, normal‑ordered intensity correlations in position space (near field)
\begin{equation}
G^{(2)}(x,x')\equiv \langle\,:\delta I(x,t)\,\delta I(x',t):\,\rangle,
\label{eq:xcorr}
\end{equation}
which, as shown in Fig.~\ref{fig:fig6} \textbf{a)}, form \textit{tongues} whose slope follows the group velocities of the involved modes at the selected analysis frequency and whose width is set by \(\max\{\xi,\) imaging point spread function\(\}\), as well as the momentum space covariance (far field)
\begin{equation}
C(k,k')\equiv \langle\,:\delta n_k\,\delta n_{k'}:\,\rangle,
\label{eq:kcorr}
\end{equation}
which shows a sharp \textit{ridge} along the frequency‑matched pair map [Fig.~\ref{fig:fig6}~\textbf{b)}]
\[
\mathcal P:\quad k\mapsto k'\ \text{such that}\ \ \omega_+(x\!\to\!-\infty,k)=\omega_-(x\!\to\!+\infty,k').
\]
Fig.~\ref{fig:fig6}~\textbf{c)}, \textbf{d)} and \textbf{e)} show that the ridge amplitude follows the low \(\omega\) thermal slope \eqref{eq:thermal-slope} at \(\omega\ll\kappa\)~\cite{unruh_sonic_1995,brout_hawking_1995,corley_hawking_1996,michel_phonon_2016}, and rolls over near \(\omega_{\max}\) where the negative norm channel closes~\cite{isoard_departing_2020}.

Using the mode expansion of Sect.~\ref{subsec:qft-scattering} one finds the following scalings for the three observable ridges/tongues (up to instrument response and smooth greybody factors away from the horizon):
\begin{itemize}[leftmargin=1.2em]
    \item \textbf{H--P (Hawking pair):} $G^{(2)}_{HP}\propto\abs{D_H D_P}\abs{S_{H\leftarrow -}\,S_{P\leftarrow -}}$ and $C_{HP}\propto\abs{A_{HP}}\,\abs{S_{H\leftarrow -}\,S_{P\leftarrow -}}$;
    \item \textbf{H--W (witness--HR):} $G^{(2)}_{HW}\propto\abs{D_H D_W}\abs{S_{H\leftarrow -}\,S_{W\leftarrow -}}$ and $ C_{HW}\propto\abs{A_{HW}}\abs{S_{H\leftarrow -}\,S_{W\leftarrow -}}$;
    \item \textbf{P--W (partner--witness):} $G^{(2)}_{PW}\propto\abs{D_P D_W}\abs{S_{P\leftarrow -}\,S_{W\leftarrow -}}$ and $C_{PW} \propto|A_{PW}|\abs{S_{P\leftarrow -}\,S_{W\leftarrow -}}$.
\end{itemize}
Here $S_{i\leftarrow -}(\omega)$ abbreviates the $S$ matrix element from the unique negative norm \emph{ingoing} channel at frequency $\omega$ to the outgoing channel $i$ (Sect.~\ref{subsec:global-scattering}). 
These equations encode a simple message: the near field (density) features follow the sums $u\!+\!v$, while the far field (number) features follow the bilinears in $u,v$.

Using the formulas above as guides for the local mode content, we find that the dominant Bogoliubov component depends on $k$:
\begin{itemize}[leftmargin=1.2em]
\item \emph{Phononic (small $k\xi$):} $u\simeq v$ so $D_i\propto (k\xi)^{-1/2}$ is large and $A_{ij}\sim \mathcal O(1)$.
The upstream $H$ channel is typically phononic $\Rightarrow$ strong density tongues and strong $k$‑space ridge with $P$~\cite{michel_phonon_2016}.
\item \emph{Particle‑like (large $k\xi$):} $u\!\to\!1$, $v\!\to\!0$ so $D_i\!\to\!1$ and $A_{ij}\!\simeq\!v_i+v_j$ is small unless one partner remains phononic.
The witness $W$ often sits in this regime~\cite{isoard_departing_2020} $\Rightarrow$ weaker $G^{(2)}_{HW}$ and a modest $C_{HW}$; by contrast $P$ may retain sizeable $v$ so $C_{PW}$ can still be visible.
\item \emph{Massive low‑$\omega$ threshold:} when $\omega\to\omega_{\min}$ (if $m_{\rm det}>0$), the upstream $H$ mode delocalises and its $k$ shifts toward $0$; $D_H$ grows but the available phase space shrinks—expect narrow, weak tongues and a suppressed $k$‑space ridge.
\end{itemize}

\newtcolorbox{takehomebox}[1][]{
  enhanced, breakable,
  colback=white, colframe=black!50,
  boxrule=0.5pt, arc=1pt,
  left=1em, right=1em, top=0.6em, bottom=0.6em,
  fonttitle=\bfseries, title={Take-home message},
  #1
}

\begin{takehomebox}
\begin{enumerate}[leftmargin=1.25em]
\item \textbf{H--P} is the primary Hawking signature: strongest when $\omega\ll\kappa$; amplitude follows the thermal slope \eqref{eq:thermal-slope}~\cite{finazzi_robustness_2011,michel_phonon_2016}; both near‑ and far‑field signals are typically robust~\cite{Gerace,jacquet_analogue_2022}.
\item \textbf{H--W} correlates two positive norm outputs; it is induced by their shared coupling to the same negative norm input. Expect weaker contrast~\cite{isoard_departing_2020}, particularly in real space (small $D_W$ when $W$ is particle‑like)~\cite{carusotto_numerical_2008,Recati_2009,jacquet_analogue_2022}. Its frequency range collapses as $\omega\to\omega_{\max}$.
\item \textbf{P--W} correlates opposite norms and can be appreciable in $k$‑space when $P$ retains a sizeable $v$ component (large $A_{PW}$); in real space it survives when $D_W$ is not too small (e.g. moderately phononic $W$ at lower $|k|$).
\end{enumerate}

Note that the phase of the measured correlator depends on the detection operator: the nearfield intensity couples to $\delta n\propto D_i b_i + D_i^\ast b_i^\dagger$, while $k$‑space number couples to $N_i b_i^\dagger b_i$ (plus small normal ordering terms).
Consequently, $G^{(2)}$ and $C$ can display opposite contrasts for the same pair~\cite{Isoard}; phase‑sensitive (homodyne) detection can select the quadrature to maximise either the $H$--$P$ or $P$--$W$ features.
\end{takehomebox}

\subsection{Gaussian quantum optics of the Hawking effect}
\label{subsec:hawking_gaussian_optics}

The quantum field theory (QFT) formulation developed above can be recast exactly in the language of Gaussian quantum optics.
This is natural for polariton experiments, where quadratures and covariance matrices are standard tools. Here we explain how the Hawking effect can be described as a compact ``symplectic circuit'' composed of a two–mode squeezer (the near–horizon mode mixing) followed by a passive beam splitter (the greybody factor). This circuit makes transparent how amplification, squeezing, and entanglement arise and how they are quantified by experimentally accessible observables.

\begin{figure*}[h]
\centering
  \begin{subfigure}[c]{0.5\textwidth}
    \centering
    \includegraphics[width=\linewidth]{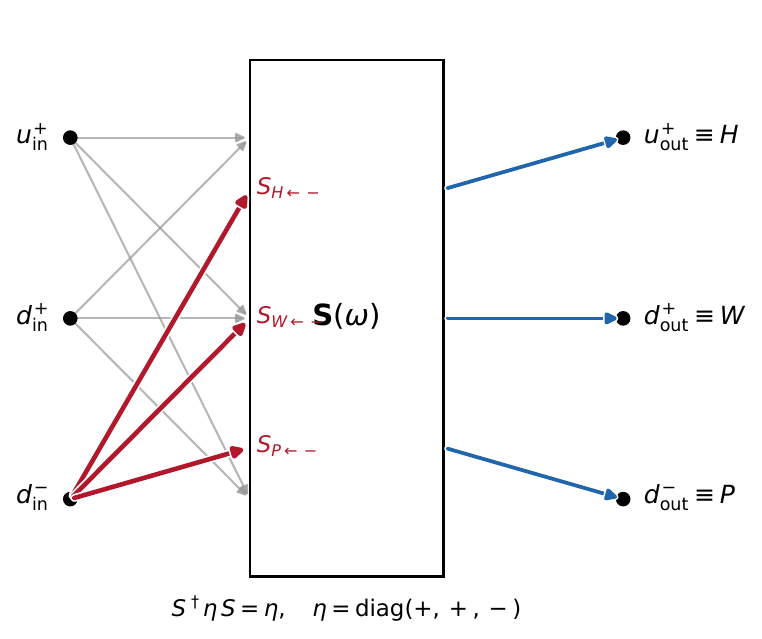}
  \end{subfigure}\hfill
  \begin{subfigure}[c]{0.5\textwidth}
    \centering
    \includegraphics[width=\linewidth]{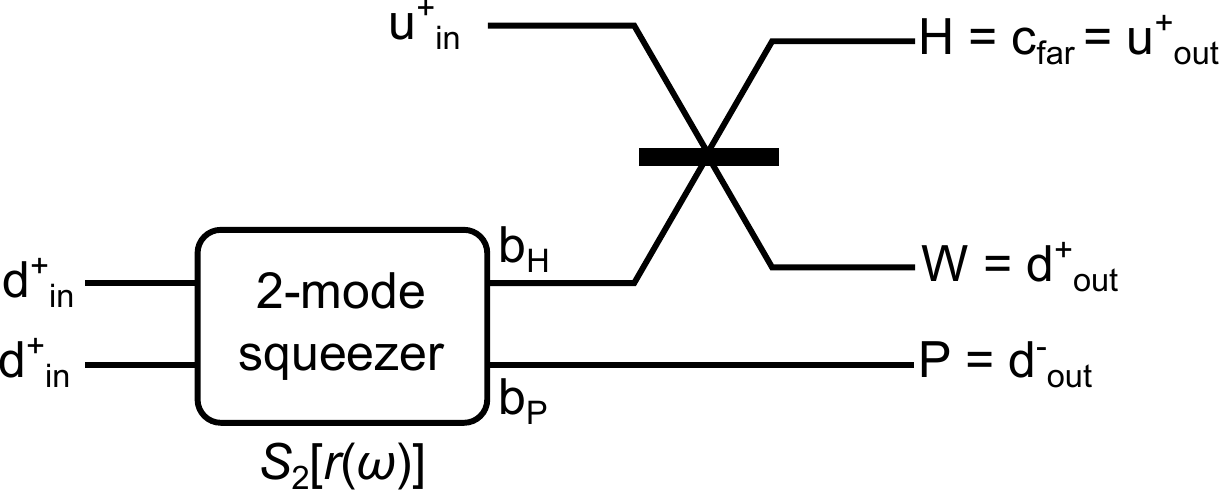}
  \end{subfigure}\hfill
  \begin{subfigure}[b]{0.5\textwidth}
    \centering
    \includegraphics[width=\linewidth]{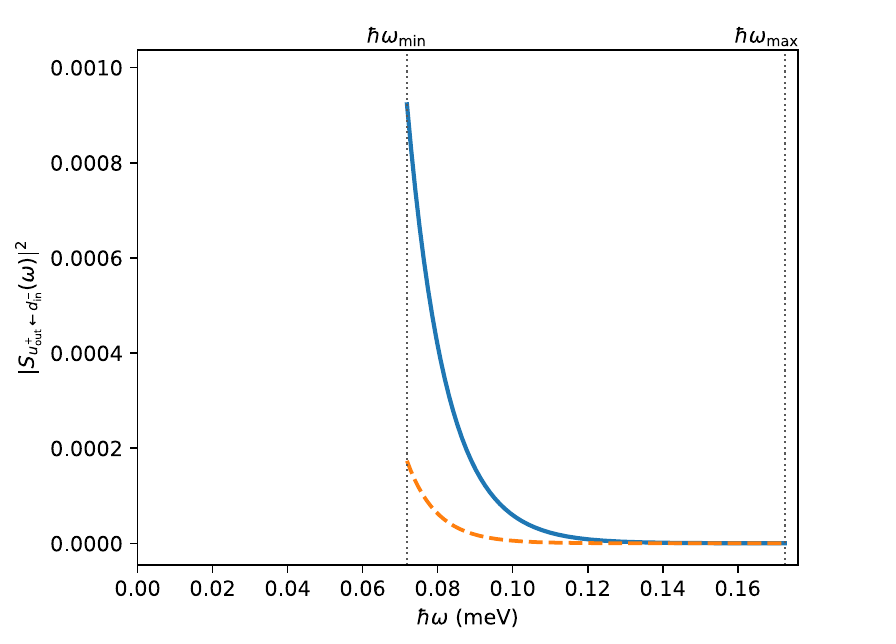}
  \end{subfigure}
    \caption{%
\textbf{Gaussian quantum optics of the Hawking effect}.
\textbf{(a)} Pseudo‑unitary $3\times3$ $S(\omega)$ connecting local in/out channels at fixed $\omega$ in the horizon frequency interval, with the negative‑norm input $d_{\rm in}^{-}$ in red; the dominant conversion amplitudes $S_{u^{+}_{\rm out}\leftarrow d^{-}_{\rm in}}$, $S_{d^{+}_{\rm out}\leftarrow d^{-}_{\rm in}}$, and $S_{d^{-}_{\rm out}\leftarrow d^{-}_{\rm in}}$ are indicated (with $H\!\equiv\!u^{+}_{\rm out}$, $W\!\equiv\!d^{+}_{\rm out}$, $P\!\equiv\!d^{-}_{\rm out}$).
\textbf{(b)} Equivalent Gaussian circuit: a two‑mode squeezer $S_{2}\!\big[r(\omega)\big]$ acting on $\big(d_{\rm in}^{+},\,d_{\rm in}^{-}\big)\mapsto (b_{H},b_{P})$ [Eq.~\eqref{eq:two_mode_squeezer}], followed by a passive greybody stage of amplitude transmission $\sqrt{T(\omega)}$ on the upstream leg, $c_{\rm far}=\sqrt{T}\,b_{H}+i\sqrt{1-T}\,v$ [Eq.~\eqref{eq:greybody_BS}]. 
\textbf{(c)} Low‑frequency Hawking law within the horizon window.
Blue, near‑horizon coefficient $\big|S_{u^{+}_{\rm out}\leftarrow d^{-}_{\rm in}}(\omega)\big|^{2}$; orange dashed, greybody‑filtered amplitude $T(\omega)\big|S_{u^{+}_{\rm out}\leftarrow d^{-}_{\rm in}}(\omega)\big|^{2}$.
For $\hbar\omega\ll k_{B}T_{H}$ the thermal slope~\eqref{eq:thermal-slope} holds; at frequencies  $\hbar\omega\gg k_{B}T_{H}$ (Wien tail), the spectrum decays exponentially as $\exp[-\hbar\omega/(k_{B}T_{H})]$.\label{fig:fig7} 
}
\end{figure*}

The in/out relations written above show that, even with vacuum inputs, the outgoing channel carries quanta and displays anomalous correlations with its partner. Concretely, Eqs.~\eqref{eq:out-occupancy}-\eqref{eq:out-anomalous} state that for vacuum in-channels scattering proceeds as in a phase–insensitive amplifier: output occupation is non-zero and anomalous (pair) correlators appear solely from mode conversion.
In Gaussian optics terms, this is the fingerprint of a two–mode squeezer seeded by vacuum.

The near–horizon mixing is captured by the two–mode squeezer~\cite{isoard_bipartite_2021}
\begin{equation}
\hat S_{2}(r)\!:\quad
\begin{cases}
\hat b_{\mathrm{H}} = \cosh r\,\hat a_{1} + \sinh r\,\hat a_{2}^{\dagger},\\[2pt]
\hat b_{\mathrm{P}} = \cosh r\,\hat a_{2} + \sinh r\,\hat a_{1}^{\dagger},
\end{cases}
\qquad r\equiv r(\omega)\ge 0,
\label{eq:two_mode_squeezer}
\end{equation}
with \(\hat a_{1,2}\) annihilating the two incoming (vacuum) modes that seed the process; the parametric gain is \(G=\cosh^{2}r\) and the pair number is \(\sinh^{2}r\). Propagation to infinity is a passive frequency-dependent beam splitter of amplitude transmission \(\sqrt{T(\omega)}\) (the greybody factor)~\cite{agullo_entanglement_2024}, which mixes \(\hat b_{\mathrm{H}}\) with a vacuum bath mode \(\hat v\)
\begin{equation}
\hat c_{\mathrm{far}}=\sqrt{T}\,\hat b_{\mathrm{H}}+i\sqrt{1-T}\,\hat v,
\qquad 0\le T(\omega)\le 1.
\label{eq:greybody_BS}
\end{equation}
The partner \(\hat b_{\mathrm{P}}\) is not affected by the greybody filter. This ``squeezer\(\to\)beam–splitter'' factorisation [Fig.~\ref{fig:fig7}~(b)] is the Gaussian counterpart of the standard factorisation of the Hawking problem into universal near-horizon Bogoliubov mixing followed by frequency–dependent greybody transmission~\cite{agullo_event_2022,agullo_entanglement_2024}.

From Eqs.~\eqref{eq:two_mode_squeezer}–\eqref{eq:greybody_BS} with vacuum inputs,
\begin{align}
\bar n_{\mathrm{far}}(\omega)
&\equiv \langle \hat c_{\mathrm{far}}^{\dagger}\hat c_{\mathrm{far}}\rangle
= T(\omega)\,\sinh^{2} r(\omega),
\label{eq:nout_vs_T_r}\\
M_{\mathrm{HP}}(\omega)
&\equiv \langle \hat c_{\mathrm{far}}\hat b_{\mathrm{P}}\rangle
=\tfrac{1}{2}\sqrt{T(\omega)}\,\sinh\! \bigl(2r(\omega)\bigr),
\label{eq:paircorr_vs_T_r}
\end{align}
which are the quantum-optical versions of the scattering moments written in Sec.~\ref{subsubsec:corrtraces}.
In particular, \(\bar n_{\mathrm{far}}>0\) for vacuum inputs expresses spontaneous amplification; \(M_{\mathrm{HP}}\neq 0\) encodes the phase–sensitive twin–beam structure (two–mode squeezing). Using either observable one can solve for \(r(\omega)\):
\[
r(\omega)=\operatorname{asinh}\!\sqrt{\frac{\bar n_{\mathrm{far}}(\omega)}{T(\omega)}} \;=\;\frac{1}{2}\,\operatorname{asinh}\!\Bigl(\frac{2\,|M_{\mathrm{HP}}(\omega)|}{\sqrt{T(\omega)}}\Bigr).
\]

\paragraph{Quadratures, squeezing and entanglement.}
Define shot–noise units (SNU) with \(\mathrm{Var}(X_{\mathrm{vac}})=1\) and the canonical quadratures
\(\hat X=(\hat a+\hat a^{\dagger})/\sqrt{2}\), \(\hat P=(\hat a-\hat a^{\dagger})/(i\sqrt{2})\).
The Einstein-Podolsky-Rosen combinations \(\hat X_{\mathrm{far}}-\hat X_{\mathrm{P}}\) and \(\hat P_{\mathrm{far}}+\hat P_{\mathrm{P}}\) are squeezed:
\begin{equation}
\mathrm{Var}\!\left(\hat X_{\mathrm{far}}-\hat X_{\mathrm{P}}\right)
=\mathrm{Var}\!\left(\hat P_{\mathrm{far}}+\hat P_{\mathrm{P}}\right)
=(1+T)\cosh(2r)-2\sqrt{T}\,\sinh(2r)+(1-T).
\label{eq:EPR_variances}
\end{equation}
For \(T\!=\!1\) this reduces to \(2e^{-2r}\) (in SNU), the textbook two–mode–squeezed vacuum. As \(T\) decreases, the greybody mixing inevitably degrades the observable squeezing.

Entanglement is best quantified by the logarithmic negativity (Gaussian)~\cite{jacquet_influence_2020,isoard_bipartite_2021,agullo_entanglement_2024}.
Let \(V\) be the \(4\times4\) covariance matrix of \((\mathrm{far},\mathrm{P})\) in the order \((X_{\mathrm{far}},P_{\mathrm{far}},X_{\mathrm{P}},P_{\mathrm{P}})\). With \(\hbar\!=\!1\) (vacuum variance \(=\tfrac{1}{2}\)), the blocks read
\begin{align}
&A=\tfrac{1}{2}\!\left(T\cosh2r+1-T\right)\mathbb{I}_2,\quad
B=\tfrac{1}{2}\cosh2r\,\mathbb{I}_2,\quad
C=\tfrac{1}{2}\sqrt{T}\,\mathrm{diag}\!\bigl(\sinh2r,-\sinh2r\bigr).
\end{align}
The smallest symplectic eigenvalue \(\tilde\nu_{-}\) of the partially–transposed covariance is
\[
\tilde\nu_{-}
=\sqrt{\frac{\tilde\Delta-\sqrt{\tilde\Delta^{2}-4\det V}}{2}},\quad
\tilde\Delta=\det A+\det B-2\det C,
\]
and the logarithmic negativity is \(E_{\mathcal{N}}=\max\{0,-\log_{2}(2\tilde\nu_{-})\}\).
For \(T\!=\!1\) this gives \(E_{\mathcal{N}}=2r/\ln2\); for \(T<1\) it decreases monotonically with loss and vanishes at and below the entanglement–breaking threshold~\cite{isoard_bipartite_2021,delhom_entanglement_2024}.

Note that,  (i) In the low–frequency regime where the near–horizon mixing is thermal, one may write \(\sinh^{2}r(\omega)=\bigl(e^{\omega/T_{\mathrm{H}}}-1\bigr)^{-1}\); the greybody simply multiplies the spectrum by \(T(\omega)\), cf.~\eqref{eq:nout_vs_T_r} and Fig.~\ref{fig:fig7}~(c). (ii) All expressions above extend trivially to a weak, frequency–independent background of incoming quanta by replacing the vacuum bath \(\hat v\) with a thermal mode of occupation \(n_{\mathrm{env}}(\omega)\)~\cite{delhom_entanglement_2024}; then \(\bar n_{\mathrm{far}}=T\sinh^{2}r+(1-T)\,n_{\mathrm{env}}\), and the pair correlator retains the \(\propto \sqrt{T}\sinh(2r)\) scaling.

\begin{equation}
\begin{aligned}&\text{squeezer:}\ \ (b_H, b_P) \sim S_2[r(\omega)]\cdot(a_1, a_2) \\[-2pt]&\text{greybody:}\ \ c_{\rm far}=\sqrt{T}\,b_H+i\sqrt{1-T}\,v\end{aligned}
\end{equation}

\newtcolorbox{HEbox}[1][]{
  enhanced, breakable,
  colback=white, colframe=black!50,
  boxrule=0.5pt, arc=1pt,
  left=1em, right=1em, top=0.6em, bottom=0.6em,
  fonttitle=\bfseries, title={The Hawking effect in brief},
  #1
}
\begin{HEbox}
\begin{itemize}[leftmargin=1.2em]
\item \textbf{Hawking frequency window.} Horizon scattering is allowed kinematically only for $\omega\in\big(\omega_{\min},\omega_{\max}\big)$, where $\omega_{\min}$ (if any) is the mass gap and $\omega_{\max}$ is the dispersive cutoff where the downstream negative norm branch closes; see Eqs.~\eqref{eq:omega-min}, \eqref{eq:omega-max}, and \eqref{eq:window}.

\item \textbf{In/out channel.} In this window there are three relevant outgoing channels:
\begin{align*}
    &H\equiv u_{\rm out}^{+}\ (\text{upstream, positive norm}),\\
    &P\equiv d_{\rm out}^{-}\ (\text{downstream, negative norm}),\\
    &W\equiv d_{\rm out}^{+}\ (\text{downstream, positive norm})
\end{align*}
fed by the corresponding incoming channels (one of which is negative norm downstream, depending on $\omega$).

\item \textbf{Hawking effect $=$ amplification.} Mixing of positive and negative norm channels at the horizon makes the $S$-matrix quasi-unitary so even vacuum inputs produce amplified outgoing fluxes:
$\langle b_i^{\rm out\,\dagger} b_i^{\rm out}\rangle=\sum_{j\in\mathcal N}|S_{ij}(\omega)|^2$.
This is a phase-insensitive two-mode squeezer (and the witness mode enters as greybody factors via a beam splitter effect).
At low frequencies, the mode conversion coefficient has a thermal slope set by the surface gravity~\eqref{eq:thermal-slope}.

\item \textbf{Main observables.}
\begin{itemize}[leftmargin=1.1em]
\item \emph{Near field:} $G^{(2)}(x,x')=\langle:\delta I(x)\delta I(x'):\rangle$ shows \emph{correlation tongues} (lines connecting upstream/downstream points) with slopes given by group velocities.
\item \emph{Far field:} the momentum‑space covariance $C(k,k')=\langle:\delta n_k\,\delta n_{k'}:\rangle$ exhibits sharp \emph{correlation ridges}, in particular along the pair map matching $\omega_+(k_{\rm up})=\omega_-(k'_{\rm down})$ (\textit{H}-\textit{P} ridge).
\item \emph{Phase choice / complex conjugates:} because the negative‑norm channel enters the Bogoliubov transform as a \emph{creation} operator, the strongest signal sits in \emph{anomalous} correlators (e.g.\ $\langle b_H b_P\rangle$). Practically, when working at the field level in $k$‑space, correlating a mode with the \textbf{complex conjugate} of its partner’s field (e.g.\ $\langle \delta\psi_k\,\delta\psi^\ast_{k'}\rangle$) maximises the \textit{H}-\textit{P} ridge; number/intensity covariances still show the ridge but with reduced contrast.
\end{itemize}
\end{itemize}
\end{HEbox}

\section{Experimental methods}
\label{sec:experimental-methods}

This section reviews practical techniques for probing microcavity polariton fluids. We begin with photodetection, noise and quadratures, then describe coherent pump-probe spectroscopy, off–axis interferometry (digital holography), balanced detection for correlation measurements, and camera–based diagnostics in real and momentum space.
\emph{Conventions, units and typical GaAs numbers are collected in App.~\ref{app:units}.}

\subsection{Excitation and all–optical control}
\label{subsec:excitation-control}

\begin{figure*}[ht]
    \centering
    \includegraphics[width=.9\linewidth]{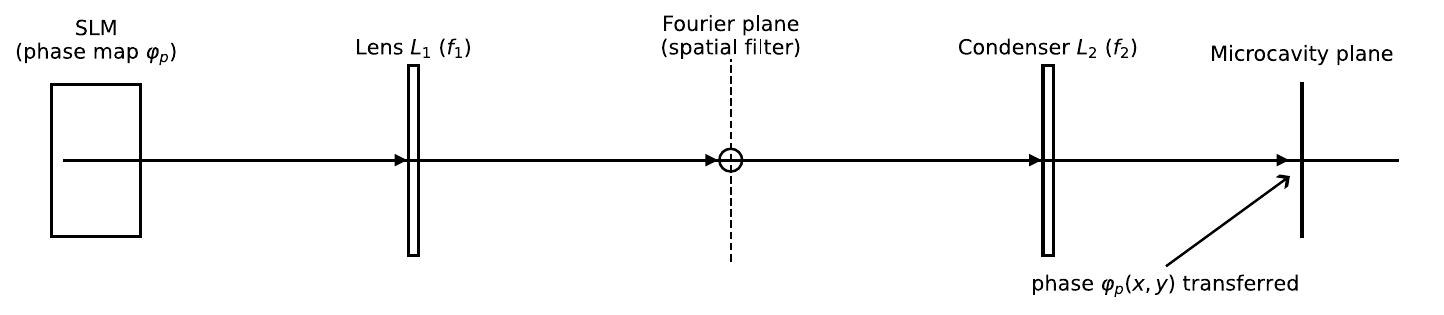}
    \caption{\textbf{Excitation scheme}. 2f-2f relay enabling 1:1 imagine of amplitude of phase.}\label{fig:2f2f}
\end{figure*}

Throughout, we excite the lower polaritons (LP) quasiresonantly.
In steady conditions near resonance, the polariton field inherits the pump phase and momentum,
\[
\psi_0(\mathbf r,t)=\sqrt{n_0(\mathbf r)}\,e^{i[\phi_\tp(\mathbf r)-\omega_\tp t]},
\qquad
\mathbf v_0(\mathbf r)=\frac{\hbar}{m^\ast}\nabla\phi_\tp(\mathbf r),
\]
while the density $n_0$ follows the (local) equation of state~\eqref{eq:eos}.
This expresses the central experimental fact used throughout these lecture notes: the amplitude and phase of the optical pump directly set the density and flow of the polariton fluid.

\paragraph{Local phase–locking approximation} (LPA).
When $|\nabla\phi_\tp|\xi\ll 1$ and the pump envelope varies smoothly on the healing length $\xi$, one may treat the steady state locally as if it were homogeneous. Writing $k_\tp(\mathbf r)=|\nabla\phi_\tp(\mathbf r)|$, the local detuning $\delta(\mathbf r)=\hbar\omega_\tp-\hbar\omega_0-\hbar^2 k_\tp(\mathbf r)^2/2m^\ast$ enters the algebraic state equation
\begin{equation}
\Big[\delta(\mathbf r)-\hbar g\,n_0(\mathbf r)-\hbar g_{\rm res}n_{\rm res}(\mathbf r)\Big]^2+\Big(\tfrac{\hbar\gamma}{2}\Big)^2
=\frac{|F_\tp(\mathbf r)|^2}{n_0(\mathbf r)}.
\label{eq:lpa-eos}
\end{equation}
Equation~\eqref{eq:lpa-eos} conveniently organises bistability and operating points in patterned pumps, and is our starting point for mean–field design [see also Appendix~\ref{app:inhomogeneouspumping}].

\paragraph{Shaping $\phi_\tp(\mathbf r)$ and $|F_\tp(\mathbf r)|$.}
Phase and intensity control are realised optically, e.g. with a reflective spatial light modulator (SLM).
A standard $2f$–$2f$ relay images the SLM plane onto the cavity [Fig.~\ref{fig:2f2f}], with Fourier–plane spatial filtering to remove unwanted diffraction orders; Off-axis (tilted) interferometry on a CCD/CMOS camera recovers the in–plane phase of the emitted field and thus verifies $\phi_\tp\mapsto\arg\psi_0$ transfer~\cite{maitre_dark_soliton_2020}.
These are routine, alignment–robust tools; the $2f$–$2f$ chain also sets the demagnification so that the SLM pixel map covers the desired in–cavity scale.

\subsection{Mean–field characterisation}
\label{subsec:meanfield-characterisation}

\begin{figure*}[h]
    \centering
    \includegraphics[width=.6\linewidth]{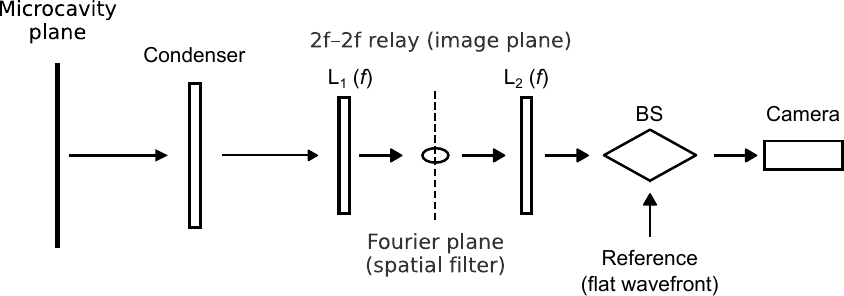}
    \caption{\textbf{Mean-field characterisation}. Light exiting the cavity is imaged on a camera placed in the image plane. Interference with a reference beam (with small tilt $\theta$) enables phase measurements by off-axis interferometry, yielding $I(\mathbf{r})=\abs{\psi_0(\mathbf{r})}^2$ and $\text{arg}(\psi_0(\mathbf{r}))=\phi_\tp(\mathbf{r})$.}
    \label{fig:meanfieldmeas}
\end{figure*}

\paragraph{Near field (real space).}
Real-space imaging of transmitted / reflected light gives $n_0(\mathbf r)\propto |\psi_0(\mathbf{r})|^2$ up to a known optical response. Combining this with calibrated power–to–intensity conversion, one maps $n_0$ against $|F_\tp|$ to reconstruct the local branch of the state equation~\eqref{eq:lpa-eos}. In plane wave drives, the S curve and hysteresis are obtained by sweeping $|F_0|$ at fixed $\delta$; in patterned drives the same analysis applies point-wise under the LPA (with the Doppler shift through $k_\tp$).

\paragraph{Far field (momentum space).}
Fourier–plane imaging directly measures the in–plane momentum content. It is used to confirm that the fluid occupies the intended wave vector(s), to quantify residual parasitic scattering, and to separate the mean field from fluctuations in later steps.

\paragraph{Phase and velocity metrology.}
Off–axis digital holography with a flat reference reconstructs the complex field in a single-camera shot~\cite{phaseutils}, see Fig.~\ref{fig:meanfieldmeas}. The pump–imposed velocity $\mathbf v_0=\hbar\nabla\phi_\tp/m^\ast$ is therefore obtained from the measured phase map and compared to the design target (e.g., for transcritical flows ~\cite{falque_polariton_2025} or vortex pumps with $\phi_\tp(\mathbf r)=D(r)+C\theta$~\cite{guerrero_multiply_2025}).

\subsection{Linear response: coherent pump–probe spectroscopy}
\label{subsec:pump-probe}

We now move from mean-field to elementary excitations.

\paragraph{Principle of coherent pump–probe spectroscopy.}
Excite the steady fluid with a weak phase-coherent probe tone that adds a small sideband to the pump~\cite{claude2021highresolution}:
\begin{equation}
F(\mathbf r,t)=F_\tp(\mathbf r)e^{-i\omega_\tp t}
+\epsilon\,F_{\rm pr}\,e^{i(\mathbf q\cdot\mathbf r-\Omega t)}e^{-i\omega_\tp t}
+\epsilon\,F_{\rm pr}^\ast e^{-i(\mathbf q\cdot\mathbf r-\Omega t)}e^{-i\omega_\tp t},
\end{equation}
with $0<\epsilon\ll 1$ and the analysis frequency $\Omega$ defined in the pump frame. Linearising yields a \emph{forced} BdG problem,
\begin{equation}
\mathcal L(\mathbf q,\Omega)
\begin{pmatrix}u\\ v\end{pmatrix}
=
i\hbar\sqrt{n_0}\,F_{\rm pr}\begin{pmatrix}1\\ -1\end{pmatrix},
\qquad
\mathcal L=\begin{pmatrix}
\Delta_\mathbf q -\hbar\Omega - i\frac{\hbar\gamma}{2} & \hbar g n_0\\[2pt]
-\hbar g n_0 & \Delta_\mathbf q +\hbar\Omega - i\frac{\hbar\gamma}{2}
\end{pmatrix},
\label{eq:forced-bdg}
\end{equation}
where $\Delta_\mathbf q=\tfrac{\hbar^2 q^2}{2m^\ast}-\delta+2\hbar g n_0+\hbar g_{\rm res}n_{\rm res}+\hbar\,\mathbf v_0\!\cdot\!\mathbf q$. The probe response is resonant when $\det\mathcal L$ is minimal; Scanning $\{\mathbf q,\Omega\}$ maps the two sheets $\omega_\pm$ of~\eqref{eq:forced-bdg}.
Because the probe coincides with the Bogoliubov spectrum, one can resolve not only the absorption line but also the anomalous conversion ($u\!\leftrightarrow\!v$) that signals the mixing of positive and negative norm components.
In practice, this method gives high–resolution dispersion maps, local group velocities, and the mass gap~\cite{claude_spectrum_2023}.

\paragraph{Geometry and filtering.}
The angle of incidence of the probe is chosen $\mathbf q$. The pump–set mean field at $\mathbf k=\mathbf k_\tp$ is blocked in the far field; Two small pinholes at $\mathbf k_\tp\!\pm\!\mathbf q$ isolate the outputs of the conjugate probe. A real space iris suppresses the residual background~\cite{claude2021highresolution}. This isolation is essential both for clean line shapes and for the correlation measurements discussed below.

\begin{figure*}
    \centering
    \includegraphics[width=.7\linewidth]{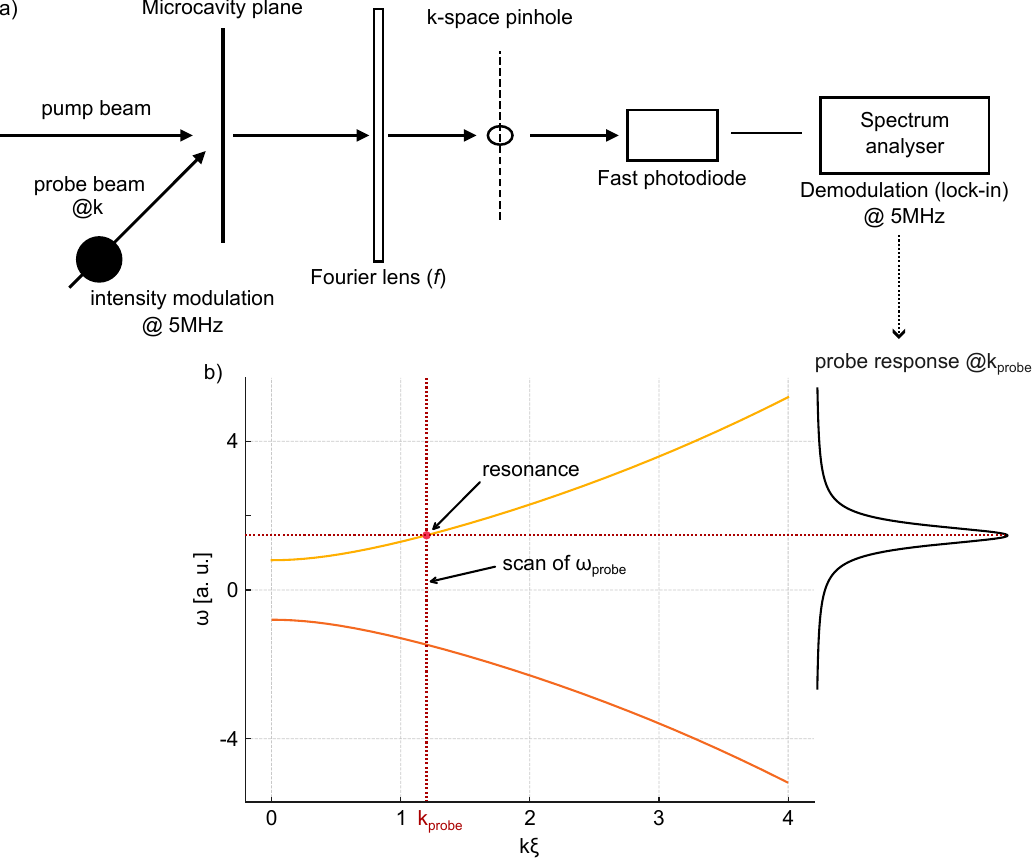}
    \caption{\textbf{Pump-probe spectroscopy}. A pump beam, that may be structured in its phase and intensity profiles, illuminates the sample, creating a polariton fluid. A weak amplitude probe beam (that is not coherent with the pump) is shone upon the sample at set $k$. Its intensity is modulated in time and, for each angle, its frequency is scanned. When it resonates with the fluid, its response peaks, which is detected in light coming out of the cavity via the demodulated photocurrent of a photodiode.}\label{fig:pumpprobe}
\end{figure*}

\paragraph{Readout options.}
(i) \emph{Camera (CMOS/CCD) spectroscopy}: a spectrally narrow probe is stepped across $\Omega$ (or, in tunable laser instruments, the optical frequency), while a camera records near– or far–field intensities; interfering light coming out of the cavity with a reference pick-up of the probe (homodyne detection) gives full 2D maps $I(\mathbf r,\Omega)$ or $I(\mathbf k,\Omega)$~\cite{guerrero_multiply_2025}. (ii) \emph{Heterodyne at radio frequency (RF)}: the probe is modulated in amplitude or phase at an RF $\Omega$ and the transmitted field is demodulated on a fast photodiode to extract components of the complex susceptibility in phase / quadrature~\cite{claude2021highresolution}, which directly resolves the content of $u/v$.

\subsection{Photodetection, noise and quadratures}
\label{subsec:photodetection}

Consider a single optical mode with an annihilation operator \(\hat a\) at frequency \(\omega\). The electric field operator at a fixed point can be written as~\cite{bachor_guide_2004}
\begin{equation}
\hat E(t)=E_0\big(\hat a\,e^{-i\omega t}+\hat a^\dagger e^{+i\omega t}\big)=E_0\big(\hat X\cos\omega t+\hat Y\sin\omega t\big),
\end{equation}
with field quadratures \(\hat X=\hat a+\hat a^\dagger\) and \(\hat Y=i(\hat a^\dagger-\hat a)\), obeying \([\hat X,\hat Y]=2i\). A photodiode of quantum efficiency \(\eta\) produces a mean photocurrent \(\bar i=\eta e P/(\hbar\omega)\). For a shot–noise–limited input, the variance of the integrated photocurrent in a time window \(\Delta t\) is \(\Delta i^2=e\,\bar i/\Delta t\). In the frequency domain, a spectrum analyser with resolution bandwidth \(\delta f\) centred at the analysis frequency \(\omega\) returns
\begin{equation}
\Delta i^2=2\,\delta f\,S_i(\omega),
\end{equation}
where \(S_i\) is the (two–sided) photocurrent power spectral density.

\paragraph{Shot–noise calibration and operating point.}
A simple, robust calibration is the linear scaling of photocurrent variance with optical power on a fast photodiode, performed at an RF analysis frequency high enough to avoid slow technical noise and well within the detector bandwidth. This sets the shot–noise unit (SNU) and verifies that the lasers are shot–noise limited under the measurement settings~\cite{bachor_guide_2004}.

\begin{figure*}[h]
    \centering
    \includegraphics[width=.6\linewidth]{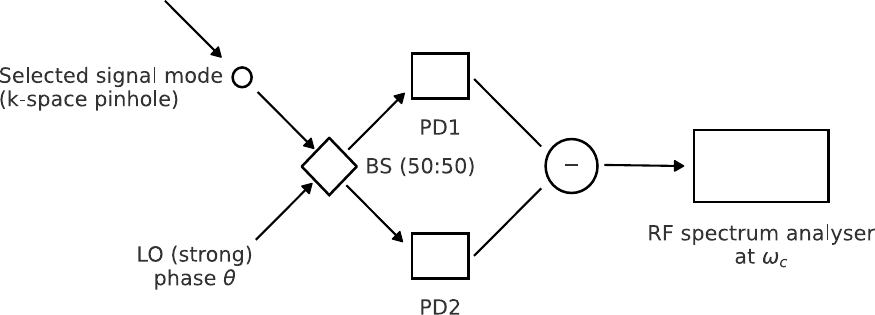}
    \caption{\textbf{Homodyne detection scheme}. A signal at given angle $k$ is interfered with a mode-matched local oscillator (LO) of high intensity and tunable phase $\theta$. The interferences amplitudes are detected by photodiodes and the difference in electrocurrents is measured by a spectrum analyser returning $\Delta_i^2=2\delta f S_i(\omega)$.
    In order to measure (anti)correlations in commuting quadratures between two optical modes, homodyne detection must be preformed on each of them, with the corresponding LO.}
    \label{fig:homodyne}
\end{figure*}

\paragraph{Quadrature selection (homodyne).}
Homodyne detection [Fig.~\ref{fig:homodyne}] with a strong local oscillator (LO) of phase \(\theta\) measures the rotated quadrature \(\hat X_\theta=\hat X\cos\theta+\hat Y\sin\theta\). In our convention \([\hat X,\hat Y]=2i\), the vacuum noise is \(\langle\Delta \hat X_\theta^2\rangle=1\). For a single–mode squeezed state with squeezing parameter \(r\) and squeezing angle \(\theta_s\), the quadrature noise \(V(\theta)=\langle\Delta \hat X_\theta^2\rangle\) obeys
\begin{equation}
V_{\min}=e^{-2r},\qquad V_{\max}=e^{+2r},
\end{equation}
with the minimum at \(\theta=\theta_s\).
Equivalently,
\[
V_\theta=\langle X_\theta^2\rangle
=1+2\,\mathrm{Re}\!\big[e^{-2i\theta}\langle \hat a\hat a\rangle\big]+2\langle \hat a^\dagger\hat a\rangle,
\]
so that the scan of \(V_\theta\) directly gives the entries of the $2\times 2$ covariance submatrix for that mode~\cite{bachor_guide_2004}. In our context, phase-sensitive detection at the emission angle \(\mathbf{k}\) can be used to reconstruct the Bogoliubov content by scanning \(\theta\) and fitting \(V(\theta)\) to extract \(r\) and \(\theta_s\). In massive or weakly lossy regimes, the quadrature that maximises the Hawking pair signal aligns with the anomalous weight \(A_{HP}\) at the analysis frequency (Sec.~\ref{subsec:balanced}).

\paragraph{Balanced homodyne (temporal).}
To resolve quadratures at RF, interfere the selected output mode with a strong LO on a 50:50 beam splitter and subtract the two photodiode currents. With the LO phase~$\theta$ one measures \(X_\theta\) as above. Using two balanced receivers simultaneously on conjugate outputs at \(\mathbf k_\tp\!\pm\!\mathbf q\) gives access to inter–mode covariances and hence to two–mode squeezing parameters and entanglement criteria. Practical implementation requires spatial filtering (iris and $k$–space pinholes) so that only the targeted modes reach each photodiode, plus single arm (LO blocked) noise checks.

\subsection{Balanced detection for intensity correlations}
\label{subsec:balanced}

Balanced detection rejects common classical noise and directly measures the correlations between two optical channels selected in \(k\)–space. After spatial filtering to isolate the Bogoliubov signals at \(\pm\mathbf{k}_{\mathrm{pr}}\) from the mean field, the two beams are sent to matched photodiodes; Their photocurrents are subtracted and analysed at a fixed radio frequency \(\omega_c\). In a shot–noise calibration, the variance of the difference signal must scale linearly with optical power; this is best performed at MHz frequencies to avoid mechanical noise from the cryostat~\footnote{In practice, MHz analysis frequencies (e.g. \(\SI{1.5}{\mega\hertz}\)) provide a good compromise between avoiding mechanical lines and staying within the photodiode bandwidth; typical settings are resolution bandwidth \(\mathrm{RBW}\sim \SI{100}{\kilo\hertz}\); video bandwidth $\mathrm{VBW}\sim\SI{100}{\kilo\hertz}$ and sweep times of order \(\SI{0.1}{\second}\)~\cite{Falquethesis2025}.}.

\begin{figure*}[h]
    \centering
    \includegraphics[width=.55\linewidth]{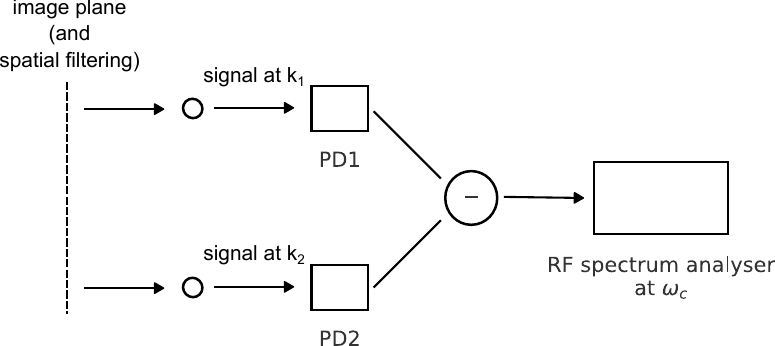}
    \caption{\textbf{Balanced detection}. The intensity in two beams coming from the cavity ($\mathbf{r}$ and $k$ resolution) is recorded by two different photodiodes. The difference in the resulting photocurrent is measured with a spectrum analyser, yielding intensity correlations.}
    \label{fig:balanceddet}
\end{figure*}

\paragraph{Filtering and sanity checks.}
Real–space masking removes edges that eject high–\(k\) light; a second filter in the Fourier plane isolates the two target modes and eliminates residual mean–field photons, preventing spurious correlations from the single–mode pump background. A cross–check uses two independent probe wavevectors \((k_x,0)\) and \((0,k_y)\): Only conjugate pairs should display noise reduction in the difference channel.

\paragraph{From spectra to correlation coefficients.}
Let \(S_{a}(\omega)\), \(S_{b}(\omega)\) be the single–channel photocurrent PSDs and \(S_{-}(\omega)\) be the PSD of the difference current. Assuming identical gains and quantum efficiencies, the (normalised) correlation coefficient at the analysis frequency is
\begin{equation}
C(\omega_c)=\frac{S_{a}(\omega_c)+S_{b}(\omega_c)-S_{-}(\omega_c)}{2\sqrt{S_{a}(\omega_c)\,S_{b}(\omega_c)}}.
\label{eq:C-from-PSD}
\end{equation}
For independent channels \(C=0\); positive correlations give \(0<C\le 1\), and a perfect classical correlation would drive \(S_{-}\!\to 0\). The shot–noise calibration pins the unity level.

\paragraph{Connection to Bogoliubov weights, amplification and squeezing.}
Starting from Eqs.~\eqref{eq:out-occupancy}–\eqref{eq:out-anomalous}, vacuum inputs produce finite outgoing flux and anomalous pair correlators because the negative norm input enters the Bogoliubov transform as a \emph{creation} operator. This mixing is an amplifier: the relevant `gains' are the mode–conversion amplitudes $S_{i\leftarrow -}(\omega)$ of the unique negative norm ingoing channel at the analysis frequency. When selecting conjugate outputs $H\equiv u^{\rm out}_{+}$ (upstream, positive norm) and $P\equiv d^{\rm out}_{-}$ (downstream, negative norm) at fixed laboratory frequency $\omega$, the dominant observable scalings are
\begin{align}
\text{near field (density):}\quad & G^{(2)}_{HP}\;\propto\; |D_H D_P|\,\big|S_{H\leftarrow -}\,S_{P\leftarrow -}\big|,\\
\text{far field (number):}\quad & C_{HP}\;\propto\; |A_{HP}|\,\big|S_{H\leftarrow -}\,S_{P\leftarrow -}\big|,
\end{align}
with detection weights $D_i=u_i+v_i$ and anomalous bilinear $A_{ij}=u_i v_j+v_i u_j$ defined in Sec.~\S\ref{subsec:observables}. Thus, the signal of near–field difference is maximised when both channels are phononic ($u\simeq v\Rightarrow |D|$ large), whereas anomalous momentum–space correlations are controlled by $A_{ij}$. In homodyne detection, the quadrature that minimises noise aligns with the squeezing axis of the two–mode state; operationally, that axis is set by the complex phase of $A_{HP}\,S_{H\leftarrow -}\,S_{P\leftarrow -}$ at the chosen $\omega$.

\paragraph{Camera–based correlation imaging.}
For stationary fluids, cameras enable parallel, spatially resolved measurements. In the near field, one forms$
G^{(2)}(\mathbf r,\mathbf r')=\langle :\delta I(\mathbf r)\delta I(\mathbf r'):\rangle$, 
whose long, narrow tongues follow the group–velocity geometry of correlated Bogoliubov pairs [see Fig.~\ref{fig:fig6}~(a)]. In the far field, the momentum-space covariance $
C(\mathbf k,\mathbf k')=\langle :\delta n_{\mathbf k}\,\delta n_{\mathbf k'}:\rangle$
exhibits a sharp ridge along the frequency–matched partner map $\mathbf k\mapsto\mathbf k'$ determined by~\eqref{eq:WKB-local} [see Fig.~\ref{fig:fig6}~(b)].
Off–axis digital holography promotes these intensity–only observables to phase–sensitive ones by reconstructing the complex field; choosing the detection quadrature then maximises either density–like ($u+v$) or anomalous ($uv$) features.

\paragraph{Noise budgets and rejection.}
Residual mean-field noise (e.g. amplitude induced by self-phase modulation / phase fluctuations in the emitted light) can spuriously correlate spatially separated regions if filtering is insufficient. Spatial isolation of the two conjugate probe outputs and common–mode rejection in the balanced detector are, therefore, crucial for unambiguous fluctuation measurements. Diagnostics include single–arm spectra, swapping pinholes between arms, and verifying the disappearance of correlations when the probe is detuned from the Bogoliubov resonance.

\newtcolorbox{howtobox}[1][]{
  enhanced, breakable,
  colback=white, colframe=black!50,
  boxrule=0.5pt, arc=1pt,
  left=1em, right=1em, top=0.6em, bottom=0.6em,
  fonttitle=\bfseries, title={How to: a schematic workflow},
  #1
}
\begin{howtobox}
\label{subsec:practical-sequence}

\begin{enumerate}[leftmargin=2em]
\item \textbf{Design the mean field.} Specify $\phi_\tp(\mathbf r)$ and $|F_\tp(\mathbf r)|$ for the target $\mathbf v_0$ and $n_0$ (or $c_{\rm s}$). Use the LPA and Eq.~\eqref{eq:lpa-eos} to choose a robust operating point (detuning, power) on a stable branch.
\item \textbf{Impose and verify.} Upload SLM maps; align the $2f$–$2f$ relay and Fourier filter; verify $\arg\psi_0\simeq\phi_\tp$ by off–axis interferometry; confirm $\mathbf k$ content in the far field and $n_0(\mathbf r)$ in the near field.
\item \textbf{Map the Bogoliubov spectrum.} Run the coherent pump–probe: step $\{\mathbf q,\Omega\}$; isolate $\mathbf k_\tp\!\pm\!\mathbf q$; record line shapes (camera or RF demodulation) and fit to \eqref{eq:forced-bdg} to extract $\omega_\pm$, group velocities and anomalous weights.
\item \textbf{Measure noise and correlations.} Calibrate SNU; acquire balanced–detector quadrature spectra and/or camera–based covariances; from the measured covariances reconstruct squeezing parameters and (when relevant) entanglement witnesses for the selected output modes.
\end{enumerate}

Within the Gaussian framework of Sec.~\S\ref{subsec:hawking_gaussian_optics}, the full information about the detected state at a given analysis frequency is contained in the covariance matrix $\sigma$. Homodyne data $V(\theta)$ reconstruct marginals in a single mode (diagonal blocks of $\sigma$); balanced detection of a pair produces cross-covariances and logarithmic negativity or an entanglement witness. In the horizon frequency interval, the reduced state of $(H,P)$ is a two–mode squeezed thermal state; The measured amplification and squeezing are set by the same $S$–matrix elements that control the flux (Sec.~\S\ref{subsec:observables}).
\end{howtobox}

\section{Where do we go from here?}

\subsection*{Beyond the Hawking effect}

\begin{figure*}[h]
    \centering
    \includegraphics[width=.7\linewidth]{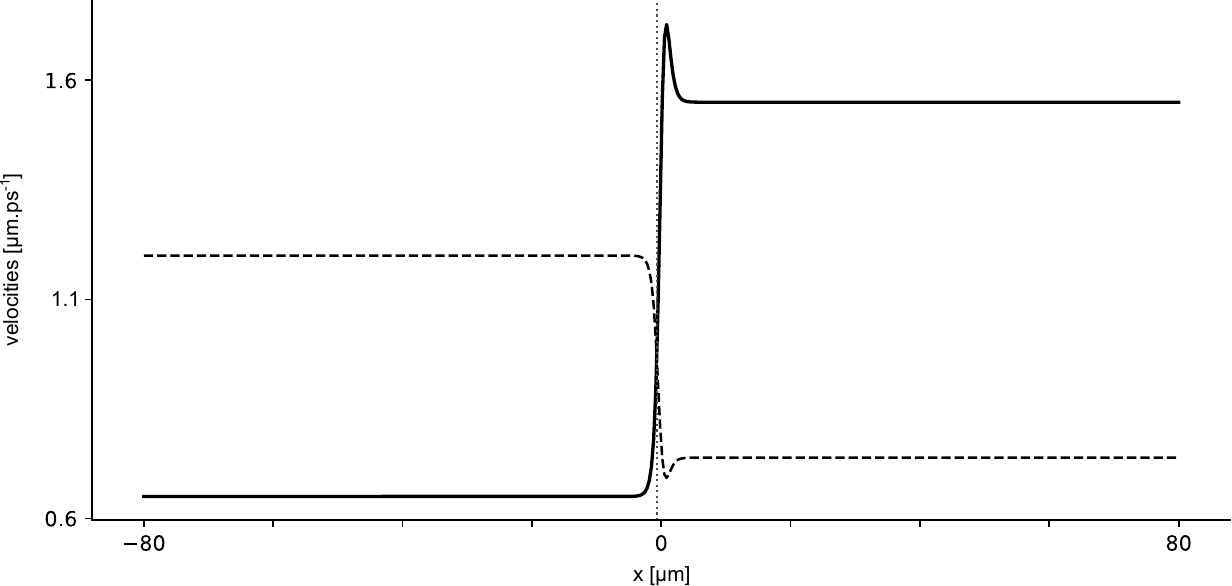}
    \caption{\textbf{Resonator inside the horizon}. A leaky resonator is formed by the dip in density in between two regions of higher density. This geometry supports precursors of dynamically unstable modes known as QNMs~\cite{jacquet_quantum_2023}. The dynamical instability is quenched by the homogeneous polariton losses. $\xi\approx\SI{2}{\micro\meter}$ inside the horizon. The width of the density dip is $\sim 2\xi$. \label{fig:qnm}}
\end{figure*}

\paragraph{Hawking effect and QNMs} Waterfall geometries can also create a short, leaky resonator just inside the horizon~\cite{jacquet_analogue_2022,falque_polariton_2025}, see Fig.~\ref{fig:qnm}; its quasi-normal mode (QNM) at \(\tilde\omega_{\mathrm{qnm}}=\Omega_{\mathrm{qnm}}-i\Gamma_{\mathrm{qnm}}/2\) typically lies just above \(\omega_{\max}\)~\cite{jacquet_quantum_2023}.
In frequency resolved \emph{stimulated} scattering, this is predicted to yield a narrow Breit-Wigner-like enhancement in the downstream positive–norm transmission (with a \(\pi\)-phase slip) at \(\omega=\Omega_{\mathrm{qnm}}\)~\cite{burkhard2025stimulatedhawkingeffectquasinormal}.
Below \(\omega_{\max}\), where the negative norm channel is open, the same near‑horizon structure mediates Hawking amplification and governs the strength of \(H\)–\(P\) correlations~\cite{jacquet_quantum_2023}.
Thus, the QNM serves both as a resonant fingerprint of near-horizon geometry and as an active mediator of amplification.

\begin{figure*}[h]
    \centering
    \includegraphics[width=.9\linewidth]{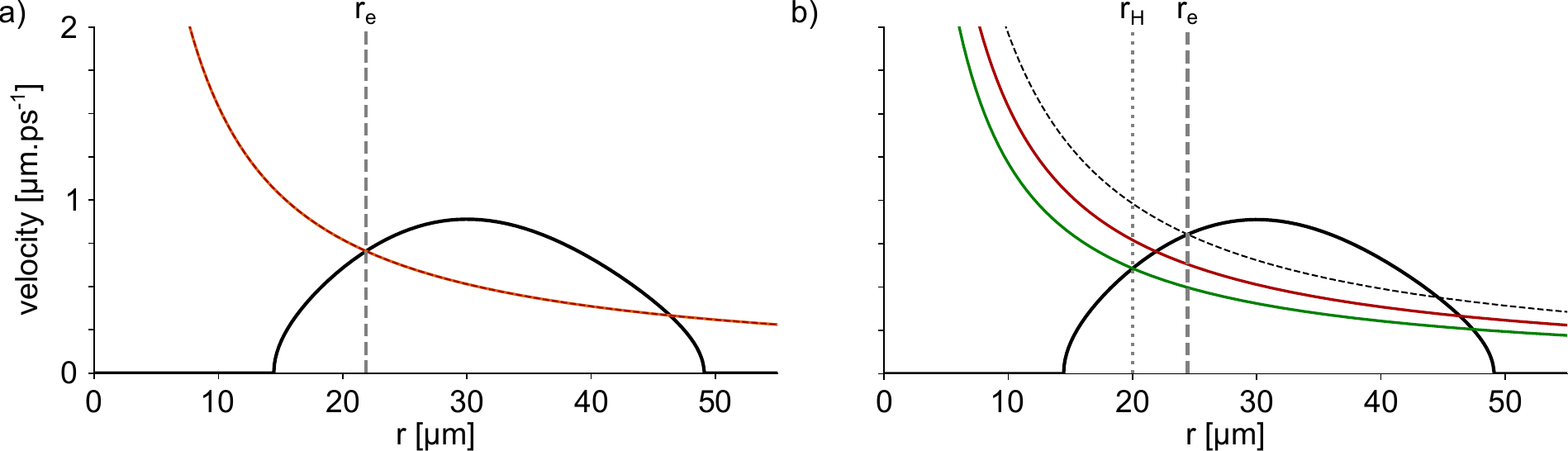}
    \caption{\textbf{Bathtub flow}. A Laguerre-Gauss pump with or without convergence can implement a rotating flow geometry with or without a horizon. Black, $c_\tb$; red, $v_\theta$; green, $v_\perp$; dashed black, $\abs{\mathbf v}$.
    There is an ergosurface at $r_\mathrm{e}$ where $\abs{\mathbf v}=c_\tb$ and a horizon at $r_\tH$ where $v_\perp=c_\tb$.
    \textbf{a)} horizonless ergosurface.
    \textbf{b)} ergosurface with a horizon inside.\label{fig:rota}}
\end{figure*}
\paragraph{Rotating geometries and superradiance.}
Axisymmetric flows add an azimuthal drift with angular velocity \(\Omega(r)\).
Two radii organise the rotating problem: the \emph{horizon} \(r_\tH\) defined by \(|v_{\!\perp}|=c_\tb\), and the \emph{ergosurface} \(r_\mathrm{e}\) defined by \(|\mathbf v|=c_\tb\) (where the Killing vector of time translation becomes space-like)~\cite{visser_acoustic_1998,fischer_riemannian_2002}.
The unstable circular null orbit defines a \emph{light ring} \(r_\mathrm{LR}\) (when present).
For modes with azimuthal number \(m\), \emph{rotational superradiance} occurs when the co‑rotating frequency becomes negative in the fluid frame, \(\omega-m\Omega(r)<0\) across the ergoregion~\cite{delhom_entanglement_2024}.
With a horizon, this is predicted to produce a phase-insensitive gain for co-rotating modes and a \(m\)-dependent redistribution of graybody factors~\cite{agullo_entanglement_2024}.
In \emph{horizonless} configurations (purely rotational flows with \(r_H\) absent but \(r_e\) present), the same condition yields amplified reflection off the ergoregion~\cite{delhom_entanglement_2024}; with additional feedback a long‑lived \emph{ergoregion resonance} (or instability) can arise~\cite{giacomelli_ergoregion_2020}.
In polariton fluids, dissipation regularises the latter into sharp, controllable, \emph{superradiant} lines set by \(\Omega(r)\) and the impedance of the surrounding profile~\cite{guerrero_multiply_2025}.

In rotating, transcritical flows, horizon mixing (Hawking) and ergoregion mixing (superradiance) act in cascade.
A convenient Gaussian picture is a two-mode near-horizon squeezer with gain \(r(\omega,m)\) followed by an \(m\)–dependent passive network that encodes rotation (\(\propto m\Omega\)), light ring filtering, and far-field greybody transmission, as in Kerr black holes~\cite{agullo_entanglement_2024}.
This cascade predicts: (i) an \(m\) selective enhancement of pair creation and entanglement for corotating channels; (ii) phase rotations that shear the \(H\)–\(P\) covariance ridge in \(k\)–space; and (iii) \emph{superradiant} shoulders or resonant peaks outside the Hawking window, which are absent in nonrotating flows.
Frequency‑unresolved intensity noise can miss these effects; they become clear in the frequency- and \(m\)-resolved pump-probe and homodyne metrology.

Optical access allows \emph{tomography} by spatial filtering and holographic reading. Four regions are operationally distinct:
\begin{enumerate}\setlength\itemsep{2pt}
    \item \(r<r_\tH\) (inside the horizon): detect the negative norm content of \(P\) and its phase–locked pairing with \(H\) via \(A_{HP}\) at negative frequencies; QNM–assisted ringdown appears as a narrow feature (accompanied by gain) at positive frequency.
    \item \(r_H<r<r_\mathrm{e}\) (between horizon and ergosurface): both Hawking conversion and rotational phase accumulation act; expect \(m\)–selective squeezing axes and a tilt of the \(H\)–\(P\) ridge.
    \item \(r_e<r<r_\mathrm{LR}\) (between ergosurface and light ring): the superradiant amplification is maximal; seek \emph{horizonless} resonant features in the corotating channels \(m\) and a sign flip of the energy flux in the rotating frame.
    \item \(r>r_\mathrm{LR}\) (far outside): extract greybody factors and total gain; at \(\omega\gtrsim\omega_{\max}\), look for narrow, phase-rotating transmission peaks (QNM and / or superradiant resonances).
\end{enumerate}

\subsection*{Outlook}
On the numerical side, map \((\kappa,\Omega,\text{gap})\) to \((\Omega_{\mathrm{qnm}},\Gamma_{\mathrm{qnm}})\) and the superradiant window, and quantify the cascade squeezer\(\to\)rotation\(\to\)greybody by fits to \(r(\omega,m)\) and phase shifts.
Experimentally, programmable curved spacetimes in polariton fluids already provide: (a) downstream support and pump–probe spectroscopy to resolve the Hawking window and \(H\)–\(P\) pairing~\cite{falque_polariton_2025}; (b) vortex and shear pumps to engineer \(r_e\), \(r_H\) and \(r_L\) independently~\cite{guerrero_multiply_2025}; and (c) covariance imaging and homodyne tomography~\cite{Boulier_squeezing_2014}, that need to be generalised to access \(m\) resolved gains and entanglement.
This puts \emph{rotational superradiance} (with or without horizons), its interplay with the Hawking effect~\cite{agullo_entanglement_2024}, and dumb hole spectroscopy~\cite{torres_analogue_2019} within immediate reach, with region‑resolved measurements that connect laboratory observables to open questions on the near-horizon~\cite{balbinot_hawking_2006,zapata_resonant_2011,finazzi_black_2010,finazzi_spectral_2011,finazzi_robustness_2011,coutant_dynamical_2016,bermudez_resonant_2018,torres_quasinormal_2020,jacquet_quantum_2023,porrotunneling2024} and ergoregion physics that connect astrophysics and condensed-matter systems~\cite{press_floating_1972,Dolan_bhbomb_2007,giacomelli_ergoregion_2020,patrick_backreaction_2021,geelmuyden_sound-ring_2021,patrick_origin_2022,patrick_quantum_2022,svancara_rotating_2024}.

\subsection*{Conclusion}
To conclude, in these notes we have shown that a driven polariton fluid of light realises a controllable effective spacetime in which long‑wavelength fluctuations obey (massive) Klein–Gordon dynamics and horizon scattering is fully captured by a pseudo unitary stationary S matrix.
The Hawking process then reduces to a universal two-mode vacuum squeezer, filtered by graybody transmission, with contrasts in real and momentum space fixed by simple detection weights that are directly obtained from the Bogoliubov theory.
Coupled with the practical quantum optics toolkit -- phase-imprinted flows, coherent pump-probe spectroscopy, and balanced or homodyne detection -- this provides a complete quantitative workflow, from mean-field design to the extraction of amplification, squeezing, and entanglement.
The resulting metrology, organised by the horizon frequency window bounded by the mass threshold and dispersive cut‑off, is sufficiently general to interrogate near‑horizon physics, quasinormal modes, and the interplay with rotation and superradiance, and sufficiently precise to enable a programme of 'dumbhole spectroscopy' in the laboratory.
In this sense, the platform provides a calibratable, optically addressable field theory in which predictions of quantum fields on curved spacetimes can be tested with high fidelity.

\section*{Acknowledgments}
    We are thankful to Luca Giacomelli, Adria Delhom and Ivan Agullo for countless profound discussions that helped shaping the concepts presented here. We are indebted to Alberto Bramati for his support and for sharing his insights on polariton physics with us. MJJ was supported by a DIM Sirteq postdoctoral fellowship and acknowledges funding from the EU Pathfinder 101115575 Q-One as well as from the Schwinger Foundation that supported the ``Analogue Gravity 20203'' workshop in Benasque.
    These lecture notes could not have been written without invaluable input from K\'evin Falque and Killian Guerrero. This manuscript is but a rendering of their work and what could be learnt from it to date.

\newpage

\newpage
\appendix
\section{Reader's guide for students with different backgrounds}
\subsection{For the Relativist}

\paragraph{Dictionary of concepts.}
Long–wavelength phase fluctuations obey a Klein–Gordon equation on an acoustic spacetime with drift \(v_0\) and light–cone speed \(c_B\). Acoustic (Killing) horizons occur where \(|v_{\perp}|=c_B\); the surface gravity is
\[
\kappa=\frac{1}{2c_B}\,\partial_n\!\left(c_B^2-v_{\perp}^2\right)\Big|_{r_H},
\]
setting the low–\(\omega\) Hawking slope. The mapping is \emph{kinematic}: it captures null cones, Doppler shifts, positive/negative KG norm, and mode mixing, while dispersion and loss limit exact KG conservation and introduce biorthogonality.

\paragraph{Scattering and pseudo–unitarity.}
At fixed \(\omega\) within the Hawking window there are three relevant outgoing channels \((H,P,W)\). The stationary scattering matrix \(S(\omega)\) is pseudo–unitary with respect to the KG metric, and mixing with a negative–norm input guarantees spontaneous amplification even from vacuum. The Gaussian “squeezer\(\rightarrow\)greybody” factorisation is the analogue of the standard near–horizon Bogoliubov mixing followed by frequency–dependent transmission.

\paragraph{Frequency window and ``universality''}
Unlike the massless, dispersionless textbook field, the present scalar field has (i) a mass threshold \(\omega_{\min}\) in gapped regimes, and (ii) a dispersive cut–off \(\omega_{\max}\) where the downstream negative–norm branch closes; thus Hawking scattering is kinematically allowed only for \(\omega\in[\omega_{\min},\omega_{\max}]\). Low–\(\omega\) thermality is recovered for \(\omega\ll \kappa\) modulo smooth greybody factors.

\paragraph{Beyond horizons: QNMs and rotation}
Waterfall geometries may host a short, leaky resonator inside the horizon, yielding a quasinormal mode just above \(\omega_{\max}\), which also mediates Hawking amplification below \(\omega_{\max}\). Axisymmetric flows realise ergoregions (\(|v|=c_B\)), light rings, and superradiance with the usual condition \(\omega-m\Omega<0\); operational consequences include \(m\)–selective gains and phase shears of the \(H\)–\(P\) ridge.

\paragraph{How to read the equations here with a GR/QFT lens}
Start from the hydrodynamic KG equation and its acoustic metric; identify horizons and \(\kappa\); construct the in/out basis and the pseudo–unitary \(S\); then use the Gaussian circuit to read off observables—fluxes, pair correlators, squeezing, and entanglement—exactly as one would compute greybody–filtered Hawking spectra, but now with laboratory access to quadratures and covariances.

\paragraph{Take–home} The notes implement an effective laboratory realisation of the \emph{kinematics} of QFT on curved spacetime—horizons, surface gravity, Hawking mixing, greybody factors, QNMs, and superradiance—translated into Gaussian–optical observables accessible with standard quantum–optics metrology.

\subsection{For the Optician}

\paragraph{Core picture}
In this platform the Hawking process is a \emph{two–mode squeezing} mechanism seeded by vacuum fluctuations: a near–horizon Bogoliubov mixing between a positive–norm and a negative–norm channel produces correlated outputs \(H\) (upstream, \(+\) norm) and \(P\) (downstream, \(-\) norm), while the additional downstream \(+\)–norm channel \(W\) plays the role of a frequency–dependent greybody pathway. The same physics may be viewed as a compact Gaussian circuit: (i) a two–mode squeezer of gain \(r(\omega)\) followed by (ii) a passive beam splitter of amplitude transmission \(\sqrt{T(\omega)}\) on the upstream leg. This factorisation directly connects to the measurable flux \(\bar n_{\rm far}(\omega)=T(\omega)\sinh^2 r(\omega)\), the anomalous correlator \(M_{HP}(\omega)\propto \sqrt{T(\omega)}\sinh 2r(\omega)\), quadrature squeezing, and entanglement criteria. 

\paragraph{Detection weights \& where contrast lives}
Near–field density correlations scale with \(D_i=u_i+v_i\) (phononic content), while far–field momentum–space covariances scale with the anomalous weights \(A_{ij}=u_i v_j+v_i u_j\). Expect the strongest signals in \(H\)–\(P\), typically weaker in \(H\)–\(W\), and conditional visibility in \(P\)–\(W\) depending on the residual \(v\)–content of \(P\) and \(W\). 

\paragraph{Optimising homodyne readout}
Use LO–phase scans to align with the squeezing axis set by the complex phase of \(A_{HP}\,S_{H\leftarrow -}\,S_{P\leftarrow -}\) at the analysis frequency. This maximises either density–like (near field) or anomalous (far field) features depending on the chosen quadrature.  

\paragraph{Regimes across the horizon frequency interval}
Correlations are kinematically allowed only for \(\omega\in[\omega_{\min},\omega_{\max}]\), with a low–frequency threshold set by the mass gap and an upper cut–off where the downstream negative–norm branch closes. In gapless (phononic) regimes, infrared enhancement yields wide tongues in \(x\)–\(x'\); in massive regimes, correlations switch on above \(\omega_{\min}\) and tongues narrow. All contrasts roll off as \(\omega\to\omega_{\max}\).  

\paragraph{What to do first in the lab.}
(i) Verify the mean–field and \(k\)–space filtering; (ii) run pump–probe to locate \(\omega_\pm(k)\), group velocities and the mass gap; (iii) calibrate SNU; (iv) measure balanced/homodyne spectra and camera covariances to reconstruct \(r(\omega)\) and \(T(\omega)\).  

\paragraph{Worked examples.}
For explicit \(x\)–\(x'\) correlation patterns in the waterfall geometry, see Jacquet \emph{et al.}, EPJ~D \textbf{76}, 152 (2022). For detailed \(k\)–\(k'\) correlation maps in the same spirit, see the forthcoming manuscript by Olivera \emph{et al.}

\paragraph{Take–home message} Treat the horizon as a tunable, lossy two–mode squeezer. Use the Gaussian toolbox (homodyne, balanced detection, covariance imaging) to extract amplification, squeezing, and entanglement from the measured \(D,N,A_{ij}\)–weighted observables.

\section{Dark‑exciton reservoir: minimal model, consequences for spectra, and practical use}
\label{app:reservoir}

\subsection{Physical origin and phenomenology}
In III–V quantum wells, the optically active (``bright'') heavy‑hole excitons carry spin projection $J_z=\pm 1$ and couple to circularly polarised light, while $J_z=\pm 2$ excitons are ``dark'' and do not couple directly to light.`
Carrier‑exchange processes between bright excitons convert population into dark states, so even under resonant, coherent pumping a long‑lived population of excitons not phase‑locked to the laser can build up~\footnote{For a concise spin picture and the role of exchange in creating dark excitons, see the discussion leading to Eqs.~(2.48)–(2.49) and the paragraph \emph{Dark excitons creation through carrier exchange} in~\cite{Falquethesis2025}.} .
This incoherent reservoir interacts repulsively with bright excitons (and hence with lower polaritons via the excitonic fraction) and may also act as a loss channel out of the LP mode; its long decay time makes it dynamical on the timescales of collective excitations, thus altering both the equation of state and the Bogoliubov spectrum.

\subsection{Minimal coupled model}
At the level required for these notes, we model the lower‑polariton (LP) field $\psi(\vb{r},t)$ and a scalar dark‑exciton reservoir density $n_{\rm res}(\vb{r},t)$ by~\cite{amelio_galilean_2020}
\begin{align}
i\hbar\,\partial_t\psi
&=\Big[\hbar\omega_0 - \frac{\hbar^2\nabla^2}{2m^\ast} + \hbar g\,|\psi|^2 
+ \hbar g_{\rm res} n_{\rm res} - i\hbar\,\tfrac{\gamma_{\rm LP}+\gamma_{\rm in}}{2}\Big]\psi
\;+\;\hbar F_\tp(\br,t),
\label{eq:R-ddGPE}\\
\partial_t n_{\rm res}
&= -\gamma_{\rm res}\,n_{\rm res} + \gamma_{\rm in}\,|\psi|^2.
\label{eq:R-rate}
\end{align}
Here $m^\ast$ is the LP effective mass, $g$ the LP–LP interaction, $g_{\rm res}$ the LP–reservoir (cross‑Kerr) interaction, $\gamma_{\rm LP}$ the LP radiative loss, and $\gamma_{\rm in}$ the scattering rate from LPs into the reservoir; $\gamma_{\rm res}$ is the reservoir decay rate. The coherent pump is $F_\tp(\br,t)=|F_\tp(\br)|e^{i(\phi_\tp(\br)-\omega_\tp t)}$. Equations \eqref{eq:R-ddGPE}–\eqref{eq:R-rate} are the working model we will use whenever the reservoir is relevant.

Under a plane‑wave pump~\eqref{eq:pump}, a stationary solution 
$\psi=\sqrt{n}\,e^{i(\bk_\tp\!\cdot\!\br-\omega_\tp t)}$ obeys the cubic equation of state~\eqref{eq:eos} with $\gamma=\gamma_{\rm LP}+\gamma_{\rm in}$. In the steady state of \eqref{eq:R-rate}, $n_{\rm res}=(\gamma_{\rm in}/\gamma_{\rm res})\,n$, so that the reservoir appears as a static renormalisation of the interaction:
\begin{equation}
g_{\rm eff} \equiv g + g_{\rm res}\,\frac{\gamma_{\rm in}}{\gamma_{\rm res}},\qquad
(\text{adiabatic‑reservoir limit } \partial_t n_{\rm res}\!\approx\!0).
\label{eq:R-geff}
\end{equation}
The reservoir reduces the optical bistability range and increases the interaction induced blueshift at fixed $n$.

\paragraph{Memory kernel (beyond adiabatic elimination).}
If $\gamma_{\rm res}$ is comparable to the eigenfrequencies of interest, eliminating $n_{\rm res}$ produces a retarded, frequency dependent nonlinearity~\cite{amelio_galilean_2020}:
\begin{equation}
n_{\rm res}(t)=\frac{\gamma_{\rm in}}{\gamma_{\rm res}}\!\int_{-\infty}^t\!dt'\;
e^{-\gamma_{\rm res}(t-t')}\,n(t'),\qquad
\Rightarrow\qquad
\hbar g_{\rm eff}(\omega)=\hbar g+\hbar g_{\rm res}\,\frac{\gamma_{\rm in}}{\gamma_{\rm res}-i\omega}.
\label{eq:R-memory}
\end{equation}
Equation \eqref{eq:R-memory} is the simplest way to account for non‑barotropic effects (delayed pressure response) in hydrodynamic interpretations; it smoothly reduces to \eqref{eq:R-geff} for $|\omega|\ll\gamma_{\rm res}$.

\subsection{Reservoir‑dressed Bogoliubov spectrum}
Linearising Eqs.~\eqref{eq:R-ddGPE}–\eqref{eq:R-rate} around a plane‑wave steady state gives a pair of excitation~\eqref{eq:WKB-local}.
The WKB is often accurate enough outside of regions of strong gradients in inhomogeneous flows~\cite{guerrero_multiply_2025}.

\paragraph{Hydrodynamic consequences:}
\begin{itemize}
    \item The effective compressibility is set by $g_{\rm eff}$ only in the adiabatic limit; otherwise it becomes frequency‑dependent via \eqref{eq:R-memory}, which damps/retards density responses and slightly shifts resonance conditions in spectroscopy. 
    \item In practice, treating $g_{\rm res}n_{\rm res}$ as a static blueshift is adequate for mean‑field profiles, while linear‑response comparisons to data should include \eqref{eq:WKB-local} and the finite $\gamma$~\cite{stepanov_dispersion_2019,claude_spectrum_2023,guerrero_multiply_2025}.
\end{itemize}

\subsection{Practical modelling and typical numbers}
\paragraph{One‑number summary of the reservoir.}  
It is convenient to parameterise $g_{\rm res}=\beta\,g$. Values $\beta\sim 0.3$–$0.7$ are typical in GaAs systems; fits to \((\hbar\omega,k)\) probe maps in near‑resonant configurations give, e.g., $\beta\simeq 0.5$ in representative devices~\footnote{See the calibration procedure and fit illustrated in SM of~\cite{falque_polariton_2025}, yielding $\beta=0.49\pm 0.02$ for the data analysed there.}.
With this convention, $g_{\rm eff}=g(1+\beta\,\gamma_{\rm in}/\gamma_{\rm res})$ in the adiabatic limit.

\paragraph{When to keep reservoir dynamics.}
Retain \eqref{eq:R-rate} explicitly (or the memory kernel \eqref{eq:R-memory}) if any of the following hold: (a) the spectral features of interest have linewidths or spacings $\lesssim\gamma_{\rm res}$; (b) slow drifts in the mean field are observed upon changing the pump, indicative of reservoir build‑up/decay; (c) fits with a static $g_{\rm eff}$ systematically require different $g$ for mean‑field and linear‑response observables.

In summary, we adopt Eqs.~\eqref{eq:R-ddGPE}–\eqref{eq:R-rate} as the reference model whenever a dark reservoir is relevant. In homogeneous settings we will often use the adiabatic reduction \eqref{eq:R-geff}; for linear‑response calculations we use the full dispersion \eqref{eq:WKB-local} and indicate when the gapless limit is enforced by tuning $gn+g_{\rm res}n_{\rm res}=\delta(v)$.

\section{Spatially inhomogeneous coherent pumping.}\label{app:inhomogeneouspumping}
Let the pump be \(F_{\mathrm p}(\mathbf r)=|F_{\mathrm p}(\mathbf r)|\,e^{i\phi_{\mathrm p}(\mathbf r)}\) at fixed frequency \(\omega_{\mathrm p}\).
When \(|\nabla \ln |F_{\mathrm p}||\xi\ll 1\) and \(|\nabla\phi_{\mathrm p}|\xi\ll 1\), the steady state admits an
adiabatic, locally phase-locked description:
\begin{equation}
\theta(\mathbf r)\simeq \phi_{\mathrm p}(\mathbf r)+\theta_0,\qquad
\mathbf v(\mathbf r)\equiv \frac{\hbar}{m^*}\nabla\theta(\mathbf r)\simeq \frac{\hbar}{m^*}\nabla\phi_{\mathrm p}(\mathbf r).
\label{eq:phase-lock}
\end{equation}
In this \emph{local-density/phase-locking approximation} (LPA), the density profile obeys the
homogeneous equation of state pointwise, with a velocity-dependent detuning~\cite{falque_polariton_2025}
\(\delta(\mathbf v)\equiv \omega_{\mathrm p}-\omega_0-\frac{m^*|\mathbf v(\mathbf r)|^2}{2\hbar}\) replacing the plane-wave detuning:
\begin{equation}
\Big(\delta(\mathbf k_\tp)-\hbar g\,n(\mathbf r)\Big)^2+\Big(\tfrac{\hbar\gamma}{2}\Big)^2
=\frac{|F_{\mathrm p}(\mathbf r)|^2}{n(\mathbf r)}\!,
\qquad
\text{(LPA: local equation of state)}.
\label{eq:local-eos}
\end{equation}
Equation~\eqref{eq:local-eos} implies a \emph{local} bistability structure for \(n(\mathbf r)\) whenever
\(\delta(\mathbf k_\tp)/\gamma\) enters the bistable regime, generalising the homogeneous case.
Numerical solutions of the driven--dissipative GPE show that, near resonance and under smooth pumping,
\(\theta(\mathbf r)\approx\phi_{\mathrm p}(\mathbf r)\) and the function \(n(\mathbf r)\) versus \(|F_{\mathrm p}(\mathbf r)|\) remains
qualitatively identical to the homogeneous equation of state~\cite{guerrero_multiply_2025}; the main new ingredient is the
Doppler shift \(\propto |\mathbf v(\mathbf r)|^2\) in \(\delta(\mathbf v)\).%
\footnote{See the plane-wave equation of state and its bistability in the homogeneous case,
and its extension to \(\delta(v)\) in the inhomogeneous case, as discussed in~\cite{guerrero_multiply_2025}, main text and Supplemental Sections II–IV.
For the homogeneous cubic state equation in Langevin/GPE form, see Eq.~(19) in~\cite{busch_spectrum_2014}.}

\emph{Validity and limitations.} The LPA breaks down where the pump has sharp features
(\(\ell\sim\xi\)), near zeros of \(|F_{\mathrm p}|\), or where the imposed phase pattern creates strong
velocity gradients (e.g., at the core of phase vortices \(\phi_{\mathrm p}=C\theta\) with \(C\in\mathds{Z}\)).
In such regions non-local current redistribution and the additional convective/divergence terms
from the hydrodynamic form of the driven--dissipative GPE must be retained and one should solve
the full equation numerically~\cite{guerrero_multiply_2025}.

\section{Effective field theory for Bogoliubov excitations in a polariton fluid}
\subsection{From the ddGPE to the Klein-Gordon equation}\label{app:metric}
\noindent\emph{This section discusses material first introduced in~\cite{falque_polariton_2025}.}

We begin from the driven-dissipative GPE for the lower polariton field~\eqref{eq:ddgpe-hydro-start} and assume the monochromatic pump~\eqref{eq:pump}.
We use the Bogoliubov ansatz inspired by the Madelung decomposition \(\psi=e^{i[\phi_0(\mathbf r)-\omega_\tp t]}\!\left(\sqrt{n_0}+e^{-i\gamma/2}\delta\psi\right)\) and compute the time derivative \[\partial t\psi=e^{i(\phi_\tp - \omega_\tp t)}[-i\omega_\tp(n\theta+\delta\psi)+\partial t \delta\psi],\]
and Laplacian
\[\nabla^2\psi=\nabla^2[(\sqrt{n_0}+\delta\psi)+\partial_t\delta\psi]. \]
We assume $\phi_\tp(x)$ varies in space (that is, has gradients) and define the velocity of the fluid as~\eqref{eq:polaritonv}.
We now apply the product rule.
\[ \nabla\psi = e^{i(\phi_\tp-\omega_\tp t)} [ i(\nabla\phi_\tp)(\sqrt{n_0}+\delta\psi) + \nabla\delta\psi ]. \]
Then
\[ \nabla^2 \psi = e{i(\phi_\tp-\omega_\tp t)} [ -(\nabla\phi_\tp)^2(\sqrt{n_0}+\delta\psi) +2i(\nabla\phi_\tp)\cdot\nabla\delta\psi+i(\nabla^2\phi_\tp)(\sqrt{n_0}+\delta\psi) + \nabla^2\delta\psi ]. \]

Now substitute everything back into Eq.~\eqref{eq:ddgpe}, drop terms beyond linear in $\delta\psi$, and collect terms multiplying $e^{i(\phi_\tp-\omega_\tp t)}$.
On the LHS, the leading-order term $i\hbar\partial_t\Phi$ becomes
\[ i\hbar\partial_t\psi = e^{i(\phi_\tp-\omega_\tp t)}[-\hbar\omega_\tp(\sqrt{n_0}+\delta\psi)+i\hbar\partial_t\delta\psi].\]
On the RHS, expand the non-linear term $g|\Phi|^2\Phi$ to linear order:
\begin{align*}
    |\psi|^2\psi&=(\sqrt{n_0}+\delta\psi)(\sqrt{n_0}+\delta\psi)^\star(\sqrt{n_0}+\delta\psi)\\
    &\approx n_0\sqrt{n_0}+2n_0\delta\psi+n_0\delta\psi^\star\\
    &\rightarrow g|\psi|^2\psi\approx gn_0\sqrt{n_0}+2gn_0\delta\psi+n_0\delta\psi^\star.
\end{align*}
The term in $F_\tp$ balances the zeroth-order steady-state solution, so what remains at linear order in fluctuations is
\begin{equation}\label{eq:A3}
i\hbar (\partial_t + \vec{v}_0 \cdot \nabla) \delta\psi = \left( -\frac{\hbar^2}{2m^*} \nabla^2 + \rho - i\sigma \right) \delta\psi + g n_0 \delta\psi^*,
\end{equation}
where
\[
\rho := 2 g n_0 - \delta(\vec{v}_0) + g_rn_r, \quad \sigma := \frac{\hbar}{2} \nabla \cdot \vec{v}_0.
\]

Solving Eq.~\eqref{eq:A3} for $\delta\psi^*$, we obtain Eq. A4:
\begin{equation}\label{eq:A4}
\delta\psi^* = \frac{1}{g n_0} \left[ i\hbar (\partial_t + \vec{v}_0 \cdot \nabla) + \frac{\hbar^2}{2m^*} \nabla^2 - \rho + i\sigma \right] \delta\psi
\end{equation}

Substitute~\eqref{eq:A4} into the complex conjugate of~\eqref{eq:A3},
\[ -i\hbar (\partial_t + \vec{v}_0 \cdot \nabla) \delta\psi^\star = \left( -\frac{\hbar^2}{2m^*} \nabla^2 + \rho + i\sigma \right) \delta\psi + g n_0 \delta\psi. \]
LHS of (\eqref{eq:A3}-conjugate):
\[ -i\hbar (\partial_t + \vec{v}_0 \cdot \nabla) \delta\psi^\star = -i\hbar(\partial_t + \vec{v}_0 \cdot \nabla)\left(\frac{1}{gn_0}[\cdot\cdot\cdot]\delta\psi\right). \]
Assuming $gn_0=cst$,
\begin{align*}
    -i\hbar (\partial_t + \vec{v}_0 \cdot \nabla) \delta\psi^\star &=-\frac{i\hbar}{gn_0}(\partial_t + \vec{v}_0 \cdot \nabla)\left(i\hbar (\partial_t + \vec{v}_0 \cdot \nabla)+\frac{\hbar^2}{2m^\star}\nabla^2-\rho+i\sigma\right)\delta\psi\\
    &=\frac{\hbar^2}{gn_0}i\hbar (\partial_t + \vec{v}_0 \cdot \nabla)^2\delta\psi+\frac{i\hbar}{gn_0}i\hbar (\partial_t + \vec{v}_0 \cdot \nabla)\left(\frac{\hbar^2}{2m^\star}\nabla^2-\rho+i\sigma\right)\delta\psi
\end{align*}
Substitute $\delta\psi^\star$ from Eq.~\eqref{eq:A4} into RHS of (\eqref{eq:A3}-conjugate):
\begin{equation*}
\begin{split}
    \left(-\frac{\hbar^2}{2m^\star}\nabla^2-\rho+i\sigma\right)\delta\psi^\star+gn_0\delta\psi=\\\left(-\frac{\hbar^2}{2m^\star}\nabla^2-\rho+i\sigma\right)\cdot\frac{1}{gn_0}\left(i\hbar (\partial_t + \vec{v}_0 \cdot \nabla)+\frac{\hbar^2}{2m^\star}\nabla^2-\rho+i\sigma\right)\delta\psi+gn_0\delta\psi
=\\ \frac{1}{gn_0}\left(-\frac{\hbar^2}{2m^\star}\nabla^2+\rho+i\sigma\right)\left(i\hbar (\partial_t + \vec{v}_0 \cdot \nabla)+\frac{\hbar^2}{2m^\star}\nabla^2-\rho+i\sigma\right)\delta\psi+gn_0\delta\psi.
\end{split}
\end{equation*}
Putting LHS and RHS together to obtain
\begin{equation}
\begin{split}
i\hbar (\partial_t + \vec{v}_0 \cdot \nabla) \frac{1}{g n_0}
\left[
i\hbar (\partial_t + \vec{v}_0 \cdot \nabla) + \frac{\hbar^2}{2m^*} \nabla^2 - \rho + i\sigma
\right] \delta\psi = \\
\left[
\left( \frac{\hbar^2}{2m^*} \nabla^2 - \rho - i\sigma \right) \frac{1}{g n_0}
\left( i\hbar (\partial_t + \vec{v}_0 \cdot \nabla) + \frac{\hbar^2}{2m^*} \nabla^2 - \rho + i\sigma \right) - g n_0
\right] \delta\psi
\end{split}
\label{eq:A5}
\end{equation}

Consider a homogeneous background --- $\nabla \cdot \vec{v}_0 = 0\rightarrow\sigma = 0$, $n_0$, $\rho$, and $\vec{v}_0$ are constant.
Define $\mathcal{D} := \partial_t + \vec{v}_0 \cdot \nabla$. Then, Eq.~\eqref{eq:A5} simplifies to:
\[
i\hbar \mathcal{D} \left[
\frac{1}{g n_0}
\left(
i\hbar \mathcal{D} + \frac{\hbar^2}{2m^*} \nabla^2 - \rho
\right)
\right] \delta\psi
=
\left[
\left( \frac{\hbar^2}{2m^*} \nabla^2 - \rho \right)
\frac{1}{g n_0}
\left(
i\hbar \mathcal{D} + \frac{\hbar^2}{2m^*} \nabla^2 - \rho
\right)
- g n_0
\right] \delta\psi
\]

Assuming $gn_0$ constant, expand the LHS:
\[
\text{LHS} = \frac{1}{g n_0} \left[
(i\hbar \mathcal{D})^2 + i\hbar \mathcal{D} \left( \frac{\hbar^2}{2m^*} \nabla^2 - \rho \right)
\right] \delta\psi
= \frac{1}{g n_0} \left[
- \hbar^2 \mathcal{D}^2 + i\hbar \left( \frac{\hbar^2}{2m^*} \mathcal{D} \nabla^2 \right)
- i\hbar \rho \mathcal{D}
\right] \delta\psi.
\]
Now expand the RHS:
\[
\text{RHS}= \frac{1}{g n_0} \left[ (\alpha - \rho)^2 + i\hbar \mathcal{D} (\alpha - \rho) \right] \delta\psi - g n_0 \delta\psi,
\]
where $\alpha := \frac{\hbar^2}{2m^*} \nabla^2$, thus $(\alpha - \rho)^2 = \left( \frac{\hbar^2}{2m^*} \nabla^2 \right)^2 - \frac{\hbar^2}{m^*} \rho \nabla^2 + \rho^2.$
Putting LHS and RHS together,
\[
- \hbar^2 \mathcal{D}^2 \delta\psi = \left[
\left( \frac{\hbar^2}{2m^*} \nabla^2 \right)^2
- \frac{\hbar^2}{m^*} \rho \nabla^2
+ \rho^2
- g^2 n_0^2
\right] \delta\psi,
\]
and dividing by $\hbar^2$,
\begin{align*}
&- \mathcal{D}^2 \delta\psi = \left[
\frac{\hbar^2}{4m^{*2}} \nabla^4
- \frac{\rho}{m^*} \nabla^2
+ \frac{\rho^2 - g^2 n_0^2}{\hbar^2}
\right] \delta\psi\\
&\rightarrow\left[
\mathcal{D}^2
+ \frac{\hbar^2}{4m^{*2}} \nabla^4
- \frac{\rho}{m^*} \nabla^2
+ \frac{\rho^2 - g^2 n_0^2}{\hbar^2}
\right] \delta\psi = 0
\end{align*}

Finally, define $\xi = \frac{\hbar}{\sqrt{m^* \rho}}$ (corresponding to Eq.~\eqref{eq:healinglength}), then
\[
\frac{\hbar^2}{4m^{*2}} \nabla^4 = \left( \frac{\xi^2}{4} \nabla^2 \right) \frac{\rho}{m^*} \nabla^2.
\]
Thus we arrive at
\begin{equation}
\left[
(\partial_t + \vec{v}_0 \cdot \nabla)^2
+ \left( \frac{\xi^2}{4} \nabla^2 - 1 \right) \frac{\rho}{m^*} \nabla^2
+ \frac{\rho^2 - g^2 n_0^2}{\hbar^2}
\right] \delta\psi = 0.
\label{eq:A6}
\end{equation}
(Note: the term $\xi^2\nabla^2/4-1$ arises from expanding the nested differential operators).

In the long-wavelength limit, $\nabla^2 \ll \xi^{-2}$ so
\[
\left( \frac{\xi^2}{4} \nabla^2 - 1 \right) \approx -1
\]
Plugging this into~\eqref{eq:A6}, we have
\[ [(\partial_t+\vec{v}_0\cdot\nabla)^2-\frac{\rho}{m^\star}\nabla^2+\frac{\rho^2-g^2n_0^2}{\hbar^2} ]\delta\psi=0. \]
Taking $c_\tb^2:=\rho/m^\star$ and $m_\text{det}^2c_\tb^2:=\rho^2-g^2n_0^2$, we arrive at the massive Klein-Gordon equation in flat spacetime with drift $\vec{v}_0$:
\begin{equation}
\left[ -(\partial_t + \vec{v}_0 \cdot \nabla)^2 + c_\tb^2 \nabla^2 - \frac{m_{\text{det}}^2 c_\tb^4}{\hbar^2} \right] \delta\psi = 0
\label{eq:A8}    
\end{equation}

Now, using Eqs.~\eqref{eq:cB} and~\eqref{eq:mdet}, Eq.~\eqref{eq:A8} becomes the covariant Klein-Gordon equation~\eqref{eq:KG-covariant}.
\subsection{Line Element and physical interpretation}\label{app:linelement}
\noindent\emph{This section discusses material first introduced in~\cite{falque_polariton_2025}.}

Expanding the drift operator
\[
(\partial_t + \vec{v}_0 \cdot \nabla)^2 = \partial_t^2 + 2 (\vec{v}_0 \cdot \nabla) \partial_t + (\vec{v}_0 \cdot \nabla)^2,
\]
Eq.~\eqref{eq:A8} becomes
\[
\left[
- \partial_t^2 - 2 (\vec{v}_0 \cdot \nabla) \partial_t - (\vec{v}_0 \cdot \nabla)^2 + c_\tb^2 \nabla^2 - \frac{m_{\text{det}}^2 c_\tb^4}{\hbar^2}
\right] \delta\psi = 0
\]

Given the general form of the d'Alembertian operator in curved spacetime,
\[
\square_q \delta\psi = \frac{1}{\sqrt{|q|}} \partial_\mu \left( \sqrt{|q|} q^{\mu\nu} \partial_\nu \delta\psi \right),
\]
the structure of Eq.~\eqref{eq:A8} suggests that $q^{\mu\nu}$ must reproduce the kinetic terms:
\begin{itemize}
    \item $q^{00} = -1$,
    \item $q^{0i} = q^{i0} = -v_0^i$,
    \item $q^{ij} = c_\tb^2 \delta^{ij} - v_0^i v_0^j$.
\end{itemize}

As the acoustic metric $g_{\mu\nu}$ (up to a conformal factor) is
\[
q_{\mu\nu} = c_\tb^2 \begin{pmatrix}
v_0^2 - c_\tb^2 & -v_0^j \\
- v_0^i & \delta_{ij}
\end{pmatrix},
\]
the line element becomes
\begin{equation}
    ds^2 = c_\tb^2 \left[
(v_0^2 - c_\tb^2) dt^2
- 2 \vec{v}_0 \cdot d\vec{x} \, dt
+ d\vec{x} \cdot d\vec{x}
\right].\label{eq:A11}
\end{equation}
Therefore, Eq.~\eqref{eq:A8} can be recast into the covariant form~\eqref{eq:KG-covariant}, where $q_{\mu\nu}$ is the acoustic metric associated with the moving fluid.
This defines the effective spacetime geometry seen by the field $\delta\psi$, with horizon formation at $v_0^2 = c_\tb^2$.
Remarks: The metric reduces to flat Minkowski space when $\vec{v}_0=0$ and $c_\tb=const$.
The term $-2\vec{v}_0\cdot d\vec{x}dt$ shows mixing of time and space, i.e., the frame is non-inertial in the laboratory.

\subsection{Open EFT for the hydrodynamic mode}\label{app:openEFT}
Let $\Phi\propto\delta\theta$ denote the scalar hydrodynamic mode (Sec.~\ref{subsubsec:KG-analogy}).
Choose the acoustic inverse metric density $q^{\mu\nu}(\mathbf r)$ of Eq.~\eqref{eq:qmunu} and define the convective derivative
$D_t\equiv\partial_t+\mathbf v_0(\mathbf r)\!\cdot\!\nabla$.
In the long‑wavelength, barotropic limit, a convenient conservative action is
\begin{equation}
S_0[\Phi]=\frac12\!\int\!dt\,d^2\mathbf r\;\sqrt{|q|}\,
\Big[q^{\mu\nu}(\mathbf r)\,\partial_\mu\Phi\,\partial_\nu\Phi
-\frac{m_{\rm det}^2(\mathbf r)}{\hbar^2}\,\Phi^2\Big],
\label{eq:KG-action}
\end{equation}
whose Euler–Lagrange equation is the covariant KG equation \eqref{eq:KG-covariant}.

Radiative losses add a friction term linear in the convective time derivative.
Introducing the Rayleigh functional
\begin{equation}
\mathcal R[\Phi]=\Gamma\!\int\!dt\,d^2\mathbf r\;\sqrt{|q|}\,\big(D_t\Phi\big)^2,
\qquad \Gamma\equiv\frac{\gamma}{2},
\label{eq:Rayleigh}
\end{equation}
the damped Euler–Lagrange equation becomes (neglecting $\mathcal O(\Gamma^2)$)
\begin{equation}
\underbrace{\frac{1}{\sqrt{|q|}}\partial_\mu\!\big(\sqrt{|q|}\,q^{\mu\nu}\partial_\nu\Phi\big)
-\frac{m_{\rm det}^2}{\hbar^2}\,\Phi}_{\text{conservative KG}}
\;+\;2\Gamma\,D_t\Phi
\;=\;\Xi.
\label{eq:KG-Langevin}
\end{equation}
The source $\Xi$ is a Langevin noise representing the input bath; for white (Markovian) noise one may take
\begin{equation}
\big\langle \Xi(\mathbf r,t)\,\Xi(\mathbf r',t')\big\rangle
=2\,\Gamma\,\mathcal N(\mathbf r)\,\delta(t{-}t')\,\delta^{(2)}(\mathbf r{-}\mathbf r') ,
\label{eq:noise-corr}
\end{equation}
with $\mathcal N(\mathbf r)$ set by the optical environment (no fluctuation–dissipation relation is assumed a priori).
For plane waves $\propto e^{i(\mathbf k\cdot\mathbf r-\omega t)}$, Eq.~\eqref{eq:KG-Langevin} gives
$\omega\to\mathbf v_0\!\cdot\!\mathbf k\pm\Omega_k - i\Gamma$, reproducing the BdG linewidth $-\gamma/2$.

Equivalently, defining classical/quantum fields $\Phi_{\rm cl},\Phi_{\rm q}$ and the retarded/advanced operators
\[
\hat{\mathcal L}_{R/A}\;\equiv\;
\frac{1}{\sqrt{|q|}}\partial_\mu\!\big(\sqrt{|q|}\,q^{\mu\nu}\partial_\nu\big)
-\frac{m_{\rm det}^2}{\hbar^2}\ \pm\ i\,2\Gamma\,D_t ,
\]
allows to write the quadratic Keldysh action
\begin{equation}
S_K=\int\!dt\,d^2\mathbf r\;\sqrt{|q|}\;\Big[\,
\Phi_{\rm q}\,\hat{\mathcal L}_R\,\Phi_{\rm cl}
+\Phi_{\rm cl}\,\hat{\mathcal L}_A\,\Phi_{\rm q}
+\;i\,\Phi_{\rm q}\,\mathcal N\,\Phi_{\rm q}\Big].
\label{eq:Keldysh}
\end{equation}
Integrating out $\Phi_{\rm q}$ enforces the Langevin equation \eqref{eq:KG-Langevin} with noise kernel $\mathcal N$.
In thermal equilibrium $\mathcal N$ and $\Gamma$ obey an FDR; for coherently pumped microcavities, $\mathcal N$ is set by the input–output vacuum/technical noise and need not satisfy an equilibrium FDR.

Equations \eqref{eq:KG-action}–\eqref{eq:Keldysh} constitute an \emph{effective field theory} for the hydrodynamic mode, valid for $k\xi\ll1$ and, if a reservoir is present, $|\omega|\ll\gamma_{\rm res}$ (barotropic limit).
Beyond this range, quantum‑pressure ($k^4$) terms and frequency‑dependent compressibility $g\!\to\!g_{\rm eff}(\omega)$ should be reinstated; they modify the retarded operator $\hat{\mathcal L}_{R/A}$ while leaving the general open‑EFT structure intact.
\section{Units, sign conventions \& typical numbers}
\label{app:units}

This appendix fixes Fourier conventions, power–spectral–density (PSD) normalisation, analysis–bandwidth usage, and lists representative GaAs microcavity parameters. We keep $\hbar$ explicit; fields are written in the pump’s rotating frame.

\subsection*{Fourier conventions (time and space)}
We use the $e^{-i\omega t}$ time dependence for physical fields. The Fourier transform pair is
\begin{equation*}
\tilde f(\omega)=\int_{-\infty}^{+\infty}\! dt\, e^{+i\omega t}\, f(t),\qquad
f(t)=\int_{-\infty}^{+\infty}\! \frac{d\omega}{2\pi}\, e^{-i\omega t}\, \tilde f(\omega).
\end{equation*}
Spatial transforms (2D in–plane) follow the same sign convention as in the field expansion,
\begin{equation*}
\tilde f(\mathbf{k})=\int d^2\mathbf{r}\, e^{-i\mathbf{k}\cdot\mathbf{r}} f(\mathbf{r}),\qquad
f(\mathbf{r})=\int \frac{d^2\mathbf{k}}{(2\pi)^2} e^{+i\mathbf{k}\cdot\mathbf{r}} \tilde f(\mathbf{k}).
\end{equation*}
Angles in the far field map to in‑plane momenta via $k_\parallel=(n_{\rm ext}\omega/c)\sin\theta$.

\subsection*{Photodetection: PSDs, shot–noise units, quadratures}
For a single optical mode of mean photocurrent $\bar i$ on a linear detector of quantum efficiency~$\eta$, we use the \emph{two‑sided} current PSD $S_i(\omega)$ (units A$^2$/Hz) defined such that the variance measured in a resolution bandwidth centred at analysis frequency $\omega_c$ is~\cite{bachor_guide_2004}
\begin{equation*}
\langle \Delta i^2\rangle \simeq 2\,\mathrm{RBW}\, S_i(\omega_c).
\end{equation*}
For a shot–noise–limited beam the \emph{two‑sided} white level is
\begin{equation*}
S^{\rm (shot)}_i(\omega)= e\,\bar i,
\end{equation*}
so $\langle \Delta i^2\rangle \simeq 2 e\,\bar i\,\mathrm{RBW}$ (consistent with the time‑domain $\langle \Delta i^2\rangle = e\bar i/\Delta t$ for an effective noise bandwidth $1/(2\Delta t)$). If a one‑sided spectrum $S^{(1)}_i(f)$ is used ($f\ge 0$), then $S^{(1)}_i(f)=2\,S_i(\omega=2\pi f)$.

Quadratures follow the convention of Sec.~\ref{subsec:photodetection}:
\begin{equation*}
\hat X=\hat a+\hat a^{\dagger},\quad \hat Y=i(\hat a^{\dagger}-\hat a),\quad [\hat X,\hat Y]=2i,
\end{equation*}
and we normalise to \emph{shot–noise units} (SNU) with ${\rm Var}(\hat X_{\rm vac})={\rm Var}(\hat Y_{\rm vac})=1$. A homodyne with LO phase~$\theta$ measures $\hat X_\theta=\hat X\cos\theta+\hat Y\sin\theta$; squeezing appears as ${\rm Var}(\hat X_\theta)<1$.

\subsection*{Analysis bandwidths (instrument settings)}
On spectrum analysers we use: centre frequency $f_c=\omega_c/2\pi$ in the $10^6$Hz range, $\mathrm{RBW}$ in the $10^4$–$10^5$Hz range, and $\mathrm{VBW}$ in the $10^4$-$10^5$Hz range and sweep times $\sim\!0.1$\,s.
\newpage
\newtcolorbox{numberbox}[1][]{
  enhanced, breakable,
  colback=white, colframe=black!50,
  boxrule=0.5pt, arc=1pt,
  left=1em, right=1em, top=0.6em, bottom=0.6em,
  fonttitle=\bfseries, title={Representative GaAs microcavity numbers (order of magnitude)},
  #1
}
\begin{numberbox}
Device‑dependent (here we use values of the sample of experiments~\cite{maitre_dark_soliton_2020,claude2021highresolution,claude_spectrum_2023,falque_polariton_2025,guerrero_multiply_2025}); quoted at low in‑plane $k$ and near zero exciton-photon detuning.

\medskip
\begin{tabular}{@{}ll@{}}
\textbf{Photon mass} $m_{\rm ph}$ & $10^{-5}-10^{-4}\,m_e$ \\
\textbf{LP mass} $m^\ast$ & $(1-5)\times 10^{-5}\,m_e$ (depending on $|C_0|^2$) \\
\textbf{Rabi splitting} $\hbar\Omega_\mathrm{R}$ & $3-\SI{15}{\milli\electronvolt}$ (multi‑QW boosts upper end) \\
\textbf{LP linewidth} $\hbar\gamma$ & $0.05-\SI{0.2}{\milli\electronvolt}$ ($\Rightarrow$ intensity lifetime $1/\gamma \sim 3-\SI{13}{\pico\second}$) \\
\textbf{LP–LP interaction} $\hbar g$ & $0.1-\SI{3}{\micro\electronvolt\micro\meter\squared}$ (GaAs; detuning dependent) \\
\textbf{Density} $n_0$ & $1-\SI{10}{\per\micro\meter\squared}$ (steady, near‑resonant drive) \\
\textbf{Healing length} $\xi$ & $1-\SI{5}{\micro\meter}$ \\
\textbf{Acoustic speed} $c_\tb$ & $0.5-\SI{2}{\micro\meter\per\pico\second}$ \\
\textbf{Surface gravity} $\kappa$ & $0.02-\SI{0.2}{\per\pico\second}$ (typical transcritical shears) \\
\end{tabular}
\end{numberbox}
\noindent\textbf{Notes.} (i) We define the loss term as $-i\hbar\gamma/2$ in the ddGPE~\eqref{eq:ddgpe}, so the intensity lifetime is $1/\gamma$; the amplitude decays at rate $\gamma/2$. (ii) Away from resonance a finite low‑$k$ gap appears~\eqref{eq:mdet}; when quoting $c_\tb$~\eqref{eq:cB} and $\xi$~\eqref{eq:healinglength}, specify the operating detuning~\eqref{eq:detuning}. (iii) For camera‑based covariance, bandwidths are set by exposure time and optics; for RF homodyne, by $\mathrm{RBW}$/$\mathrm{VBW}$.
(iv) Record the detector transimpedance to convert between V$^2$/Hz and A$^2$/Hz when needed.

\end{document}